\documentclass[aps,prd,twocolumn,superscriptaddress,amsmath,amssymb,amsfonts,nofootinbib]{revtex4-2}

\newcommand{\aap}{A\&A}

\usepackage[varg]{txfonts}
\usepackage{lineno}
\usepackage{bbold}
\usepackage{bm}
\usepackage{graphicx}
\usepackage{float}
\usepackage{amssymb}
\usepackage{amsfonts}
\usepackage{amsmath}
\usepackage{float}
\usepackage{xspace}
\usepackage[english]{babel}
\makeatletter
\adddialect\l@en\l@english
\makeatother

\usepackage{hyperref}
\usepackage{longtable}
\usepackage{dirtytalk}
\usepackage{multirow}
\usepackage{hhline}
\usepackage{soul}
\usepackage{booktabs}
\usepackage{orcidlink}

\usepackage{mathtools}

\newcommand{\lint}{\lambda_{\rm int}}

\def\helens{{\rm HE}0435$-$1223\xspace}
\def\pglens{{\rm PG}1115$+$080\xspace}
\def\wfilens{{\rm WFI}2033$-$4723\xspace}

\usepackage{physics}

\newcommand{\e}[1]{\texttt{#1}}

\newcommand{\hunit}{km s$^{-1}$ Mpc$^{-1}$}
\newcommand{\hval}{73.5$_{-2.8}^{+2.7}$}
\newcommand{\hvalpg}{71.8$_{-7.0}^{+9.2}$}
\newcommand{\hvalhe}{74.2$_{-4.2}^{+4.2}$}
\newcommand{\hvalwfi}{73.4$_{-4.4}^{+3.4}$}
\newcommand{\hvalold}{73.6$_{-2.6}^{+2.6}$}

\usepackage{xcolor}
\usepackage{enumitem}

\usepackage[normalem]{ulem}

\setlist[enumerate,1]{label=\arabic*., left=0.25cm}

\usepackage[nameinlink]{cleveref}
\usepackage{cleveref}
\crefname{section}{Sect.}{Sects.}
\Crefname{section}{Sect.}{Sects.}

\crefname{subsection}{Sect.}{Sects.}
\Crefname{subsection}{Sect.}{Sects.}

\crefname{subsubsection}{Sect.}{Sects.}
\Crefname{subsubsection}{Sect.}{Sects.}

\crefname{figure}{Fig.}{Figs.}
\Crefname{figure}{Fig.}{Figs.}

\crefname{table}{Table}{Tables}
\Crefname{table}{Table}{Tables}

\crefname{equation}{Eq.}{Eqs.}
\Crefname{equation}{Eq.}{Eqs.}

\newcommand{\farcs}{.\!\!^{\prime\prime}}
\newcommand{\degr}{^\circ}

\crefname{appendix}{Appendix}{Appendices}
\Crefname{appendix}{Appendix}{Appendices}

\usepackage[thinc]{esdiff}

\usepackage{textcomp}
\usepackage{placeins}

\usepackage{threeparttable}

\newcommand{\kms}{km\,s$^{-1}$}
\let\oldAA\AA
\renewcommand{\AA}{\textup{\oldAA}}

\newcommand{\lowup}[2]{\raisebox{0.5ex}{\tiny$^{+#2}_{-#1}$}}

\begin{document} 
	
	\title{TDCOSMO. XXVII. JWST-based Lens Models and H$_0$ Measurement of WFI2033, HE0435, and PG1115}
    \author{D.~M.~Williams\,\orcidlink{0000-0002-8386-0051}}
    \email{devon@astro.ucla.edu}
    \author{T.~Treu\,\orcidlink{0000-0002-8460-0390}}
    \affiliation{Department of Physics and Astronomy, University of California, Los Angeles, CA 90095, USA}

    \author{D.~P.~Johnson\,\orcidlink{0000-0002-0311-2513}}
    \affiliation{Laboratoire Univers et Particules de Montpellier (LUPM), CNRS \& Universit\'e Montpellier (UMR-5299), Parvis Alexander Grothendieck, F-34095 Montpellier Cedex 05, France}

    \author{P.~Mozumdar\,\orcidlink{0000-0002-8593-7243}}
    \author{S.~Knabel\,\orcidlink{0000-0001-5110-6241}}
    \affiliation{Department of Physics and Astronomy, University of California, Los Angeles, CA 90095, USA}

    \author{S.~Birrer\,\orcidlink{0000-0003-3195-5507}}
    \affiliation{Department of Physics and Astronomy, Stony Brook University, Stony Brook, NY 11794, USA}

    \author{C.~D.~Fassnacht\,\orcidlink{0000-0002-4030-5461}}
    \affiliation{Department of Physics and Astronomy, UC Davis, 1 Shields Ave., Davis, CA 95616 USA}

    \author{A.~Galan\,\orcidlink{0000-0003-2547-9815}}
    \affiliation{Department of Astronomy, University of Geneva, ch. d’Ecogia 16, 1290 Versoix, Switzerland}

    \author{A.~J.~Shajib\,\orcidlink{0000-0002-5558-888X}}
    \affiliation{Department of Astronomy \& Astrophysics, University of Chicago, Chicago, IL 60637, USA}
    \affiliation{Kavli Institute for Cosmological Physics, University of Chicago, Chicago, IL 60637, USA}
    \affiliation{Center for Astronomy, Space Science and Astrophysics, Independent University, Bangladesh, Dhaka 1229, Bangladesh}

    \author{K.~C.~Wong\,\orcidlink{0000-0002-8459-7793}}
    \affiliation{Research Center for the Early Universe, Graduate School of Science, The University of Tokyo, 7-3-1 Hongo, Bunkyo-ku, Tokyo 113-0033, Japan}

    \author{M.~Cappellari\,\orcidlink{0000-0002-1283-8420}}
    \affiliation{Sub-Department of Astrophysics, Department of Physics, University of Oxford, Denys Wilkinson Building, Keble Road, Oxford OX1 3RH, UK}

    \author{F.~Courbin\,\orcidlink{0000-0003-0758-6510}}
    \affiliation{ICC-UB Institut de Ci\`encies del Cosmos, Universitat de Barcelona, Mart\'i Franqu\`es 1, 08028 Barcelona, Spain}
    \affiliation{Instituci\'o Catalana de Recerca i Estudis Avan\c{c}ats (ICREA), Pg.\ Llu\'is Companys 23, 08010 Barcelona, Spain}
    \affiliation{Institut d'Estudis Espacials de Catalunya (IEEC), Edifici RDIT, Campus UPC, Castelldefels, 08860 Barcelona, Spain}

    \author{T.~Morishita\,\orcidlink{0000-0002-8512-1404}}
    \affiliation{Astronomical Institute, Tohoku University, 6-3 Aramaki, Aoba-ku, Sendai 980-8578, Japan}

    \author{V.~Motta\,\orcidlink{0000-0003-4446-7465}}
    \affiliation{Instituto de F\'{\i}sica y Astronom\'{\i}a, Universidad de Valpara\'{\i}so, Avda. Gran Breta\~na 1111, Valpara\'{\i}so, Chile}
    
    \author{D.~Sluse\,\orcidlink{0000-0001-6116-2095}}
    \affiliation{STAR Institute, University of Li\`ege, Quartier Agora, All\'ee du six Ao\^ut 19c, 4000 Li\`ege, Belgium}

   \author{M.~Stiavelli\,\orcidlink{0000-0001-9935-6047}}
    \affiliation{Space Telescope Science Institute, Baltimore, MD 21218, USA}

	\date{Received 00 Month 0000; accepted 00 Month 0000}
	
	\begin{abstract}
		Time-delay cosmography offers a one-step, distance-ladder-independent route to the Hubble-Lemaître constant, H$_0$. In this work, we present new cosmography-grade lens models of three quadruply imaged quasars based on JWST-NIRCam/F115W imaging (\wfilens, \helens, \pglens). We use the STARRED modeling technique, introduced in our previous analysis of \wfilens, to reconstruct the complex JWST-NIRCam Point Spread Function (PSF) at high fidelity. We then combine the NIRCam-based lens models with improved external convergence estimates, published time delays, and aperture-integrated stellar velocity dispersions measured from JWST NIRSpec, to infer the Hubble constant H$_0$. The analysis was carried out blindly for \helens and \pglens, while it was not blind for \wfilens, as we build upon the previous published model. For comparison with previous HST-based work, we limit our analysis to the case of no internal mass-sheet degeneracy ($\lint=1$). We quantify the impact of improved imaging, single-aperture kinematics, and environment measurements on both central values and uncertainties, comparing to prior HST-based inferences for these systems. Within flat $\Lambda$ cold dark matter (CDM) and assuming a uniform prior on $\Omega_{\rm m}$,  we find H$_0$ = \hvalpg $\lint$ \hunit\ for \pglens, \hvalhe $\lint$ \hunit\ for \helens, and \hvalwfi $\lint$ \hunit\ for \wfilens. Combining the three lenses yields H$_0$=\hval $\lint$ \hunit, consistent with the value obtained from HST imaging (\hvalold$\lint$ \hunit), but with reduced scatter between the three systems. Our JWST-NIRCam lens models will be incorporated in the next TDCOSMO-2026 milestone paper, in which we will carry out the analysis with free $\lint$ in a hierarchical fashion. We close by situating these results in the current TDCOSMO landscape and outlining how 16 forthcoming JWST NIRCam targets (11 also having NIRSpec) will further tighten uncertainties and drive toward percent-level precision on H$_0$.
	\end{abstract}


	\maketitle
	
	\section{Introduction}
	\label{sec:intro}
        
	While the standard cosmological model, \text{$\Lambda \text{CDM}$}, has been able to reproduce a wide number of cosmological phenomena over the years, evidence in opposition to this model has steadily grown \citep[for a comprehensive review, see e.g.,][]{divalentino2025a}. A core prediction for cosmological models is the current expansion rate of the Universe, H$_0$, and over the years a troubling discrepancy has only grown worse: our prediction for how quickly the Universe should be expanding, based on measurements of the early Universe, does not agree with what we actually observe today. Between this conflict (known as the ``Hubble tension''), suggestions of dynamical dark energy from the Dark Energy Spectroscopic Instrument (DESI) \citep{adame2025}, $\sigma_8$ tension from cosmic shear measurements \citep{descollaboration2026, dalal2023}, and questions whether early or dynamical dark energy could solve this tension \citep{vagnozzi2023}, the standard model appears challenged on multiple fronts \citep{turner2025}. Therefore, understanding if these tensions truly exist or are the result of unknown systematics is crucial for determining the next steps in cosmology.
	If the Hubble tension is independently confirmed, these next steps would require adjustments if not a complete revision of the standard model and our understanding of the Universe today \citep[e.g.,][]{knox2020, efstathiou2021,divalentino2025}.
	
	There are two approaches for confirming the Hubble tension. The first involves ``dotting our i's and crossing our t's'': checking and re-checking cosmological probes for  unaccounted or underestimated systematics. But after two decades of intense scrutiny, no significant issue has been confirmed \citep{riess2019,  riess2021, riess2022, dainotti2021, mortsell2022, h0dncollaboration2026}. To bring the Hubble tension to a close, we must rely on the second approach: making multiple independent measurements with probes of comparable precision.
	
	Time-delay cosmography offers this measurement of H$_0$ in one step, independent of early- and late-Universe measurements. First proposed by \citet{refsdal1964}, this method does not depend on the local distance ladder, the typical assumption of other late-Universe measurements, or pre-recombination physics (such as Cosmic Microwave Background measurements) which are core to early-Universe measurements. We focus on lensed quasars, specifically quadruply imaged quasars (hereafter quads), where light from these quasars is bent into four distinct images around a galaxy in between us, called the lensing galaxy. Quads are particularly valuable as they offer additional time delay information along with tight constraints on the lensing galaxy's mass distribution for accurate measurements of H$_0$.
	
	Four key ingredients are needed to make these measurements of H$_0$: time delays, imaging data for strong lensing mass models, wide-field data to constrain external convergence, and spectroscopic data for redshifts and stellar kinematics. Over the past few years, significant progress has been made on all fronts. The COSMOGRAIL project has overhauled our approach for time-delay measurements \citep[e.g.,][]{millon2020, bonvin2019, millon2020a, dux2025}. The methodology to measure the external convergence along each lens's line of sight has been improved \citep[e.g.,][Johnson et al. 2026, in prep]{greene2013a,rusu2017a,wells2023,wells2024}. Stellar kinematic measurements have improved both in methodology \citep[e.g.,][]{knabel2025b} and data quality \citep{shajib2026}. Integral field spectroscopy has delivered unprecedented measurements of spatially resolved kinematics to clamp down on residual uncertainties due to internal degeneracies in the systems \citep{shajib2018, yildirim2020, shajib2023, shajib2026, paic2026, sheu2026}.
	
	The primary goal of this paper is to improve the quality of strong lensing mass models by exploiting the advances in image resolution and signal-to-noise ratio afforded by JWST/NIRCam over HST \citep{suyu2010} and AO-assisted ground-based images \citep{chen2019}, building on previous work by our collaboration \citep{williams2025}.  
    	
    The first JWST/NIRCam modeling for time-delay cosmography was done by \citet{williams2025}, modeling \wfilens. This work focused on reconstruction of the Fermat potential and compared it with previous HST-based results, finding: JWST/NIRCam improved the precision of the Fermat potential difference (the key lens-model quantity entering the time-delay cosmographic inference) by $22\%$; the results were consistent within 1$\sigma$, showing no significant instrument-related systematic differences; and new PSF-modeling techniques enabled a $71\%$ increase in precision of the astrometry compared to HST and previous techniques. Interestingly, the shift in Fermat potential differences between the JWST- and HST-based models was almost entirely explained by an improved characterization of a satellite near the main deflector. The mass of this satellite could not be well constrained by the HST data, while JWST's increased resolution and sensitivity in the ring structure provided tight constraints. In the long term, this increase in sensitivity and resolution not only benefits previously analyzed systems, but also enables the exploitation of new and already-known quads that are too faint to model with HST data.
    
    This paper extends the lensing analysis of NIRCam images to the three systems (\helens, \pglens, and \wfilens) imaged by GTO program 1198 (PI: Stiavelli). It combines the new determination of the Fermat potential differences with improvements in external convergence (Johnson et al. 2026, in prep) and stellar velocity dispersion measurements (Knabel et al. 2026, in prep) to improve the determination of H$_0$ for these systems. While the analysis of \wfilens\ was not blinded by \citet{williams2025} to enable detailed investigations of the new data processing method for JWST, the analysis of the other two systems is carried out blindly with respect to H$_0$ and the power-law slope. Unblinding occurred on 15 July 2026 for PG1115 and 29 July 2026 for HE0435\footnote{After unblinding a coding error was discovered in the squirrel software used for kinematic measurements as part of the internal review process of the companion paper by Knabel et al.\ (2026). This coding error does not affect in any way the lens models presented here, but it affects the postprocessing weight dictated by the stellar velocity dispersions. We have updated all the relevant numbers and plots after fixing the bug. For transparency, we report in Appendix~\ref{sec:appendix_outdated} the version of Figure~\ref{fig:H0_update_comparison} obtained after unblinding prior to fixing the bug.}. After unblinding, we carried out a detailed comparison with the previous measurements by our team based on these systems, quantifying the impact of improved imaging, single-aperture kinematics, and improved characterization of the line-of-sight environment on best-fit values and uncertainties.

    \begin{figure*}[ht]
        \centering
        \includegraphics[width=\textwidth]{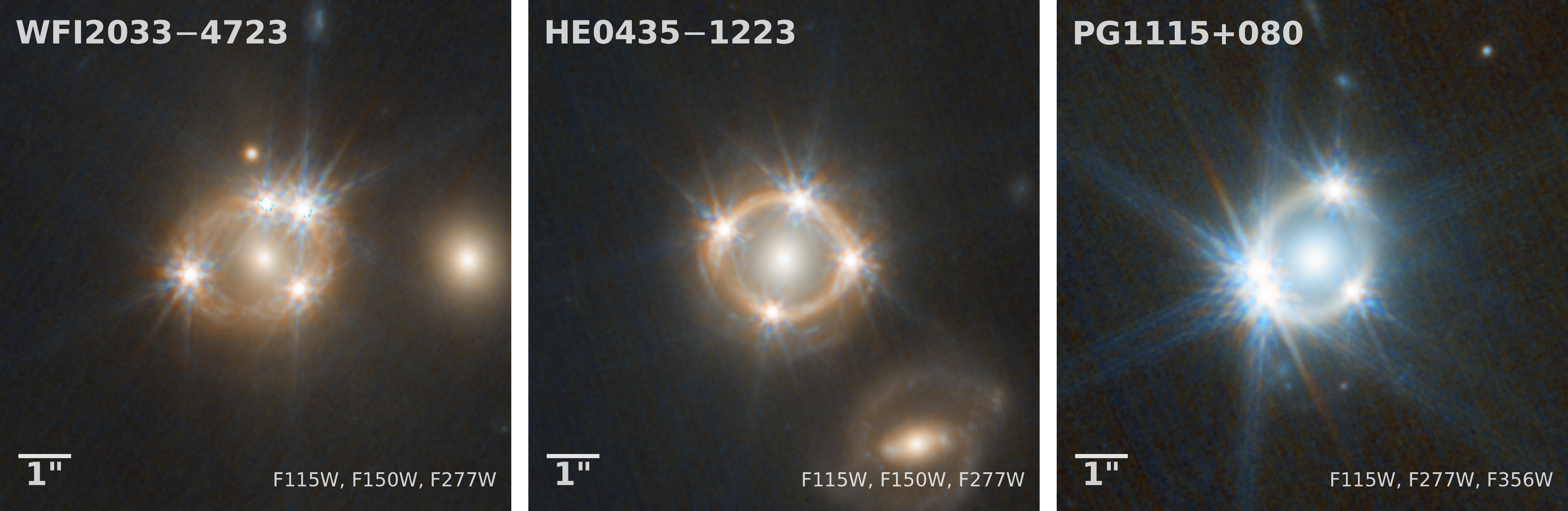}
        \caption{JWST NIRCam color images of WFI2033, HE0435, and PG1115 (GTO Program \#1198; PI: Stiavelli), the three systems modeled here. Blue arcs can be seen in the extended source light of \wfilens and \helens. These are magnified regions, likely star-forming, in the lensed quasar host galaxies (at redshifts of 1.662 and 1.693, respectively), which our models find on scales $\sim 100–500$ pc wide and $\sim 2–3$ kpc from the center of the host galaxy (\cref{fig:example_fit_HE}).}
        \label{fig:wfi2033_color_image}
    \end{figure*}

    In this paper, we adopt a single family of mass models: power-law mass models. This choice is equivalent to fixing the internal mass-sheet parameter to $\lambda_{\rm int}=1$ in the notation of \citet{collaboration2025}. Constraining $\lint$ requires combining lens models with more detailed kinematic maps and population-level information for these systems. This will be addressed in the upcoming TDCOSMO-2026 compilation of time-delay analyses, which will incorporate spatially resolved kinematics from JWST/NIRSpec observations, rather than the aperture-integrated measurements used in this work, together with axisymmetric dynamical models. We conclude by discussing the potential for improvement in time-delay cosmography afforded by approved JWST-NIRCam programs in Cycle 4 (GO-7184; PI: Millon) and 5 (GO-9637; PI: Treu).

	The paper is organized as follows. First, in \Cref{sec:ana}, we review the theory of our analysis. In \Cref{sec:data}, we present the data products used in the analysis of the three systems. In \Cref{sec:mod}, we describe the modeling choices made and systematic tests performed. In \Cref{sec:kappa_ext}, we briefly outline an update in our external convergence constraints from the lenses' environments that will be presented in full in a companion publication (Johnson et al 2026, in prep). In \Cref{sec:comb}, we describe how we combine the model variants for each system, weighting them by their goodness of fit to the imaging data and external kinematic constraints. In \Cref{sec:results_and_discussion}, we present the results for each system and compare them to previous analyses. Finally, we conclude with a summary of our findings in \Cref{sec:conclusion}.
	
	When necessary, we adopt a standard flat \text{$\Lambda \text{CDM}$} cosmology and sample $\Omega_{\rm m}\sim U(0.05, 0.5)$ for a fair comparison to the previous analysis by \citealt{chen2019}. We will consider a broader range of cosmologies and more informative priors on $\Omega_{\rm m}$ when compiling all systems together for the TDCOSMO-2026 paper, as done in our TDCOSMO-2025 paper \citep[][hereafter TDCOSMO25]{collaboration2025}.

	\section{Outline of Analysis}
	\label{sec:ana}
		
	This work improves the H$_0$ measurements for three quadruply-imaged quasars (\wfilens, \helens, \pglens) by incorporating JWST/NIRCam imaging, JWST/NIRSpec single-aperture kinematics, and updated environment analyses and propagating all these ingredients to H$_0$. The modeling approach follows our previous JWST analysis of \wfilens \citep{williams2025}, extended to two additional systems. We (i) summarize theory and observables in \Cref{ssec:ana_theory}, (ii) factor in the corrections to address degeneracies in time-delay cosmography in \Cref{ssec:corrections}, and (iii) present the Bayesian framework that delivers H$_0$ in \Cref{ssec:ana_bayes}.
	
    \subsection{Base theory and observables}
	\label{ssec:ana_theory}
	
	We extend the analysis of \citealt{williams2025} from the Fermat-potential differences $\Delta\tau$ to the full H$_0$ inference. Notation can be found in \Cref{tab:notation}. For an in-depth derivation, see recent reviews such as \citealt{treu2016}, \citealt{suyu2018}, or \citealt{birrer2024}.
	
	\begin{table}
		\footnotesize
		\caption{Notation used in \Cref{ssec:ana_theory}.}
		\label{tab:notation}
		\begin{tabular}{l l p{4.5cm}}
			\toprule
			\textbf{Symbol} & \textbf{Units} & \textbf{Meaning / Definition} \\
			\midrule
			\multicolumn{3}{l}{Time-delay cosmography theory} \\
			\midrule
			$\bm{\theta},\,\bm{\beta}$ & arcsec & Image-plane and source-plane angular positions.\\
			& & \makebox[\linewidth]{\(\bm{\beta}=\bm{\theta}-\bm{\alpha}(\bm{\theta})\)}\\
			$\bm{\alpha}(\bm{\theta})$ & arcsec & Reduced deflection angle.\\
			& & \makebox[\linewidth]{\(\bm{\alpha}(\bm{\theta})=\nabla\psi(\bm{\theta})\)}\\
			$\psi(\bm{\theta})$ & arcsec$^2$ & Two-dimensional lensing potential.\\
			& & \makebox[\linewidth]{\(\nabla^2\psi(\bm{\theta})=2\,\kappa(\bm{\theta})\)}\\
			$\kappa(\bm{\theta})$ & — & Convergence; surface mass density scaled by the critical density.\\
			& & \makebox[\linewidth]{\(\kappa=\Sigma/\Sigma_{\rm crit}\)}\\
			$\Sigma,\ \Sigma_{\rm crit}$ & ${\rm M_\odot\,Mpc^{-2}}$ & Surface mass density and the critical surface density.\\
			& & \makebox[\linewidth]{\(\displaystyle \Sigma_{\rm crit}=\frac{c^2 D_{\rm s}}{4\pi G D_{\rm d} D_{\rm ds}}\)}\\
			$D_{\rm d},\,D_{\rm s},\,D_{\rm ds}$ & Mpc & Angular-diameter distances to deflector, to source, and between them. \\
			$z_{\rm d}$ & — & Deflector (lens) redshift. \\
			$\tau$ & arcsec$^2$ & Fermat potential. \\
			$\Delta\tau_{ij}$ & arcsec$^2$ & Fermat-potential difference between images $i$ and $j$. \\
			$\Delta t_{ij}$ & days & Measured time delay between images $i$ and $j$ (speed of light, $c$, in units of Mpc/days).\\
			& & \makebox[\linewidth]{\(\Delta t_{ij}=\dfrac{D_{\Delta t}}{c}\,\Delta\tau_{ij}\)}\\
			$D_{\Delta t}$ & Mpc & Time-delay distance. \\
			\midrule
			\multicolumn{3}{l}{Kinematics and environment corrections} \\
			\midrule
			$\kappa_{\rm ext}$ & — & External convergence (LOS correction; rescales $\Delta\tau$ and links $\sigma$’s).\\
			$I(R)$ & — & Surface-brightness profile used for kinematics projection. \\
            $l(r)$ & — & Deprojected luminosity density.\\
			$\Phi(r),\,M(r)$ & — & Gravitational potential and enclosed 3D mass. \\
			$\sigma_{\rm ap,obs}$ & \kms & Observed (spectroscopic) aperture-averaged LOS velocity dispersion. \\
			$\sigma_{\rm ap,model}$ & \kms & Model-predicted (seeing-convolved) aperture-averaged LOS velocity dispersion.\\
			& & \makebox[\linewidth]{\(\sigma_{\rm ap,obs}^2=(1-\kappa_{\rm ext})\,\sigma_{\rm ap,model}^2\)}\\
			$\mathcal{S}$ & — & PSF/seeing kernel used for image/kinematics convolution. \\
			\bottomrule
		\end{tabular}
	\end{table}
	
	The lens model computes expected geometric and gravitational time delays, governed by the Fermat potential
    \begin{equation}\label{eq:tau}
		\tau (\bm{\theta}, \bm{\beta}) \equiv \left[ \frac{(\bm{\theta} - \bm{\beta})^2}{2} - \psi(\bm{\theta}) \right].
	\end{equation}
	
	Given a Fermat potential, the excess time delay between images $i$ and $j$ is described by
	\begin{equation}
		\Delta t_{ij} \;=\; \frac{D_{\Delta t}}{c}\,\Delta\tau_{ij},
		\label{eq:excess_td}
	\end{equation}
    with the speed of light $c$.
	This is a rescaling of the Fermat potential difference between the images ($\Delta\tau_{ij}$) by a parameter inversely proportional to H$_0$, called the time-delay distance:
	\begin{equation}\label{eq:ddt}
		D_{\Delta t} \equiv (1 + z_\text{d})
		\frac{D_\text{d} D_\text{s}} {D_\text{ds}}.
	\end{equation}
	Because $D_{\Delta t} \propto \mathrm{H}_0^{-1}$, after taking into account two required corrections for the environment and kinematics of the system, one can assume a cosmological model to constrain H$_0$ from measurements of $D_{\Delta t}$.
	
	Any galaxy or object along the line of sight (LOS) sufficiently massive will have an impact on the observed time delays, and these objects need not lie in the lens plane of the main deflector. To fully encapsulate these effects, we use the multi-plane lens equation (e.g., \citealt{blandford1986, kovner1987, schneider1992, collett2014, mccully2014}). For our previous assumption of \Cref{eq:excess_td} and \Cref{eq:ddt}, we define the effective time-delay distance as that between the main deflector and the source plane, obtained from the full multi-plane evaluation \citep[e.g.,][]{wong2017, rusu2020, chen2019, shajib2020}. Perturbers that are exceptionally massive or close to the deflector are explicitly included in the model, while residual effects of inhomogeneities along the LOS are accounted for with estimates of the external convergence.

	\subsection{Mass-sheet degeneracy and corrections}
	\label{ssec:corrections}
	
	The mass-sheet degeneracy \citep[MSD][]{falco1985} can be broken up into two parts: an external portion, attributable to the LOS structure and local environment of the lens, and an internal portion that physically corresponds to departures from the parametrization of the mass density profile assumed in the lens model of the main deflector. They stem from the same principle: mathematically, strong lensing configurations only constrain the surface mass density $\kappa$, up to an unknown scaling factor $\lambda$, $\kappa_{\lambda}=\lambda \kappa + (1-\lambda)$.
	
	\paragraph{External MSD:} Mass external to the primary deflector still contributes to the lensing configuration of the system. This contribution cannot be fully constrained without auxiliary data and is called the ``external'' MSD, which we must correct for. We use a term describing all the convergence resulting from the line of sight, called the ``external convergence'', which is a multiplicative corrective factor on the modeled time-delay distance:
	\begin{equation}
		D_{\Delta t} \;=\; \frac{D_{\Delta t}^{\rm model}}{1-\kappa_{\rm ext}}.
	\end{equation}
    $\kappa_{\rm ext}$ also connects the observed and model-predicted single-aperture dispersions of the main deflector,
	\begin{equation}
		\sigma_{\rm ap,obs}^2 \;=\; (1-\kappa_{\rm ext})\,\sigma_{\rm ap,model}^2,
		\label{eq:sigma_kext}
	\end{equation}
	following \citet{shajib2018}.
	
	As the external convergence cannot be directly observed, it is estimated by matching observable tracers in wide-field data to simulations within which convergence maps for different LOS $\kappa_{\rm ext}$ have already been produced. The set of statistical measures used as tracers has evolved over the years, such as number counts of galaxies \citep{suyu2010b,fassnacht2011} and weighted statistics such as redshift and angular separation of galaxies \citep{greene2013a}. Details of our measurement process and updates are described briefly in \Cref{sec:kappa_ext}, and in detail by Johnson et al. (2026, in prep), chiefly removing model-predicted external shear from these weights, drawing comparison fields from the Euclid Flagship simulation \cite{Castander_2025}, and computing the full redshift-dependent estimate of $\kappa_{\rm ext}$ in terms of the observer-source, observer-lens and lens-source terms $\kappa_{\rm s}$, $\kappa_{\rm d}$ and $\kappa_{\rm ds}$, instead of assuming $\kappa_{\rm ext}\simeq \kappa_{\rm s}$.

	\paragraph{Internal MSD:}  This can be used to describe and characterize mismatches between the model's assumed total mass-density profile of the main deflector vs.\ the true profile \citep{schneider2013, birrer2020}. This modeling degeneracy allows an infinite family of models to reproduce the same image positions, despite changing the mass distribution of the system:
	\begin{equation}
		D_{\Delta t} \;=\; \frac{D_{\Delta t}^{\rm model}}{\lambda_{\rm int}},
		\label{eq:int_mst}
	\end{equation}
    with the same impact on the observed and modeled velocity dispersions as \Cref{eq:sigma_kext}.
	
	These changes to the mass density profile require changing the mass normalization and local logarithmic slope near the Einstein radius; both can be constrained by non-lensing observables such as the deflector's aperture-averaged stellar velocity dispersion $\sigma_{\rm ap}$ (modulo anisotropy and deprojection assumptions). To facilitate comparison with previous analyses \citep[e.g.,][]{rusu2020, chen2019, bonvin2017}, we adopt a power-law elliptical mass distribution (PEMD) model family, which corresponds to fixing the internal MST ($\lambda_{\rm int}=1$) by construction. Any residual departure of the true mass profile from this family would appear as an \emph{effective} $\lambda_{\rm int}\neq 1$.
	
	\paragraph{Correction for H$_0$:} Taking both effects into account, we can correct our modeled Fermat potential to get the true value,
	\begin{equation}
		\Delta\tau_{ij}^{\rm true} \; = \; \Delta\tau_{ij}^{\rm model}\lambda_{\rm int}\,(1-\kappa_{\rm ext}).
		\label{eq:tau_true}
	\end{equation}
	With this correction, we can measure H$_0$ through the time-delay distance. From \Cref{eq:excess_td} and \Cref{eq:tau_true},
	    \begin{equation}
        D_{\Delta t}
        =
        \frac{c\ \Delta t_{ij}}
        {\lambda_{\rm int}(1-\kappa_{\rm ext})
        \Delta\tau_{ij}^{\rm model}} .
        \label{eq:td_subbed}
    \end{equation}
    In the single-plane formalism, this inference of $D_{\Delta t}$ is independent of the assumed background cosmology. To translate this distance posterior into a posterior on H$_0$, we compare $D_{\Delta t}$ to the flat $\Lambda$CDM prediction
    \begin{equation}
        D_{\Delta t}^{\rm pred}
        =
        \frac{c}{\mathrm{H}_0}
        \underbrace{\Big[(1+z_{\rm d}),
        \frac{\hat{D}_{\rm d}\hat{D}_{\rm s}}{\hat{D}_{\rm ds}}\Big]}_
        {\mathcal{J}(z_{\rm d},z_{\rm s};\Omega)} .
        \label{eq:ddt_fact}
    \end{equation}
    Here $\hat{D}\equiv (\mathrm{H}_0/c)D$ denotes the Hubble-free angular-diameter distance. Thus $\mathcal{J}$ depends on the lens and source redshifts and on the remaining cosmological parameters $\Omega$, but not on H$_0$. In practice, we marginalize over $\Omega$ and map each corrected $D_{\Delta t}$ sample to the corresponding H$_0$.
	
	\subsection{Kinematics analysis}
	\label{ssec:ana_kinematics}
    We mitigate the effects of the mass-sheet degeneracy by incorporating the luminosity-weighted projected stellar velocity dispersion, which provides an independent probe of the deflector's three-dimensional mass distribution \citep{treu2002a, shajib2018, shajib2023, knabel2025a,sheu2026}. Following \citealt{williams2025}, for each lens mass model we compute the corresponding gravitational potential $\Phi(r)$ and tracer luminosity density $l(r)$ under the assumed symmetry. We then solve the spherical Jeans equation,
    \begin{equation}
        \diff{\left( l(r) \sigma_\text{r}^2(r)\right)}{r}
        + \frac{2 \beta_\text{ani}(r) l(r) \sigma_\text{r}^2(r)}{r}
        =
        -l(r) \diff{\Phi(r)}{r},
    \end{equation}
    where $\sigma_\text{r}(r)$ is the radial velocity dispersion, $\sigma_\text{r}(r)$ is the velocity dispersion orthogonal to the radius, which is assumed to be constant in any direction by symmetry, and $\beta_\text{ani}(r) \equiv 1-\sigma_\text{t}^2(r)/\sigma_\text{r}^2(r)$ is the velocity anisotropy parameter. The Jeans solution is then projected along the line of sight and luminosity-weighted over the spectroscopic aperture to obtain the model-predicted aperture velocity dispersion, which is compared directly to the observed velocity dispersion.

	However, the projected velocity dispersion depends on both the gravitational potential and the orbital anisotropy, so the mass profile and $\beta_{\rm ani}(r)$ cannot be uniquely determined from the velocity dispersion alone. This is known as the Mass Anisotropy Degeneracy (MAD) \citep{binney1982, merritt1985}. To address this, we adopt a uniform prior on a spatially constant anisotropy
	\begin{equation}
		0.93 \le \frac{\sigma_{\rm t}}{\sigma_{\rm r}} \le 1.06
		\quad\Longleftrightarrow\quad
		-0.12 \,\lesssim\, \beta_{\rm ani} \,\lesssim\, 0.13,
	\end{equation}
	consistent with studies of nearby elliptical galaxies \citep{cappellari2007, cappellari2020, zhu2023}, and additionally verify this choice in \cref{sec:results_and_discussion}. Studies of local galaxies \citep{cappellari2023} and analyses of simulated galaxies \citep{verma2026} show that assuming constant anisotropy does not introduce any bias beyond the percent level on the inferred Hubble constant.
	
	Given \(\beta_{\rm ani}=\) const, the spherical Jeans equation admits the solution
	\begin{equation}
		\sigma_r^2(r)
		\;=\;
		\frac{G}{l(r)} \int_r^{\infty}
		\frac{M(s)\,l(s)}{s^2}
		\left(\frac{s}{r}\right)^{2\beta_{\rm ani}}
		\,ds ,
		\label{eq:sigmar_constbeta}
	\end{equation}
	with the gravitational constant $G$.
	
	The observable, the projected, luminosity-weighted LOS dispersion, then follows \citep{binney1987}:
	\begin{equation}
		\sigma_{\rm los}^2(R)
		\;=\;
		\frac{2}{I(R)}
		\int_R^{\infty}
		\left[1-\beta_{\rm ani}\frac{R^2}{r^2}\right]
		\frac{l(r)\,\sigma_r^2(r)}{\sqrt{r^2-R^2}}\; r\,dr.
		\label{eq:slos_constbeta}
	\end{equation}
	
	Next, we compute the aperture-averaged, PSF-convolved prediction in our measurement aperture:
	\begin{equation}
		\sigma_{\rm ap}^2
		\;=\;
		\frac{\displaystyle \int_{\rm ap}
			\big[I(R)\,\sigma_{\rm los}^2(R)\big]\ast \mathcal{S}\; dx\,dy}
		{\displaystyle \int_{\rm ap} I(R)\ast \mathcal{S}\; dx\,dy},
		\label{eq:sigma_ap_constbeta}
	\end{equation}
	where \(\mathcal{S}\) is the seeing/PSF kernel and \(\int_{\rm ap}\) denotes the on-sky spectroscopic aperture.
	
	Finally, we correct for the external-convergence scaling between observed and model dispersions,
	\begin{equation}
		\sigma_{\rm ap,obs}^2
		\;=\;
		(1-\kappa_{\rm ext})\,\sigma_{\rm ap,model}^2.
        \label{eq:vd_scaled}
	\end{equation}

    \subsection{Bayesian inference of \texorpdfstring{$D_{\Delta t}$}{Ddt} and \texorpdfstring{H$_0$}{H0}}
    \label{ssec:ana_bayes}
    
    Our goal is to infer the time-delay distance $D_{\Delta t}$ and the corresponding value of H$_0$ from the imaging, kinematic, and time-delay observables. Let the full data set for a system be
    \[
    O = \{O_{\rm img},\, O_{\rm kin},\, \Delta t\},
    \]
    where $O_{\rm img}$ denotes the imaging data, $O_{\rm kin}$ the spectroscopic velocity-dispersion measurement, and $\Delta t$ the measured time delays.
    
    The lens model parameters are collected into the vector
    \[
    \xi = \{\xi_{\rm mass},\, \xi_{\rm light},\, \beta_{\rm ani}\},
    \]
    describing the mass and light parameters along with the stellar anisotropy parameter used in the kinematic modeling. We note the external convergence $\kappa_{\rm ext}$, constrained by the independent spectroscopic and photometric wide-field data, is not included here as it can be accounted for with an independent prior $\Pr(\kappa_\text{ext})$.
    
    Conditioned on $(\xi,\kappa_{\rm ext}, D_{\Delta t})$, the likelihood can be written as the product of the independent data components,
    \begin{equation}
    \Pr(O \mid \xi,\kappa_{\rm ext}, D_{\Delta t})
    =
    \Pr(O_{\rm img} \mid \xi)\,
    \Pr(O_{\rm kin} \mid \xi,\kappa_{\rm ext})\,
    \Pr(\Delta t \mid \xi, D_{\Delta t}).
    \end{equation}
    The imaging likelihood constrains the lens parameters $\xi$ through the forward modeling described in \Cref{subsec:forward_mod_like_sample}, thereby producing posterior samples of the Fermat-potential differences $\Delta \tau_{ij}$. The time-delay likelihood then combines these with the observed delays to constrain $D_{\Delta t}$ through \Cref{eq:td_subbed}. The kinematic likelihood constrains the internal mass normalization through the comparison between the observed velocity dispersion and the model prediction given in \Cref{eq:vd_scaled}.
    
    Within a Bayesian framework, combining these terms with priors on $\xi$, $\kappa_{\rm ext}$, and $D_{\Delta t}$ gives the joint posterior
    \begin{equation}
    \begin{split}
    \Pr&(\xi,\kappa_{\rm ext}, D_{\Delta t} \mid O) \propto \\
    &
    \Pr(O_{\rm img} \mid \xi)\times
    \Pr(O_{\rm kin} \mid \xi,\kappa_{\rm ext})
    \times \Pr(\Delta t \mid \xi, D_{\Delta t}) \\ &
    \times \Pr(\xi)\times\Pr(\kappa_{\rm ext})\times\Pr(D_{\Delta t}).
    \end{split}
    \end{equation}
    
    Marginalizing over $(\xi,\kappa_{\rm ext})$ yields the posterior distribution of the time-delay distance,
    \begin{equation}
    \Pr(D_{\Delta t} \mid O)
    =
    \int_2
    \Pr(\xi,\kappa_{\rm ext}, D_{\Delta t} \mid O)\,
    d\xi\, d\kappa_{\rm ext}.
    \end{equation}
    
    As we adopt a uniform prior on $D_{\Delta t}$, this is equivalent to a uniform prior on $1/\mathrm{H}_0$, rather than on $\mathrm{H}_0$. We therefore apply an additional change-of-variables weight to recover a posterior corresponding to a uniform prior on $\mathrm{H}_0$. At fixed $\Omega_m$ and $\kappa_{\rm ext}$, the flat $\Lambda$CDM prediction gives
    \begin{equation}
        \mathrm{H}_0 =
        \mathrm{H}_{0,\rm fid}
        \frac{D_{\Delta t}(\mathrm{H}_{0,\rm fid},\Omega_m)}
        {D_{\Delta t}/(1-\kappa_{\rm ext})}
        =
        \frac{A(\Omega_m)(1-\kappa_{\rm ext})}{D_{\Delta t}},
    \end{equation}
    where $A(\Omega_m)=\mathrm{H}_{0,\rm fid}D_{\Delta t}(\mathrm{H}_{0,\rm fid},\Omega_m)$ is independent of $\mathrm{H}_0$. Therefore,
    \begin{equation}
        w_{\mathrm{H}_0}
        \propto
        \left|
        \frac{\partial \mathrm{H}_0}{\partial D_{\Delta t}}
        \right|
        =
        \frac{A(\Omega_m)(1-\kappa_{\rm ext})}{D_{\Delta t}^2}
        =
        \frac{\mathrm{H}_0}{D_{\Delta t}}.
    \end{equation}

    Finally, $D_{\Delta t}$ is mapped to the Hubble constant through \Cref{eq:ddt_fact}, assuming a standard, flat, \text{$\Lambda \text{CDM}$} cosmology with $\Omega_{\rm m}\sim \mathcal{U}(0.05,\,0.5)$\footnote{This broad prior is used to facilitate comparison with previous work. A more informative prior from other cosmological measurements will be adopted in future studies. We note however that the dependency of H$_0$ on $\Omega_{\rm m}$ is very weak.}. For each posterior sample, we draw $\Omega_{\rm m}$ and use the sampled $D_{\Delta t}$ to solve for the corresponding H$_0$. At fixed lens and source redshifts, the same draw also fixes the deflector distance,
    \begin{equation}
        D_d =
        \frac{D_{\Delta t}}{(1+z_d)(D_s/D_{ds})},
    \end{equation}
    where the ratio $D_s/D_{ds}$ is determined by the sampled $\Omega_{\rm m}$ and is independent of H$_0$ in flat $\Lambda$CDM. The kinematic prediction is therefore rescaled sample-by-sample using the corresponding distance ratio,
    \begin{equation}
        \sigma^2_{\rm pred}
        =
        \lint\,(1-\kappa_{\rm ext})\,
        \left(\frac{D_s/D_{ds}}{(D_s/D_{ds})_{\rm fid}}\right)
        \sigma^2_{\rm fid},
    \end{equation}
    where $\lint=1$ for our power-law mass model \citep[see][]{birrer2020}. The resulting posterior $\Pr(\mathrm{H}_0\mid O)$ is reported for each lens individually and combined across systems in~\Cref{sec:comb}.

	\section{Data}
	\label{sec:data}
	
	This work incorporates three updates relative to previous cosmographic analyses of these systems: lens modeling based on JWST/NIRCam images, updated stellar velocity dispersions measured with JWST/NIRSpec, and updated external convergence estimates derived from wide-field imaging data. We use the JWST/NIRCam F115W band datasets from the Cycle~1 GTO Program \#1198 (PI: Stiavelli). The observations have a pixel scale of $\approx 0\farcs031\,\mathrm{pixel}^{-1}$, with total effective exposure times of $\sim$30min, combining 4 individual exposures acquired in a 4-point dither pattern, which are drizzled together. The velocity dispersions are based on JWST/NIRSPec observations from the same program (Knabel et al. 2026, in prep), and supersed those published by TDCOSMO25 owing to several improvements in data reduction, and the values used in previous modeling papers for these systems based on data of lower signal to noise ratio and angular resolution. We also include information about the wide-field data used to update the external convergence measurements, referring to a companion paper for details (Johnson et al. 2026, in prep).
	
	\subsection{WFI2033}
	WFI 2033--4723 (J2000: $20^h33^m42.2^s$, $-47^\circ23'44''$; hereafter WFI2033) was discovered by \citealt{morgan2004}, and has since been modeled for cosmography with HST data \citep{rusu2020} and most recently JWST data \citep{williams2025}. The redshifts for the source and deflector were measured at $z_s = 1.662$ \citep{sluse2012} and $z_d = 0.6575$ \citep{sluse2019} (hereafter H0LiCOW X), respectively. With a new reduction of the JWST/NIRSpec data, the observed velocity dispersion for this system was updated from $\sigma = 250\pm19$ \kms, used by \citet{rusu2020} and measured with ESO MUSE by \citealt{sluse2019} to $\sigma =204.6\pm6.1$ \kms \footnote{The value used in the original unblinding, prone to the kinematic bug discovered afterwards, was $\sigma = 219.5\pm5.8$ \kms} (Knabel et al 2026, in prep). Both the previous HST and JWST cosmographic models predicted velocity dispersions significantly lower than the MUSE values, indicating a tension between the models and observational measurement.
    The time delays come from 14 years of monitoring through the COSMOGRAIL project, and were most recently updated by \citet[][COSMOGRAIL XVIII]{bonvin2019}. The time delays and uncertainties used in this analysis, with naming conventions from Fig. 3 of \citet{williams2025}, were $\Delta t_\mathrm{A2B} = 37.3 \pm 2.8$ days (7.5\% uncertainty), $\Delta t_\mathrm{A1B} = 36.2 \pm 1.95$ days (5.4\% uncertainty), and $\Delta t_\mathrm{CB} = 59.4 \pm 1.3$ days (2.2\% uncertainty). Since no covariance matrix was reported in COSMOGRAIL XVIII, we assume the time-delay measurements are independent and adopt a diagonal covariance matrix.
	
	\subsection{HE0435}
	\begin{figure}[ht]
		\centering
		\includegraphics[width=0.5\textwidth]{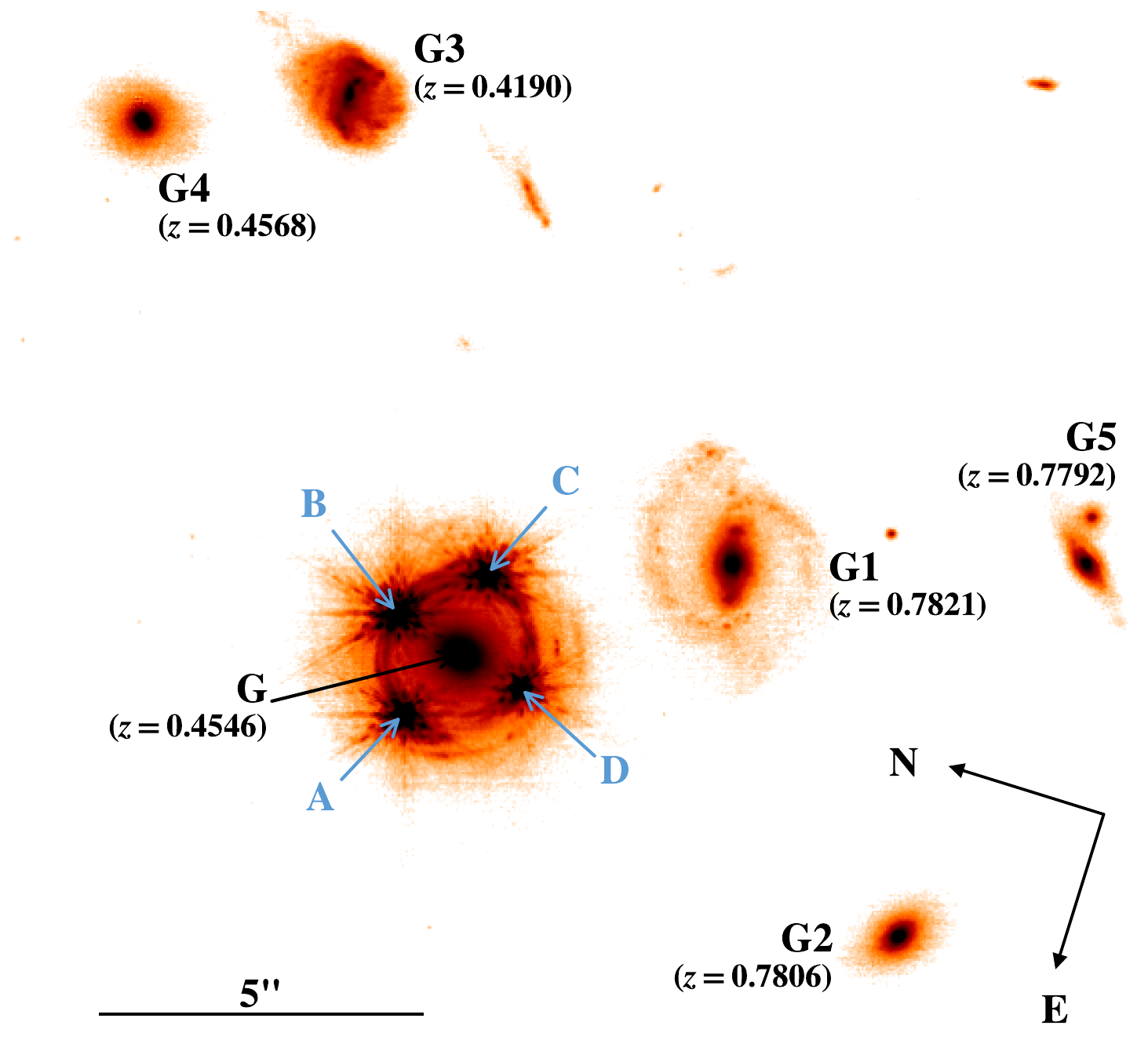}
		\caption{Environment of \helens. Black labels denote the main deflector (D) and the five galaxy perturbers (G1-G5) along with their redshifts. The quasar images are labeled in blue.}
		\label{fig:he_labeled_environment}
	\end{figure}

    HE 0435--1223 (J2000: $04^h38^m14.9^s$, $-12^\circ17'15''$; hereafter HE0435) was discovered by \citealt{wisotzki2002}. This system has been modeled for cosmography with data from HST \citep{wong2017} and adaptive optics (AO) on the Keck Telescope \citep{chen2019}. The deflector is at a redshift of $z_d = 0.4546$ \citep{morgan2005}, while the background quasar is at $z_s = 1.693$ \citep{sluse2012}. The system is in a galaxy group with at least $\sim\!$12 confirmed members with a velocity dispersion of $\sigma = 471 \pm 100$ \kms\ \citep{momcheva2006,wong2010,wilson2016,sluse2017}.  The stellar velocity dispersion of the main deflector was previously measured to be $\sigma = 222 \pm 15$ \kms\ \citep{wong2017}, and updated to $\sigma = 220.1 \pm 4.2$ \kms \footnote{The value used in the original unblinding, prone to the kinematic bug discovered afterwards, was $\sigma = 234.4 \pm 3.4$ \kms} by Knabel et al. (2026, in prep). 
    The redshifts of nearby galaxies G1-G5 were measured by \citet{sluse2017} and can be found in \cref{fig:he_labeled_environment}, while the  stellar velocity dispersions were measured from integrated spectra extracted within one effective radius ($1R_{\rm e}$) from VLT/MUSE observations, following the spectral fitting methodology of \citet{mozumdar2025}. Briefly, \texttt{pPXF} \citep{cappellari2023} was used to fit the spectra over a rest-frame wavelength range of $3600$--$5300$\,\AA, employing three independent clean stellar template libraries -- Indo-US \citep{valdes2004}, MILES \citep{falcon-barroso2011}, and XSL \citep{verro2022}, constructed following \citet{knabel2025b}. The final velocity dispersions and their associated statistical and systematic uncertainties were defined as the equally weighted mean of the measurements obtained from the three libraries, and can be found in \cref{tab:he_galaxy_dispersions}.

    \begin{table}[ht]
        \centering
        \resizebox{\columnwidth}{!}{%
        \begin{threeparttable}
        \caption{VLT/MUSE stellar velocity dispersion measurements for galaxies near HE0435, see \cref{fig:he_labeled_environment}}
        \label{tab:he_galaxy_dispersions}
        \begin{tabular}{l|ccccc}
        \hline
        Galaxy & $\sigma$ & Stat.\ err.\ & Sys.\ err.\ & Total err.\ & S/N/\AA{} \\
         & [km s$^{-1}$] & [km s$^{-1}$] & [km s$^{-1}$] & [km s$^{-1}$] & \\
        \hline
        G1 & 137.0 & 10.0 & 11.0 & 15.0 & 12 \\
        G2 & 148.0 & 17.0 &  2.0 & 17.0 &  5 \\
        G3 &  75.0 & 12.0 &  7.0 & 14.0 & 17 \\
        G4 &  90.0 & 12.0 &  9.0 & 15.0 &  7 \\
        G5 & 143.0 & 22.0 & 10.0 & 24.0 &  2 \\
        \hline
        \end{tabular}
        \begin{tablenotes}
        \footnotesize
        \item Total error is the quadrature sum of the statistical and systematic uncertainties. S/N per \AA{} is measured over the fitted spectral range.
        \end{tablenotes}
        \end{threeparttable}%
        }
    \end{table}
    
    The time delays of this system have also been updated by \citet[][COSMOGRAIL XIX]{millon2020}. The time delays used, with naming conventions from \cref{fig:he_labeled_environment}, were $\Delta t_\mathrm{BA} = 9.0 \pm 0.8$ days (8.9\% uncertainty), $\Delta t_\mathrm{CA} = 0.8 \pm 0.8$ days (100\% uncertainty), and $\Delta t_\mathrm{DA} = 13.8 \pm 0.8$ days (5.8\% uncertainty), and we assume these measurements are independent.

	\subsection{PG1115}
	\begin{figure}[ht]
		\centering
		\includegraphics[width=0.5\textwidth]{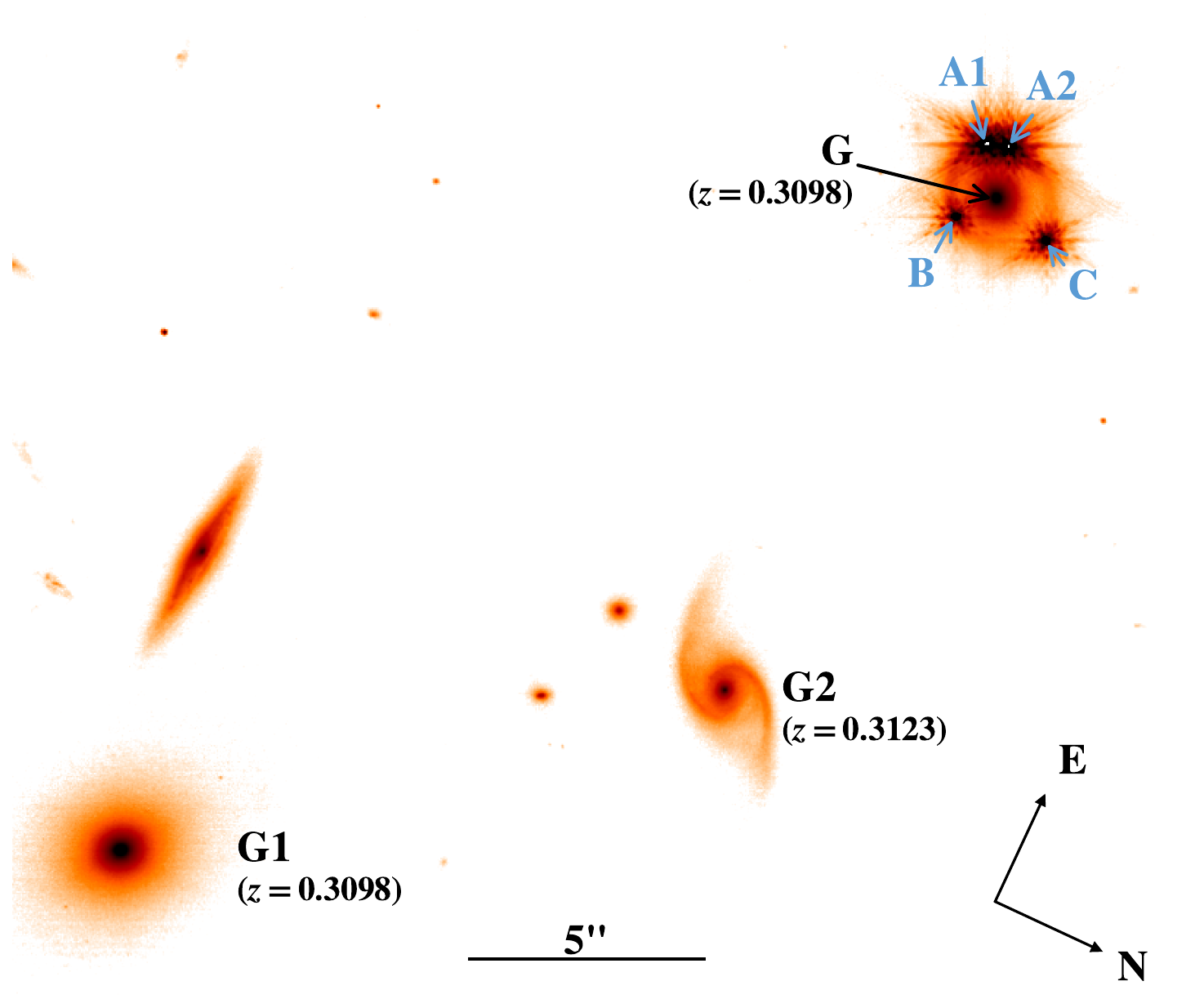}
		\caption{Environment of PG1115. Black labels denote the main deflector (G) and the two galaxy perturbers (G1 and G2) along with their redshifts. The quasar images are labeled in blue.}
		\label{fig:pg_labeled_environment}
	\end{figure}
    
	PG 1115+080 (J2000: $11^h18^m16^s.9$, $+07^\circ45'59''$; hereafter PG1115) was the second ever strong lens system discovered \citep{weymann1980}. Previously modeled for cosmography with Keck AO \citep{chen2019}, the system's deflector is at $z_d = 0.3098$ with previous measurements of its background quasar at $z_s =1.722$ \citep{tonry1997, henry1986, christian1987}.  Recent analysis of the JWST/NIRSpec data has updated this quasar redshift to $z_s = 1.727$ (Knabel et al. 2026, in prep). PG1115 is part of a group with velocity dispersion $\sigma = 390\pm 60$ \kms and at least 13 members \citep{wilson2016}, two of which are nearby galaxies G1 and G2. Redshifts of G1 and G2 were measured at $z_{G1} = 0.3098$ and $z_{G2} = 0.3123$, with their velocity dispersions at $\sigma_{G1} = 256 \pm 20$ \kms and $\sigma_{G2} = 130 \pm 60$ \kms and the deflector's velocity dispersion of $\sigma = 281 \pm 25$ \kms \citep{tonry1997}. We use the improved deflector's velocity dispersion measurement $\sigma = 244.0 \pm 5.7$ \kms \footnote{The value used in the original unblinding, prone to the kinematic bug discovered afterwards, was $\sigma = 251.0 \pm 3.3$ \kms} based on JWST/NIRSpec (Knabel et al. 2026, in prep).
    COSMOGRAIL, once again, measured the time delays with 15 years of data in \citet{bonvin2018} (COSMOGRAIL XVII). The time delays used, with naming conventions from \cref{fig:pg_labeled_environment}, were $\Delta t_\mathrm{A2A1} = 0.0 \pm 1.55$ days, $\Delta t_\mathrm{BA1} = 8.3 \pm 1.55$ days (18.7\% uncertainty), and $\Delta t_\mathrm{CA1} = -9.9 \pm 1.1$ days (11.1\% uncertainty), with these measurements again assumed independent.

	\section{Modeling}
	\label{sec:mod}

	This section describes our modeling of the JWST/NIRCam imaging and the extraction of the lens parameters used in the cosmographic analysis. All image modeling is performed with \texttt{lenstronomy}\footnote{\url{https://github.com/lenstronomy/lenstronomy}} \citep{birrer2015, birrer2018, birrer2021}, which implements a forward-modeling approach to strong-lens systems. Parameter priors are uniform over physically motivated ranges determined from preliminary fits, except when warranted by external information.

    We reiterate that -- to avoid experimenter bias -- our analyses of HE0435 and PG1115 were done blind: $D_{\Delta t}$, $D_s/D_{ds}$, $\gamma_{\mathrm{PEMD}}$ and H$_0$ were kept hidden during modeling and data processing. WFI2033 was not blinded, as expectations for the change in H$_0$ were already produced by \citealt{williams2025}.

	We first describe the fiducial modeling framework common to all three systems, then the system-specific components and exploratory models used to refine each fiducial model. We then introduce the systematic variations applied to the fiducial models -- only these controlled variations enter the final cosmographic inference -- and finish with the forward-modeling framework and likelihood evaluation.

	\subsection{Common modeling framework}
	\label{subsec:common_framework}

	The following framework is shared by all three fiducial models. For WFI2033 we adopt the models of \citealt{williams2025}, which also follow this framework; modifications are described in \cref{subsec:wfi2033}.

	\paragraph{Main deflector galaxy}
	The deflector mass is a power-law elliptical mass distribution (PEMD; \citealt{barkana1998}),
    \begin{equation}
        \kappa_\text{PEMD}\left(\theta_1 , \theta_2\right) \coloneqq \frac{3 - \gamma}{2}\left[\frac{\theta_\text{E}}{\sqrt{q_\text{m}\theta_1^2 + \theta_2^2/q_\text{m}}}\right]^{\gamma-1},
    \end{equation}
    defined by an Einstein radius $\theta_\text{E}$, logarithmic slope $\gamma$, and axis ratio $q_\text{m}$, plus an external shear component accounting for the net tidal distortion from mass along the line of sight and in the local environment. Gaussian priors on the deflector's (cartesian) ellipticity parameters limit its degeneracy with the external shear, preventing solutions where an unphysically elliptical galaxy is offset by an equally unphysical shear \citep{schmidt2022}.

    The deflector surface brightness is the sum of two elliptical Sérsic profiles,
    \begin{equation}
        I(\theta_1,\theta_2) = A \exp{
                -k \left( \left( \frac{\sqrt{q_L \theta_1^2 + \theta_2^2 / q_L}}{r_{\text{eff}}}\right)^{1/n} - 1 \right)
            },
    \end{equation}
    with amplitude $A$, effective (half-light) radius $r_\text{eff}$, axis ratio $q_\text{L}$, Sérsic index $n$, and corrective constant $k$ \citep{sersic1968}. The two Sérsic centroids are joined together but vary freely from the mass centroid.

    \paragraph{Quasar and host-galaxy light}
    The quasar is modeled as four point sources in the image plane convolved with the PSF. Their centroids are constrained by the data on a grid supersampled by a factor of 3, so that point sources falling between pixel centers are handled accurately. For details on how the PSFs are generated, see \citet[Sect.~4]{williams2025}. In addition, we allow small per-image astrometric corrections: each quasar image is given a free offset parameter relative to its lens-model-predicted position, fit to the pixel data. This offset enters the imaging likelihood as a Gaussian penalty with a 1$\sigma$ scale of 3~mas (hard-bounded within $\pm5\sigma$), absorbing residual positioning offsets and astrometric perturbations from substructure lensing.

    The light from the lensed quasar host galaxy is modeled in the source plane as a Sérsic profile supplemented by a linear basis set of shapelets \citep{refregier2003, birrer2015}, which capture galactic-scale structure such as star-forming regions or spiral arms. The shapelet spatial scale is a free parameter, subject to the maximum spatial resolution of the data \citep[see Appendix F][]{williams2025}; the maximum order $n_\text{max}$ is a modeling choice, selected for each fiducial model by exploring $n_\text{max} = 0$–$30$ and identifying the BIC turnover beyond which improved fit no longer justified the added complexity (see \Cref{sec:comb}). The centroids of the point source, Sérsic, and shapelets are joined so that the maximum information region of the shapelets coincides with the brightest region of the Sérsic, nearest the quasar. As described in \cref{subsec:pg1115}, the fiducial PG1115 host is modeled with the Sérsic component alone.

    \paragraph{Galaxy-scale perturbers}
    Perturbing galaxies near enough to the main deflector are included explicitly in the lens model, at their measured redshifts, using the multi-plane lens equation (\cref{ssec:ana_theory}); external shear alone cannot account for their effect, as their proximity (in LOS) makes higher-order perturbative terms significant \citep{sluse2017}. We select these perturbers using the flexion shift $\Delta_3 x$, a measure of the higher-order (third-order) image distortion induced by a nearby object \citep{mccully2017}, retaining those above the conventional cutoff of $\log\Delta_3 x > -4$. Each is modeled as a singular isothermal sphere (SIS), with Einstein radii not fit independently but jointly scaled, their ratios fixed by the measured velocity dispersions and redshifts (\cref{sec:data}) under the assumption that lensing and stellar velocity dispersions agree within the scatter \citep{treu2006}. This encodes the kinematic information while minimizing degeneracies between individual perturber masses, external shear, and the deflector ellipticity.

    \paragraph{Time-delay likelihood}
    For all three systems, time-delay information is included during the fitting process, so the time-delay likelihood is computed in-loop to ensure models reproduce the observed time-delay ratios. For HE0435 and PG1115 this preserves the blinding, as effectively only the time-delay ratios constrain the likelihood.

	\subsection{WFI2033}
	\label{subsec:wfi2033}

    For WFI2033 we adopt the models of \citealt{williams2025}. The main deflector is modeled with the PEMD-plus-shear mass profile and double-Sérsic light described above. In addition to the nearby galaxy-scale perturbers G2, G3, and G7, the model includes the small satellite (X) at the deflector redshift, whose mass is modeled as an SIS and whose light is captured with a Sérsic. This satellite is the key feature distinguishing the JWST models from the earlier HST analysis: only marginally detected with HST, its mass is well constrained by the NIRCam imaging, which accounts for most of the shift in Fermat potential relative to \citet{rusu2020}.

    Relative to \citealt{williams2025}, the fiducial model now also includes the time-delay information and increased astrometric flexibility described above.

    We also perform three exploratory tests probing explanations -- unexplored in \citet{williams2025} -- for the differences with the HST-based models of \citet{rusu2020}. First, we vary the stellar anisotropy prescription, comparing the prior adopted in the HST models to the updated prior motivated by TDCOSMO25 (\cref{ssec:ana_kinematics}). Second, we alternate between the previously used MUSE velocity dispersion and the new JWST/NIRSpec value to gauge the impact of improved aperture-integrated kinematics. Third, we allow larger offsets between observed and model-predicted quasar image positions, testing whether the astrometric precision offered by the JWST PSF might be overestimated or over-constrained during fitting.

	\subsection{HE0435}
	\label{subsec:he0435}
	\begin{figure*}[ht]
		\centering
		\includegraphics[width=\textwidth]{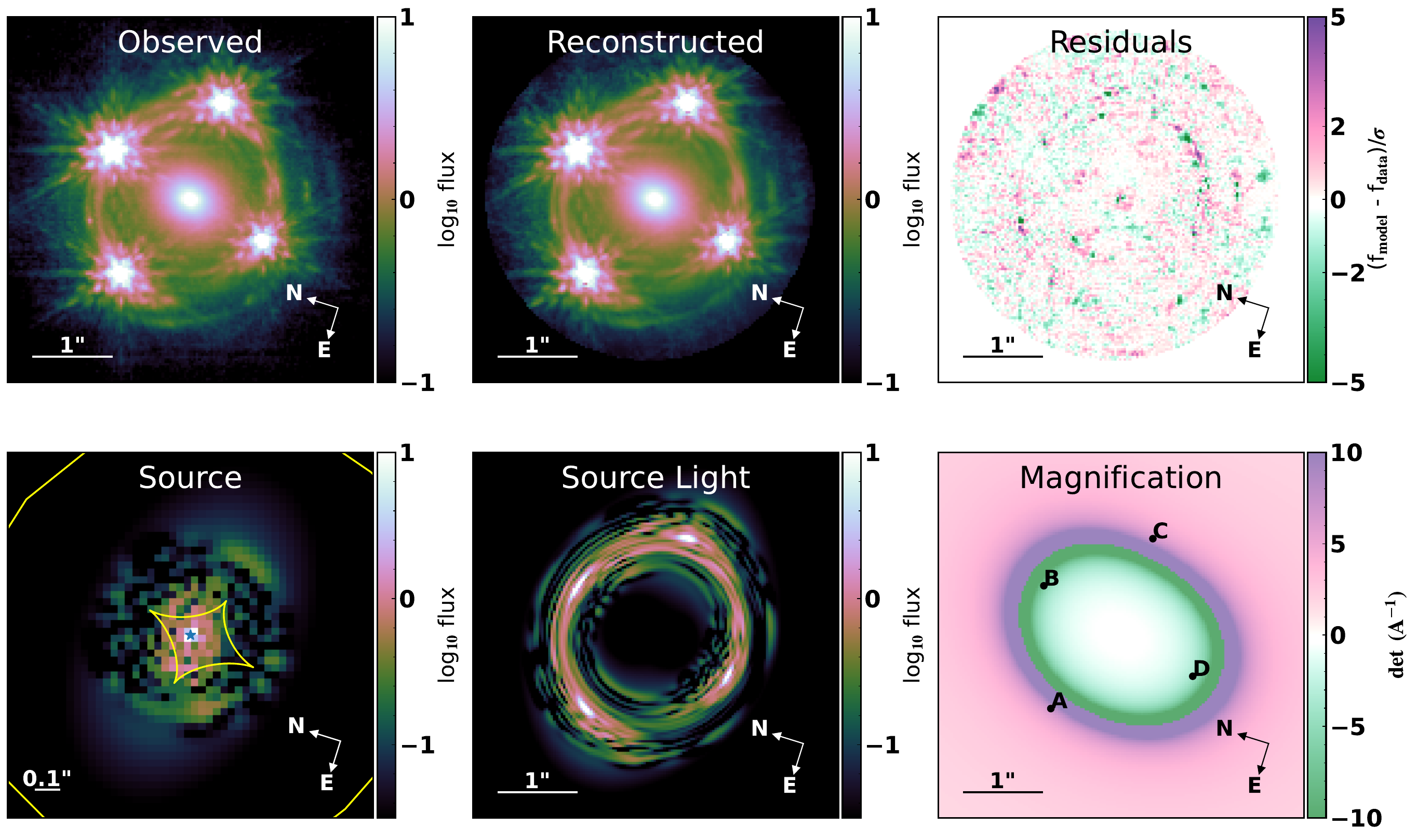}
		\caption{Example \helens fit results from the best performing model (\Cref{fig:he0435_plot_final}). The top-left panel shows the observed image, the top-middle panel shows the model-predicted reconstruction of the image, and top-right panel shows the fit's residuals normalized by the estimated uncertainty of each pixel. In the bottom left is the reconstructed source plot, wherein the star symbol denotes the location of the quasar. The bottom-middle panel shows the lensed, unconvolved, extended-source light, and the bottom-right panel shows the magnification map of the system.}
		\label{fig:example_fit_HE}
	\end{figure*}

    For the \helens lens light, fixing one Sérsic index to $n_\text{disk}=1$ (while the other freely fits the bulge) reproduced the lens light to the noise level (\cref{fig:example_fit_HE}).

    Five nearby galaxy-scale perturbers G1-G5 (\cref{fig:he_labeled_environment}) meet the flexion-shift criterion and are modeled explicitly \citep{sluse2017}. We make the deflection correction for G1, G2, G4, and G5, which lie behind the main lensing galaxy. Due to its proximity, G1's light is additionally modeled with a Sérsic (with its light and mass centers fixed together) to account for its disproportionate brightening of one side of the lensed source light.

    \helens\ also lies in a galaxy group (\cref{sec:data}). Unlike the group in \pglens, its flexion shift falls below the $\log\Delta_3 x > -4$ cutoff for explicit inclusion \citep{sluse2017}, so we do not model it as a separate halo; its leading-order convergence and tidal shear are instead absorbed by the fitted external shear and the external-convergence prior $\kappa_{\rm ext}$ (\cref{sec:data}). Moreover, since the main deflector is the brightest group member and thus likely near the group center, its mass model already absorbs much of the group halo \citep{sluse2017}.

	\subsection{PG1115}
	\label{subsec:pg1115}
	\begin{figure*}[ht]
		\centering
		\includegraphics[width=\textwidth]{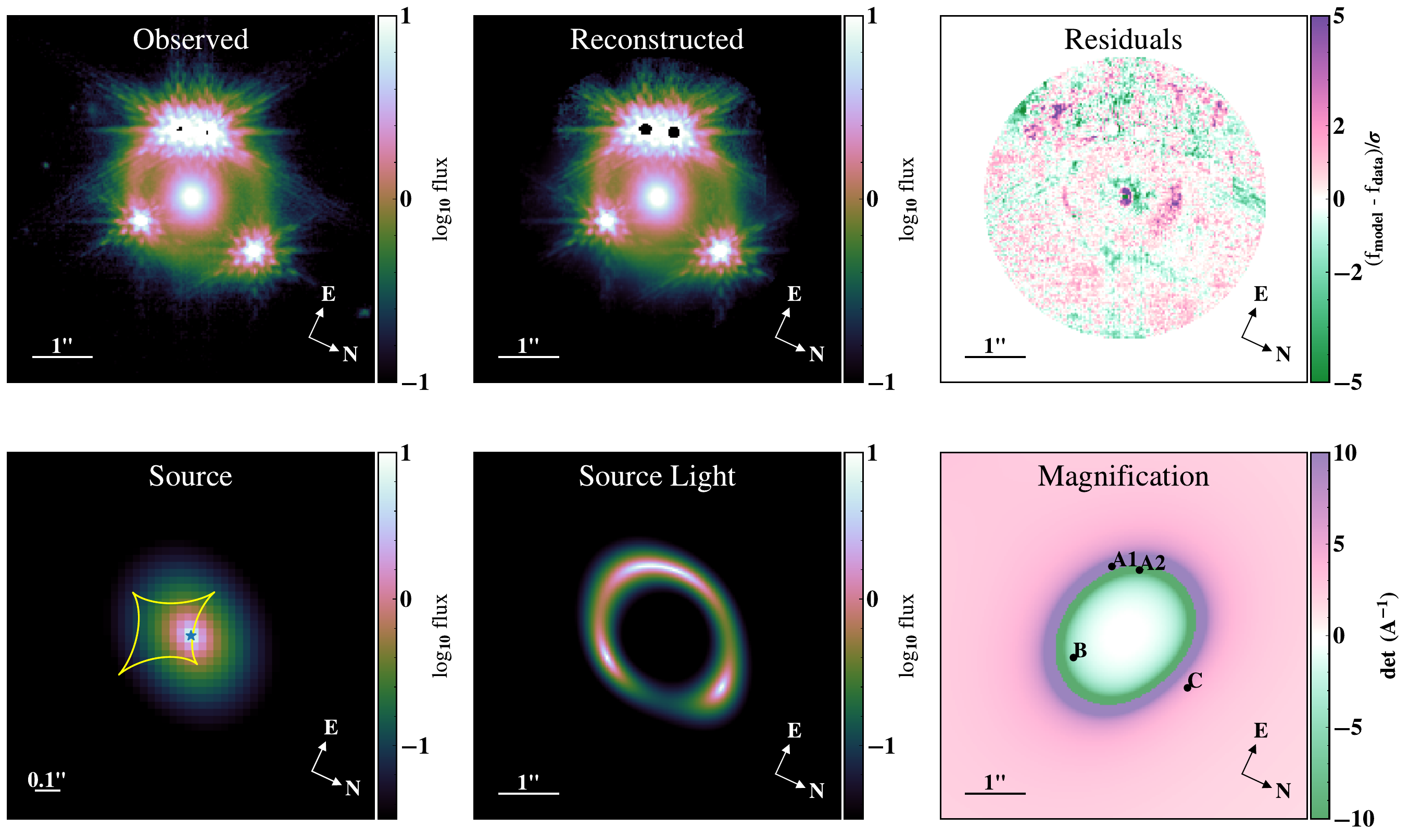}
		\caption{Same as \Cref{fig:example_fit_HE}, but the best model for \pglens. We masked the center of the two brightest images in the East to avoid the saturated pixels seen on the top left figure.}
		\label{fig:example_fit_PG}
	\end{figure*}

    For \pglens, the two nearby perturbers G1 and G2 (\cref{fig:pg_labeled_environment}) are near (or above) the flexion-shift cutoff \citep{chen2019} and are modeled explicitly, along with the halo of the galaxy group identified in \cref{sec:data}, whose center is expected to lie near these 3 galaxies and whose own flexion shift is significant \citep{mccully2017}.

    \paragraph{Main deflector galaxy}
    While the very center of the deflector is poorly fit, this region is primarily important for the kinematic interpretation, not the lens modeling. Testing additional light components in this region -- a point source, Gaussian, additional Sérsic, and fixed Sérsic indices -- left it poorly constrained; the best predictions on the external kinematic data came from models with either no additional profile or a tight central Gaussian. The residual uncertainties associated with the deflector center are propagated into the final error budget through our systematic testing marginalization.

    \paragraph{Quasar and host-galaxy light}
    Unlike the other two systems, the host of \pglens is modeled with the base Sérsic profile alone, as adding shapelets or any other source complexity produced unphysical fits to the lens light, owing to the difficulty of fitting the two saturated point sources. These two quasar images are fully saturated, so we mask their central regions; the upper half of the cutout nonetheless remains difficult to fit consistently due to this saturation combined with nearby resolved light sources. Here the extended wing structure of the JWST PSF is advantageous, allowing robust and consistent constraints on the quasar image positions despite the masking. Identifying a suitable initial PSF is still challenging, however: it must cover the full cutout, but field stars far from saturation provide limited constraints on the extended wings, while near-saturated stars typically yield poor, inconsistent constraints on the PSF core. This is particularly problematic for lens modeling, where core biases propagate into the astrometry and thus the overall results. Nearby objects, perturbers, and spatially varying exposure times further prevent stars from being modeled reliably to the required radial extent.

    \paragraph{Group halo}
    The fiducial group halo is an NFW profile \citep{navarro1996, golse2002} with Gaussian priors on the centroid, scale radius, and deflection angle at the scale radius from \citealt{chen2019}. This centroid prior corresponds to a group center projected only $\sim\!10\arcsec$ ($\sim\!45$~kpc at $z_d=0.31$) from the deflector, but with $1\sigma$ positional uncertainties of $23.4\arcsec$ and $15.8\arcsec$ in RA and Dec \citep{wilson2016} -- larger than the offset itself, so the prior is only weakly informative on the halo position. As the halo centroid and scale radius are poorly constrained, we tested SIS, NFW, and eNFW parameterizations, each with and without G1 and G2, and found the NFW models most consistent with the data. The final posterior on M200 remains consistent with our prior.

	\subsection{Primary systematic tests}
	\label{subsec:comb_sys}

	The primary systematics tested for all three systems modify (i) the flexibility of the source reconstruction, (ii) the mask size used in the imaging likelihood, and (iii) the initial PSF model seeding the iterative PSF refinement. We vary:
		\begin{itemize}
				\item Source complexity: maximum shapelet order of the source model,
				\begin{itemize}
						\item WFI2033: $n_\text{max}\in\{18, 20, 22\}$
						\item HE0435: $n_\text{max}\in\{23, 25, 27\}$
                        \item PG1115: $n_\text{max}=-1$\footnote{$n_\text{max}=-1$ corresponds to a Sérsic-only source model}
					\end{itemize}
				\item Mask size: two choices for the radial extent of the imaging mask,
				\begin{itemize}
						\item WFI2033: mask radius $\in\{2.1''\text{(68 pix)}, 2.2''\text{(71 pix)}\}$
						\item HE0435: mask radius $\in\{2.1''\text{(67 pix)}, 2.2''\text{(71 pix)}\}$
						\item PG1115: mask radius $\in\{2.3''\text{(74 pix)}, 2.4''\text{(77 pix)}\}$
					\end{itemize}
				\item Initial PSF: the two PSF modeling methods thoroughly discussed in \citealt{williams2025}:
				\begin{itemize}
						\item \e{STARRED}
						\item \e{PSFr}
					\end{itemize}
                \item Additional deflector light profile: (only for PG1115)
                \begin{itemize}
						\item PG1115: profile $\in$\{\e{None}, \e{Gaussian}\}
                    \end{itemize}
                \item Additional sampling: PG1115 models were doubled due to inconsistent lens light constraints
                \begin{itemize}
						\item PG1115: fit instance $\in$\{1, 2\}
                    \end{itemize}
			\end{itemize}

	\noindent
    This gives 12 model configurations for WFI2033 and HE0435 and 16 for PG1115. To account for the systematic errors associated with these choices, we combine the models via a weighted sampling for each system, described in \cref{sec:comb}.

    \subsection{Forward modeling and likelihood sampling}
    \label{subsec:forward_mod_like_sample}

    For a given set of nonlinear lens and light parameters, the model components are ray-traced and rendered in the image plane, convolved with the PSF, and compared to the data through a pixel-based imaging likelihood whose per-pixel variance combines the imaging noise (Poisson, readout, and flat-field terms) with a PSF error map -- an empirical PSF-uncertainty term derived from field stars -- that inflates the variance near the point sources explained further in Sec.~4 of \citet{williams2025}. Linear parameters (e.g., component amplitudes and the linear shapelet coefficients) are optimized at each likelihood evaluation via linear inversion.

    For each model we optimize the nonlinear parameters with particle swarm optimization \citep[PSO;][]{kennedy1995} using 200 particles and 400 iterations, followed by MCMC sampling with \texttt{emcee} \citep{foreman-mackey2013} using 10 walkers per free parameter. The PSO is run in a staged schedule of decreasing step scale (\texttt{sigma\_scale} $= 1.0, 0.1, 0.01$), interleaved with six iterative PSF updates and a gradual freeing of model complexity. Each chain is run for an initial 500 steps and extended in five continuation runs of 750 steps; the burn-in is removed per model after inspecting convergence.

    To minimize potential PSF biasing of modeled parameters (or vice versa), the initial PSF was generated solely from stars in the field of WFI2033, which provides the highest quality initial guess owing to the specific stars present. Each model of each system had its own final PSF, iteratively updated at six stages of the fitting process: at each stage the point-source residuals are stacked and used to correct the PSF kernel before re-optimizing. A comparison of the initial and final PSFs of the best models is shown in \cref{fig:psf_comparison}, and the method is further discussed in Sec.~4 of \citet{williams2025}. To assess whether sub-pixel structure in the quasar images or aliasing are significant, we run a fiducial model of each system that supersamples the ray-tracing grid near the point sources and convolves these regions with the PSF. We find no significant change in relevant model parameters.
    
    Imaging uncertainties are constructed from the 2D resampled Poisson, readout noise, and flat-field variance estimates summed in quadrature; following \citealt{williams2025}, per-pixel uncertainties near the point sources are additionally inflated with empirical estimates of our PSF model uncertainty based on residuals of stars fit in the field. This PSF error correction becomes even more important near saturation (e.g., near quasar image positions), where the detector response is non-linear.

    The image likelihood includes three further constraints. First, the quasar images are ray-traced back to the source plane and the RMS scatter of their inferred source positions is penalized with a Gaussian of width $0.4$~mas together with a hard bound at the same scale, enforcing a common source. Second, models with negative point-source or Sérsic amplitudes are rejected. Third, following \citet{schmidt2022, williams2025}, we reject models in which the deflector mass and light are strongly misaligned: the mass axis ratio may not fall more than $0.2$ below that of the light, and the mass and light position angles must agree to within a $q$-dependent tolerance ($\approx17$--$35\degr$ at typical ellipticities, relaxing as the mass profile rounds). To verify the radial extent of the likelihood region, the mask size is varied as one of our systematic tests.

    \subsection{Compute time}
    The final models were processed on 36-core CPUs, with approximate total computing times:
    \begin{itemize}
        \item WFI2033: $\sim$91,000 CPU hours
        \item HE0435: $\sim$82,000 CPU hours
        \item PG1115: $\sim$900 CPU hours
    \end{itemize}
    The two-order-of-magnitude difference is driven primarily by the complexity of the extended source arcs, itself limited by the data quality in the arc regions. These regions provide the strongest constraints on the key lens model parameters, including the Einstein radius and power-law slope (see \Cref{sec:param_results}).

	\section{Updated $\kappa_{\rm ext}$ measurements}
	\label{sec:kappa_ext}

    Recent studies have highlighted that model-predicted external shear in standard PEMD+$\gamma_\text{ext}$ models does not trace the true tidal shear from large-scale structure \citep[see e.g.,][\cref{sec:appendix_shear_interpretation}]{etherington2023}. Instead, it predominantly absorbs missing complexity in the lens mass model, leading to biased and physically uninterpretable shear values.

    The impact of this degeneracy on time-delay cosmography is typically subdominant, however. This is because time delays are more sensitive to the radial mass profile and external convergence, while degeneracies between shear and ellipticity largely preserve the Fermat potential differences. Consequently, external shear acts mainly as a nuisance parameter whose misinterpretation does not strongly bias H$_0$. This will be discussed further in our next work, where we test constraints on more complex mass models to see if the improved imaging of JWST can consistently constrain these profiles for the first time and quantify their impact on measurements of H$_0$.

    As a result, our collaboration no longer uses shear-weighted estimators of external convergence $\kappa_{\rm ext}$. In addition, we have incorporated the historically-omitted foreground $\kappa_{\rm d}$ and background $\kappa_{\rm ds}$ contributions to $\kappa_{\rm ext}$. The effect of these two changes is shown in \cref{tab:kappa_ext_comparison}, and discussed further in Johnson et al. (2026). They show that in general the new estimates are consistent with the old ones, however more accurate due to removing shear constraints from the matched lines of sight and explicitly taking into account background and foreground contributions.

    \begin{table}[ht]
        \centering
        \begin{threeparttable}
        \caption{External convergence estimates compared to previous work.}
        \renewcommand{\arraystretch}{1.5}
        \begin{tabular}{l|ccl}
        \hline
        System & New $\kappa_\mathrm{ext}$ & Previous $\kappa_\mathrm{ext}$ & Previous work \\
        \hline
        WFI2033 & $0.020\lowup{0.032}{0.057}$  & $0.059\lowup{0.044}{0.079}$ & \cite{rusu2020} \\
        HE0435  & $-0.006\lowup{0.017}{0.031}$ & $0.004\lowup{0.022}{0.033}$ & \cite{rusu2017a} \\
        PG1115  & $-0.006\lowup{0.012}{0.022}$ & $-0.005\lowup{0.021}{0.032}$ & \cite{chen2019} \\
        \hline
        \end{tabular}
        \begin{tablenotes}
            \item Reported values are medians with errors corresponding to the 16th and 84th percentiles.
        \end{tablenotes}
        \label{tab:kappa_ext_comparison}
        \end{threeparttable}
    \end{table}

	\section{Combined analysis}
	\label{sec:comb}
	
	To combine the different models resulting from the systematic tests, we weight each sample based on its ability to match 3 observable quantities: the imaging data, the velocity dispersion prediction of the main deflector, and the time delays, which are discussed in order. The combination of these weights allows us to disentangle diverging physical interpretations from models that may be able to reproduce one of the observables but not the others.
	
	\subsection{Imaging Weights}
	
	While increasing model complexity can easily improve a model's fit to the data, it may not be a better prediction of the real physical system we are modeling. Therefore, we use the Bayesian Information Criterion (BIC) evaluated on each model's best prediction, which we calculate as
	\begin{equation}
		\mathrm{BIC} = \ln(n)\,k \;-\; 2\,\ln(\hat{L}),
	\end{equation}
    with the number of data points $n$, the number of model parameters $k$, and the maximum joint imaging-and-time-delay likelihood of the model $\hat{L}$. 
	
	Formally, the BIC only offers a meaningful statistic when comparing models fit to the same data. This is not the case for our systematic test varying the mask size. We follow the typical approximation, where the smaller of the two masks is used to calculate the imaging likelihoods \citep{wong2017, birrer2019a, williams2025}.
	
	For any two models $M_1$ and $M_2$, we define their relative probabilities given their respective BIC values of $\mathrm{BIC}_1$ and $\mathrm{BIC}_2$ as
	\begin{equation}
		\frac{p(M_1)}{p(M_2)} \;\propto\; \exp\!\left[-\frac{\mathrm{BIC}_1 - \mathrm{BIC}_2}{2}\right].
	\end{equation}
	Given the lowest (best) BIC among all the models, $\mathrm{BIC}_\mathrm{min}$, the weight for each model is then given by
	\begin{equation} \label{bic_weight}
		f_{\mathrm{BIC}}(x) =
		\begin{cases}
			1 & \text{if } x \le \mathrm{BIC}_\mathrm{min},\\[6pt]
			\exp\left[-\frac{x - \mathrm{BIC}_\mathrm{min}}{2}\right] & \text{if } x > \mathrm{BIC}_\mathrm{min},
		\end{cases}
	\end{equation}
	with $x$ being the model's BIC.
	
	We also take into account the intrinsic variability of the BIC values to ameliorate the issue of ``lucky samples" over-weighting a particular model. Our implementation followed \citet{birrer2019a} to convolve each model's proposed $\mathrm{BIC}_{\rm raw}$ with a Gaussian with width $\sigma_\mathrm{BIC, intrinsic}$, which reflects uncertainty in the BIC due to limited sampling. To compute $\sigma_\mathrm{BIC, intrinsic}$, we computed the variance of each systematic test with the base mask size to its counterpart on the larger mask, holding all other systematic choices constant. This convolution ensures that small fluctuations in $\mathrm{BIC}_{\rm raw}$ do not randomly dominate the weighting process, while still keeping computational time low.
	
	Including this convolution, the final image-based model weights are given by
	\begin{equation}
		W_{\mathrm{abs}} = \int_{-\infty}^{\infty} \frac{1}{\sqrt{2\pi}\,\sigma_\mathrm{BIC,\ intrinsic}}
		\exp\left[-\frac{\bigl(\mathrm{BIC}_{\rm{raw}} - x\bigr)^{2}}{2\,\sigma_\mathrm{BIC,\ intrinsic}^2}\right]
		f_{\mathrm{BIC}}(x)\,dx,
	\end{equation}
	with these final relative weights being normalized such that $\max(W_n) = 1$. These weights were used as sampling probabilities from the different models, such that models with a lower (better) BIC value receive proportionally more draws in the final aggregated parameter space.
	
	\subsection{Kinematics Weights}
	
	To ensure that our models' predictions for the deflector mass are correct, we compared the model-predicted (single 0.55$''$ aperture-integrated) velocity dispersion $\sigma_\text{ap, model}$ to the observed velocity dispersion $\sigma_\text{obs,D}$, assuming the same seeing conditions and taking into account cosmological dependence. We used the Gaussian likelihood
	\begin{equation}
		f_\text{kin} = \mathcal{N}\!\bigl(\sigma_\text{ap, model} \mid \sigma_\text{obs, D},\,\sigma_\text{obs,$\sigma$}\bigr),
	\end{equation}
	including the uncertainty in the observed velocity dispersion $\sigma_\text{obs,$\sigma$}$. This model prediction takes into account both the sampled $D_d$ and $\kappa_\text{ext}$ adjustments.
	
	\subsection{Total Weights}
    
    Combining the kinematic weight with the BIC-based imaging weight, we formed the ``total weight":
    \begin{equation}
            W_{\mathrm{total}} \;=\; W_{\mathrm{BIC}} \;\times\;f_\text{kin} .
        \end{equation}
    After computing $W_{\mathrm{total}}$ for each model sample, we normalized the values such that $\max(W_{\mathrm{total}}) = 1$. For the final posteriors, MCMC samples were drawn in proportion to their $W_{\mathrm{total}}$.

	\section{Results and discussion}
	\label{sec:results_and_discussion}

	\begin{figure*}[ht]
		\centering
		\includegraphics[width=\textwidth]{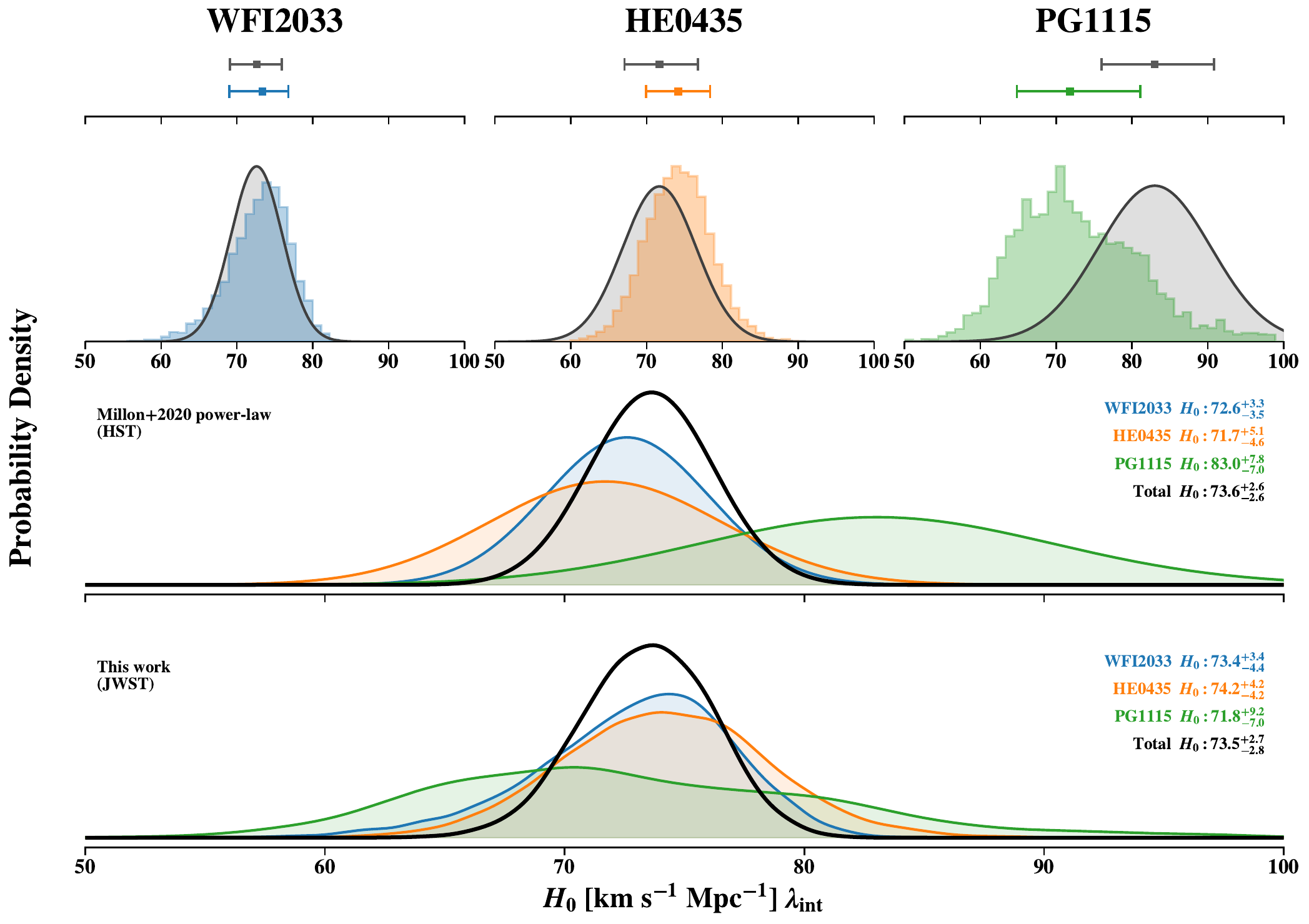}
		\caption{Comparison of our JWST-based $H_0$ results with previous HST-based power-law results \citep[TDCOSMO~I]{millon2020b}. Top row: per-system median, $16$--$84\%$ credible intervals, and posterior distribution functions for the previous HST-based models (gray pdfs) and this work (colored histograms). Second row: the corresponding $H_0$ HST-based posteriors for each system from TDCOSMO~I, with results shown as Gaussian approximations to its quoted median and $16/84\%$ uncertainties and the black curve giving the combined constraint (the product of the individual posteriors). Third row: same as the second row, but results from this work based on JWST-NIRCam imaging. Quoted values in the lower panels list the median $^{84\%}_{16\%}$ for each system and their combination. Note the improved agreement between the three systems.}
		\label{fig:H0_update_comparison}
	\end{figure*}

	\subsection{WFI2033}
	This system was not blinded, as it was already analyzed by \citet{williams2025}. The only change made to these models was the addition of a time-delay likelihood during the fitting, typical for these analyses, to ensure the model results align with our expectations for the observed time-delays.
    
\begin{figure*}[ht]
		\centering
		\includegraphics[width=\textwidth]{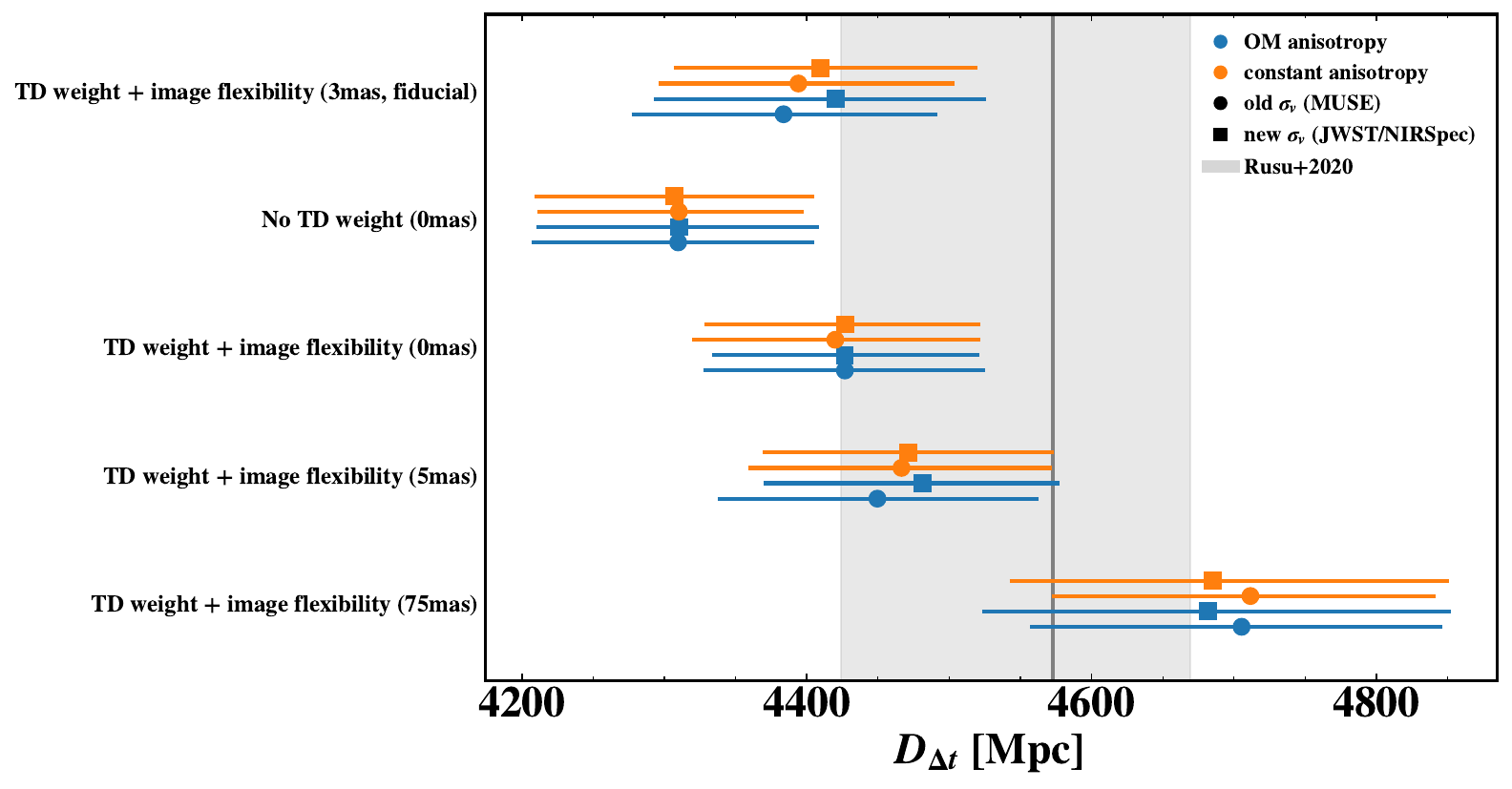}
		\caption{Comparison of the shift in the time-delay distance from changing modeling choices for the fiducial WFI2033 model. The bottom axis is the time-delay distance result for these models, and along the Y axis we have different modeling choices. The top row is our fiducial model with 3mas of flexibility between painted and formal image positions used for the time-delay computations. The second row is without using time-delay information during the modeling fit, and the third through last rows show how increasing image flexibility changes the measured time-delay distance. We also model each of these cases with the two different choice in anisotropy, Osipkov-Merritt given in blue and constant anisotropy given in orange. The last quantity we varied was the observed value used in the velocity dispersion, where the previous MUSE-based value is denoted by the circle and the new JWST/NIRSpec result by the square. We find the only modeling choice with significant discrepancy comes from the model with unphysically large image position flexibility. We note this maximally flexible model does not converge to a single solution for the image positions, and instead exploits this flexibility to better fit the time delays.}
		\label{fig:WFI2033_H_0_systematics_comp}
	\end{figure*}

    \citet{williams2025}, found a $\sim3\%$ shift in the primary Fermat potential difference compared to the previous HST-based models. However, the Fermat potential is only one ingredient of H$_0$. Anisotropy priors, other images' Fermat potentials and time delays, and new kinematic data from JWST/NIRSpec all affect the measurement. To verify if any one of these data products has a significant impact on the final result ($D_{\Delta t}$), we compute variations of our fiducial model of \wfilens (\Cref{fig:WFI2033_H_0_systematics_comp}). One possible explanation is that the HST-based models had a larger-than-usual discrepancy between the observed- and predicted-image positions ($\sim$22mas) compared to other TDCOSMO systems. This offset is significantly reduced in JWST models (to $<$4mas), but to test if this could explain the difference between the two results, we increased the image position flexibility in the JWST models to see if we reproduce the HST results. By using extremely flexible image positions---able to mismatch the observed and predicted quasar positions by up to 75mas---we are able to reproduce the HST-based results. However, these models never produce a consistent set of image positions, implying that this flexibility does not meet the astrometric constraints required for cosmology \citep{birrer2019}.
	
	We also tested using the previous Osipkov-Merritt \citep{osipkov1979,merritt1985a} anisotropy parameterization and previous velocity dispersion measurements to see if there was any significant change in H$_0$ resulting from these improvements, and we found no significant change. This is likely due to the fact that this particular system is not particularly sensitive to kinematic measurements (for $\lint=1$). Similarly, using the time delay measurements as constraints during fitting, increase the time delay distance by less than the uncertainty ($\sim\!$2\%).
	
\begin{figure*}[ht]
		\centering
		\includegraphics[width=\textwidth]{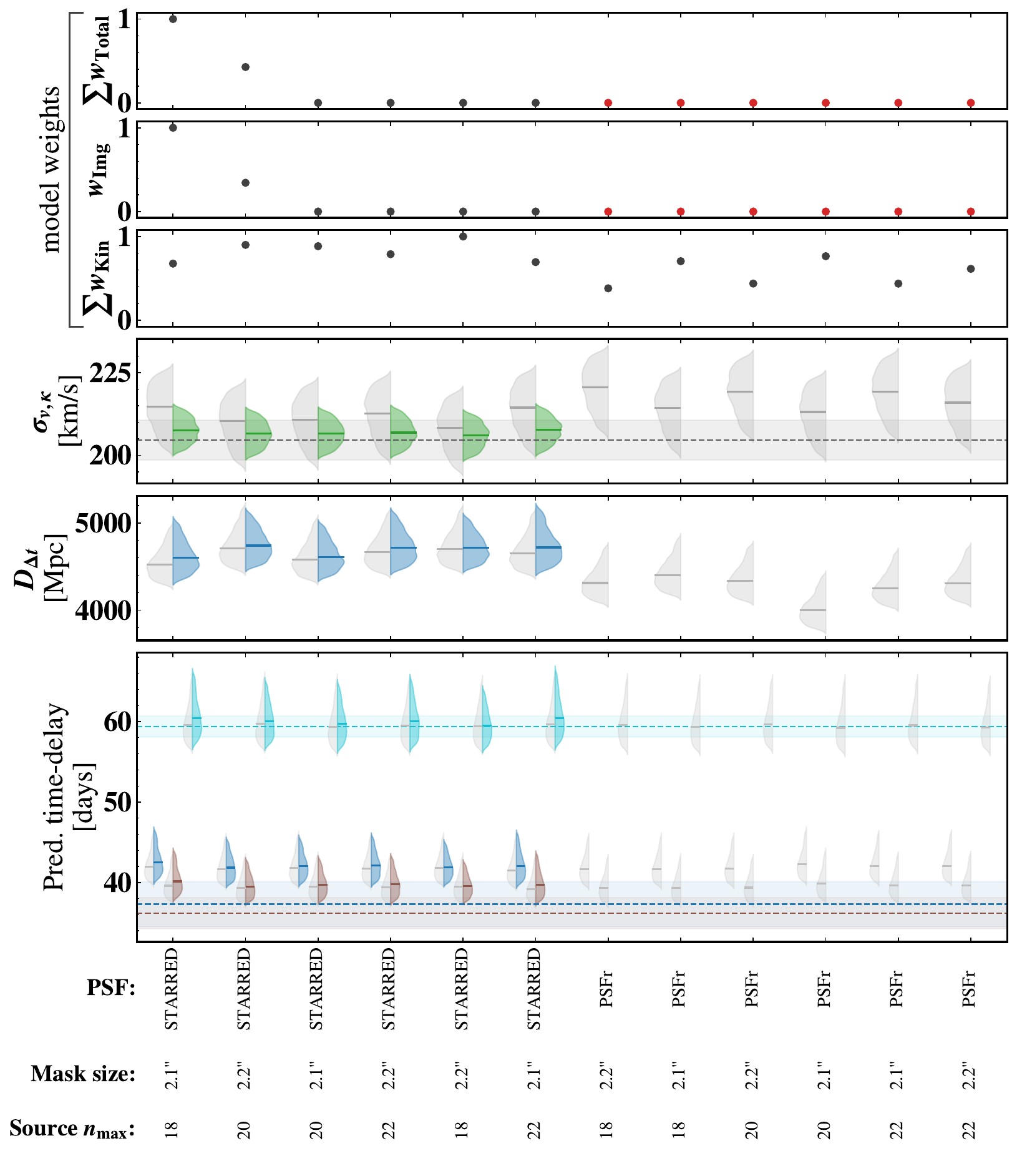}
		\caption{WFI2033 model results, with model weights at the top. First is the total model weight $w_{\rm{Total}}$, second row is the imaging-only (or BIC) weight $w_{\rm{Img}}$, and third row shows the kinematic weights $w_{\rm{Kin}}$ of each of the models, where points in red indicate a weight of 0.
        The fourth row compares the model-predicted velocity dispersions $\sigma_{v,\kappa}$ to the externally observed value (black line with gray 16/84\% region). Then the sampled time-delay distance $D_{\Delta t}$ is plotted, followed by the predicted time-delays for each of the image pairs relative to image B ($\Delta t_{CB}$ in cyan, $\Delta t_{A2B}$ in blue, $\Delta t_{A1B}$ in brown), with the externally observed values plotted in the corresponding color. The left/gray half of the violin plot shows all samples from the model, while the right/colored half shows the weighted selection within the model, selecting samples best able to reproduce the image and observed kinematics.}
		\label{fig:wfi2033_plot_final}
	\end{figure*}
	
	\begin{figure*}[ht]
		\centering
		\includegraphics[width=\textwidth]{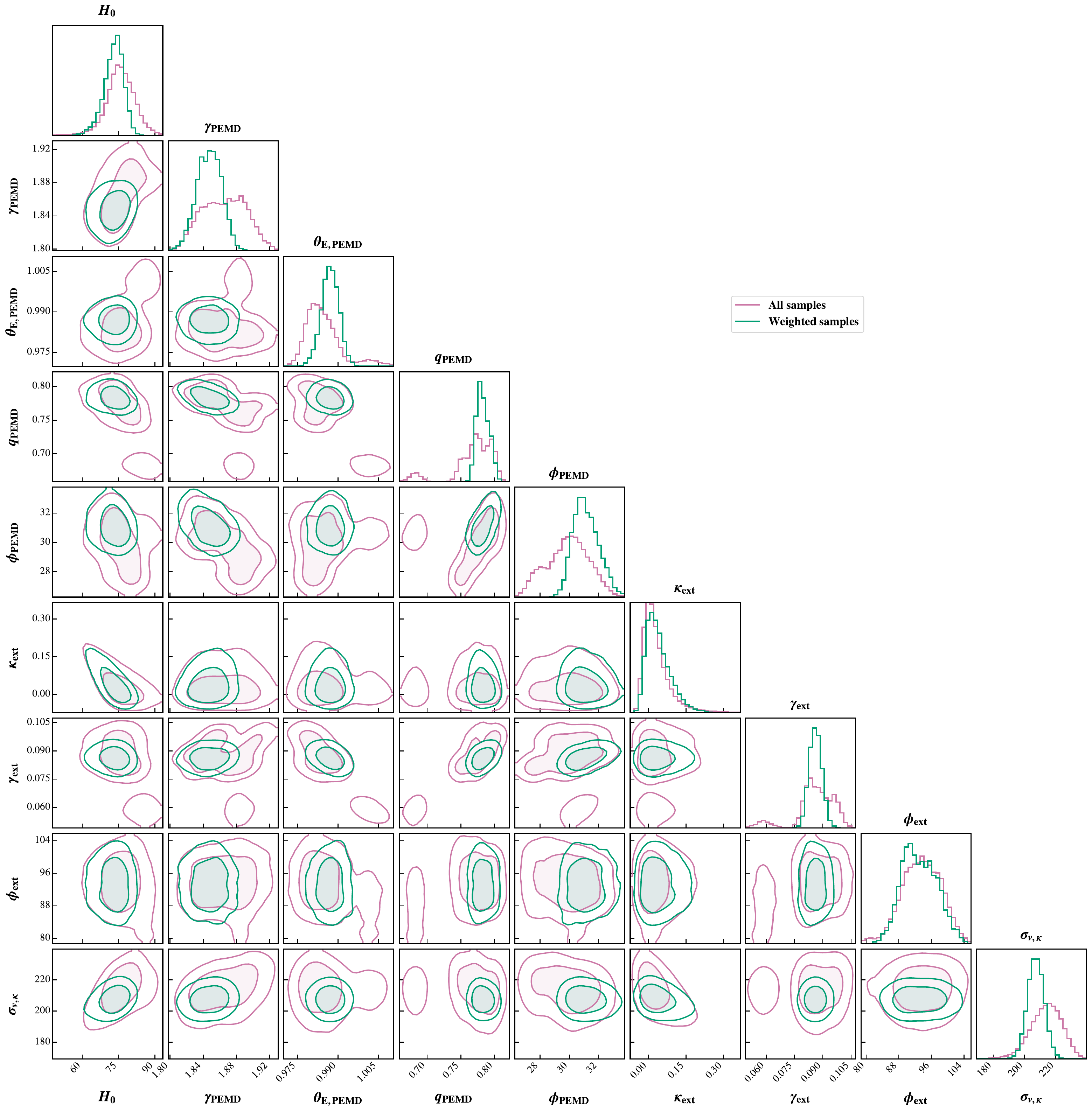}
		\caption{WFI2033 parameter distributions, with unweighted samples shown in pink and the final weighted selection shown in green. The contours represent the 68\% and 95\% highest-posterior-density credible regions. We note bimodality in some parameter distributions due to the heavy weighting of two of the 16 models, but we see this selection has no impact on H$_0$.}
		\label{fig:wfi2033_corner}
	\end{figure*}
	
	\subsection{HE0435}

\begin{figure*}[ht]
		\centering
		\includegraphics[width=\textwidth]{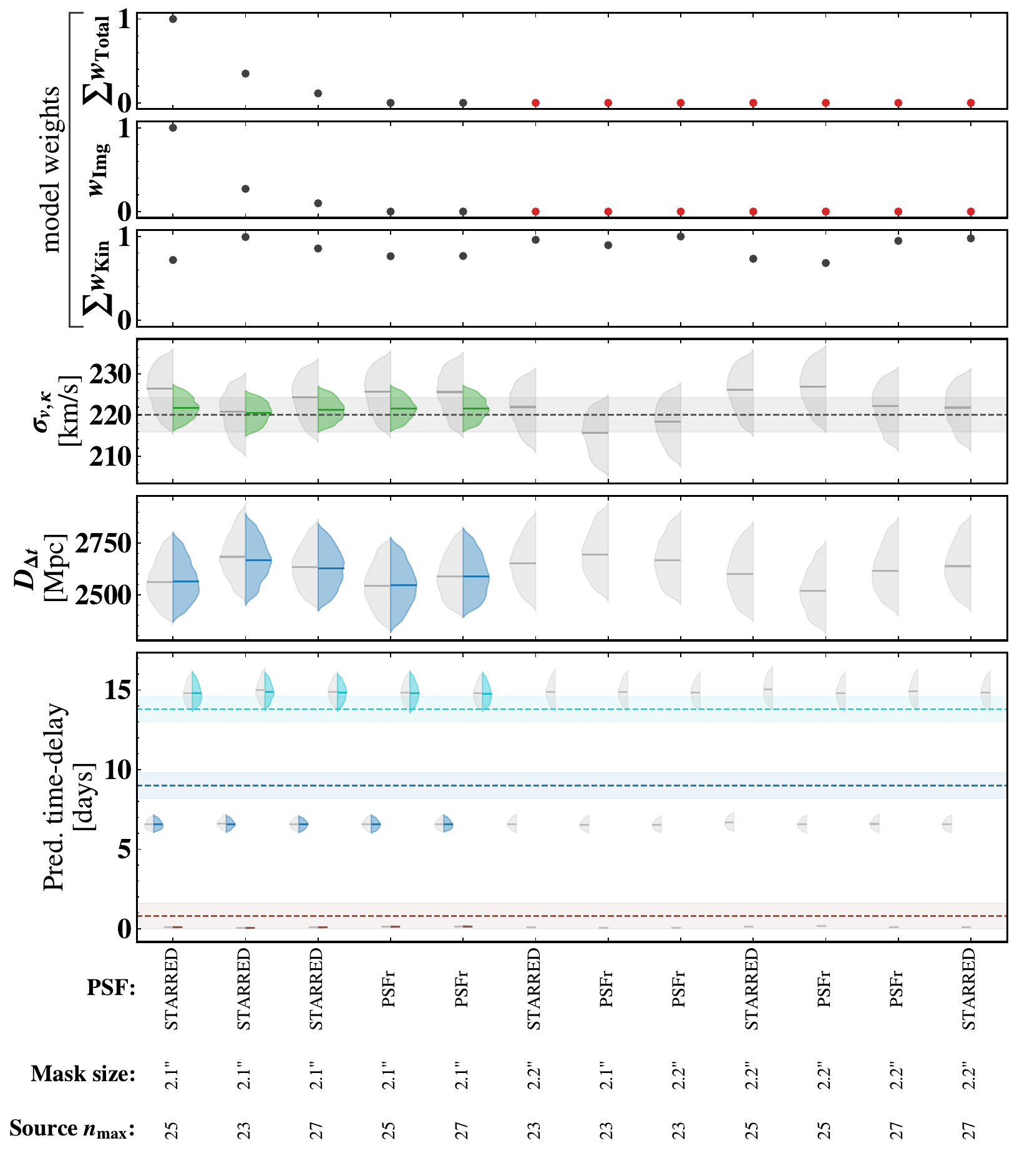}
		\caption{Same as \Cref{fig:wfi2033_plot_final} but for HE0435.}
		\label{fig:he0435_plot_final}
	\end{figure*}

\begin{figure*}[ht]
		\centering
		\includegraphics[width=\textwidth]{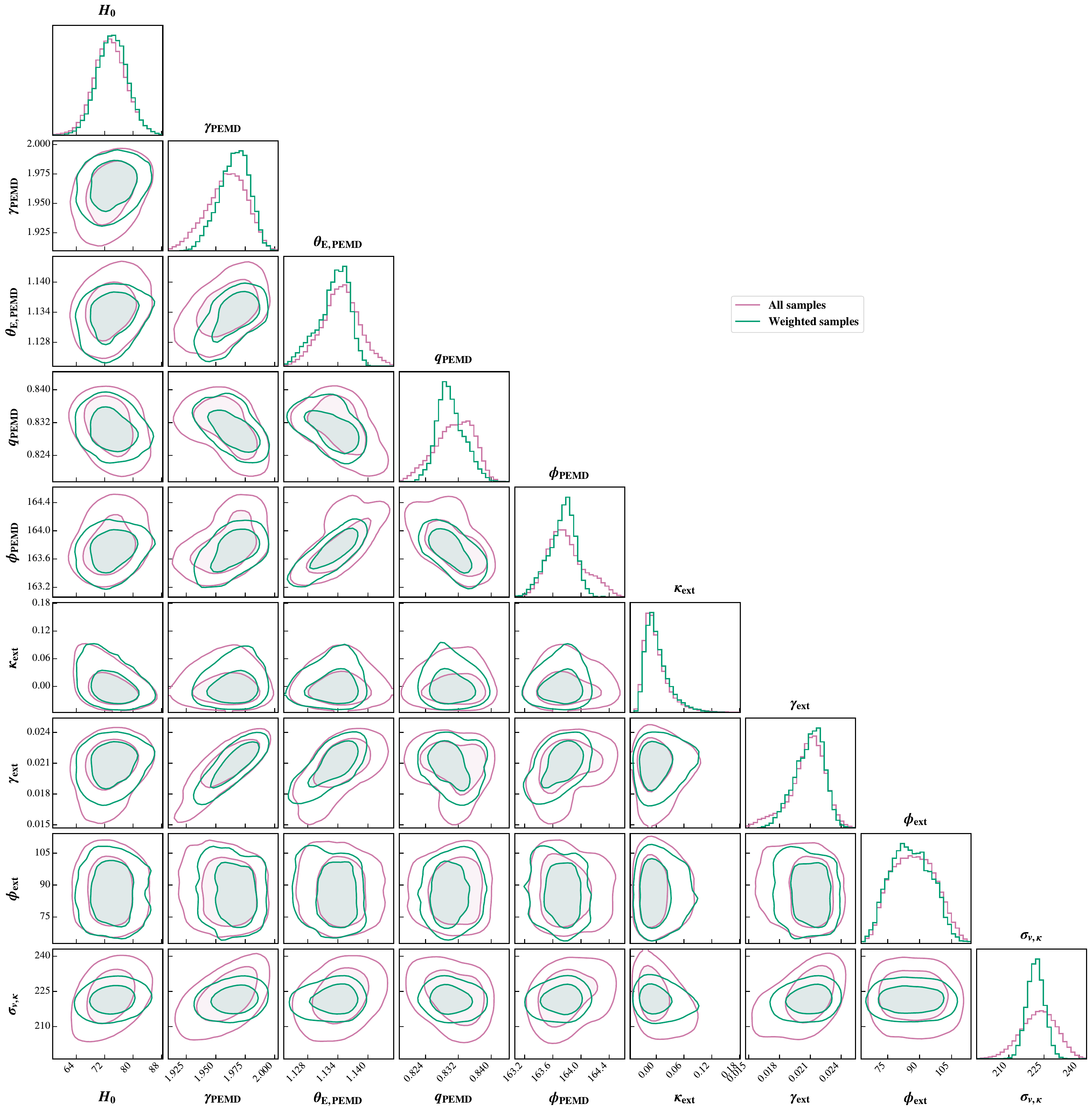}
		\caption{Same as \cref{fig:wfi2033_corner} but for HE0435.}
		\label{fig:he0435_corner}
	\end{figure*}

    The model parameters for \helens are well constrained. The data prefer the STARRED PSF model, although two PSFr models get some weight (Figure~\ref{fig:he0435_plot_final}. The predicted stellar velocity dispersion is stable and in good agreement with the measured value. The Time delay distance is stable. As a result, H$_0$ is well constrained and so are most model parameters (Figure~\ref{fig:he0435_corner}).

	\subsection{PG1115}

\begin{figure*}[ht]
		\centering
		\includegraphics[width=\textwidth]{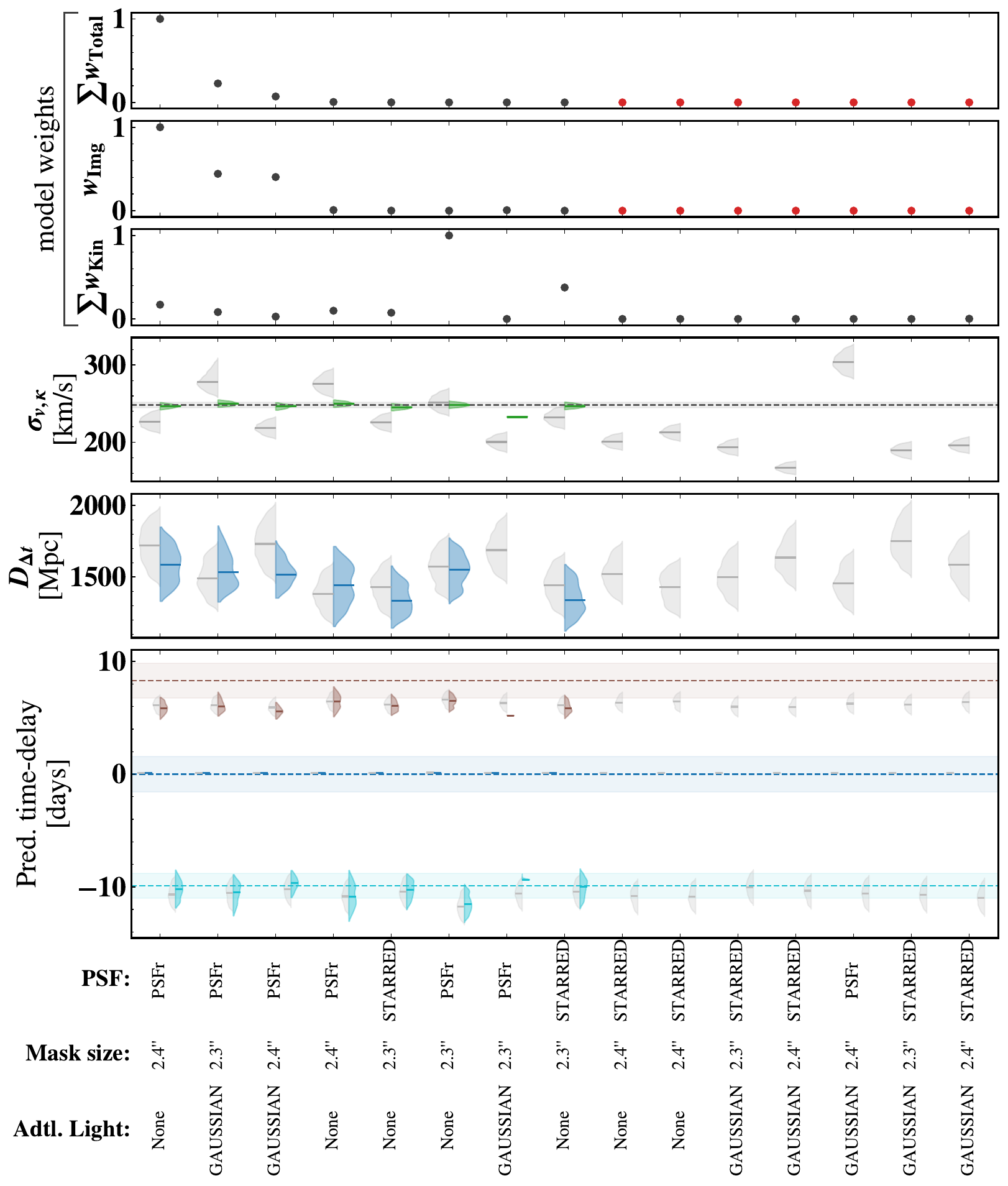}
		\caption{Same as \Cref{fig:wfi2033_plot_final} but for PG1115. We note a wide range of velocity dispersion predictions along with the time-delay distance, due to limitations in data quality. As the two eastern images are saturated and the configuration is highly symmetric, some models have significant difficulty fitting this region of the image, leading to weak constraints on some lens parameters.}
		\label{fig:pg1115_plot_final}
	\end{figure*}

\begin{figure*}[ht]
		\centering
		\includegraphics[width=\textwidth]{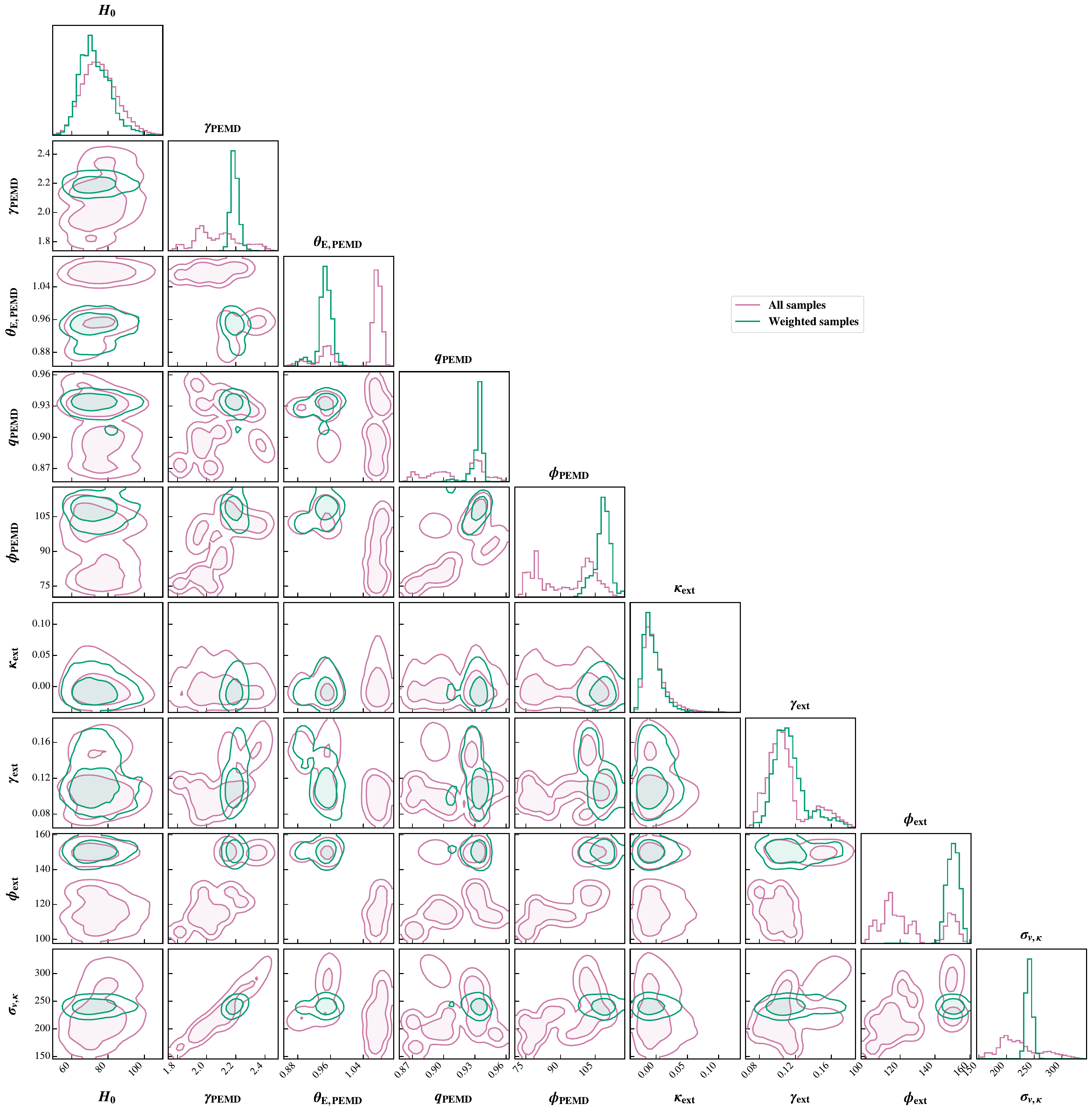}
		\caption{Same as \cref{fig:wfi2033_corner} but for PG1115.}
		\label{fig:pg1115_corner}
	\end{figure*}

    The final inference of this system is largely dominated by model-predicted kinematics. Due to the saturation of the eastern two images and fainter ring, the symmetry of the configuration, and the smooth host galaxy surface brightness, the models struggled to tightly constrain the mass profile of the deflector. This led to significant variability in model-predicted velocity dispersion (\cref{fig:pg1115_plot_final}). Therefore, once the observed velocity dispersions are compared to model predictions, these values slice a subset of these models which are subject to variability in the velocity dispersion due to the mass-sheet degeneracy (this selection can be seen most clearly in the selection of the power-law slope in \Cref{fig:pg1115_corner}). We expect the inference for this system to be most improved by the addition of spatially resolved kinematics in combination with a flexibile mass sheet ($\lint\neq1$). This will reduce the residual degeneracies between cosmological parameters and in the model parameters \Cref{fig:pg1115_plot_final}. This analysis will be done at a population-scale in our forthcoming TDCOSMO 2026 milestone paper. Despite these limitations in the data, our results are consistent with independent datasets from HST and AO observations from Keck, providing an encouraging validation of our PSF reconstruction techniques even facing saturated point sources.

    \subsection{Comparison with previous work}

The analysis of three systems provides a great opportunity to compare with previous work based on independent data and methodology. The comparison is summarized in Figure~\ref{fig:H0_update_comparison}. The top panel shows the individual results for the three systems comparing the previous analysis \citep{wong2017,chen2019,rusu2020} to the current one. As a reminder many things have changed between the two analyses: i) we used Lenstronomy \citep{birrer2018} instead of GLEE \citep{GLEE}; ii) we used JWST NIRCam imaging instead of HST/Keck-AO imaging; iii)  we reconstructed the PSF using the STARRED and PSFr algorithms; iv) we used stellar velocity dispersion from JWST NIRSpec spectra, instead of ground based measurements; v) we used new estimates of the external convergence based on Euclid Flagship simulations instead of Millennium based ones. The time delays are largely the same with minor updates, detailed earlier in this paper. The mass sheet degeneracy parameter is kept fixed $\lint=1$, and the prior on $\Omega_{\rm m}$ is the same. 

The main results are the following. First, the inference from each individual system is consistent within the errors. Second, the uncertainty per system is reduced somewhat but not dramatically, as expected since the time delay and external convergence impose a noise floor. Third, the scatter between the systems is reduced.  This is very encouraging, as the measurement was carried out blindly, and it did not have to be this way. With three systems and the current level of random uncertainties, we cannot quantify whether this is significant or a coincidence (in fact, the \citet{millon2020b} results were consistent with zero excess variance). We will soon be in a position to quantify this with future expanded samples, as we discuss below.

	\section{Conclusion}
	\label{sec:conclusion}

    This work utilizes JWST/NIRCam imaging to construct lens models and produce H$_0$ measurements of three quadruply imaged quasars. In addition to the JWST/NIRcam images, we use improved measurements of stellar velocity dispersion from JWST/NIRSpec (Knabel et al. 2026, in prep), and line of sight external convergence (Johnson et al. 2026, in prep). Analysis of two of the systems is carried out blinded, while for one of the systems we update a previous model by our team that was not blinded \citep{williams2025}.
    
    The three systems have been previously modeled using HST/Keck-AO images, ground based stellar velocity dispersions, external convergence estimates based on older cosmological simulations, and a different lens modeling software - GLEE vs Lenstronomy adopted in this work - \citep{rusu2020,chen2019,wong2017}. The comparison with previous results, provides an important test of systematic effects across telescopes and modeling choices. To facilitate comparison with previous work we adopt $\lint=1$ and a flat prior on $\Omega_{\rm m}$. 

    Our main results can be summarized as follows

\begin{itemize}
    \item Using improved PSF reconstruction techniques we show that high precision lens models can be obtained with JWST-NIRCam data, even when some of the point sources are saturated.
    \item Our lens models are more precise and consistent with those based on HST and Keck-AO data based on different software, stellar kinematics, and external convergence estimate.
    \item The three systems give independent measurements of H$_0$ in excellent mutual agreement: \hvalhe $\lint$ \hunit for \helens; \hvalpg$\lint$\hunit for \pglens; \hvalwfi$\lint$ \hunit for \wfilens.
    \item Combining the three lenses yields \hval $\lint$ \hunit.
\end{itemize}

    Our lens models will be combined with spatially resolved kinematics and the models for other lenses in a hierarchical analysis relaxing the $\lint=1$ assumption and adopting empirically motivated priors on $\Omega_{\rm m}$ in our TDCOSMO 2026 milestone paper.

    Looking ahead we will soon be able to expand our NIRcam modeling of time delay lenses to 19, based on approved Cycle-4 and Cycle-5 Programs. In addition to the increase in sample size, the new programs are designed for lens modeling and therefore the improvement should be more than just square root of 19/3. The Cycle4/5 data have twice the exposure times compared with the data discussed here. Furthermore, they take advantage of a strategy designed to deal with quasar saturation,  stacking a short exposure for constraints on the quasar images with a longer exposure for further constraints on extended emission.

    \begin{acknowledgments}
    DW and TT acknowledge support by National Science Foundation through grant NSF-AST-2407277, and from the Moore Foundation through grant 8548.
    SB acknowledges support by the Department of Physics and Astronomy, Stony Brook University.
    KCW is supported by JSPS KAKENHI Grant Numbers JP24K07089, JP24H00221.
    DS acknowledges the support of the Fonds de la Recherche Scientifique-FNRS, Belgium, under grant No.\ 4.4503.1.
    DJ acknowledges support by the First Rand Foundation, South Africa, and the Centre National de la Recherche Scientifique of France. This work has made use of CosmoHub, developed by PIC (maintained by IFAE and CIEMAT) in collaboration with ICE-CSIC. It received funding from the Spanish government (grant EQC2021-007479-P funded by MCIN/AEI/10.13039/501100011033), the EU NextGeneration/PRTR (PRTR-C17.I1), and the Generalitat de Catalunya.
    V.M.\ acknowledges support from ANID FONDECYT Regular grant number 1231418, and Centro de Astrof\'{\i}sica de Valpara\'{\i}so CIDI 21.
    This work was supported by the Agencia Estatal de Investigaci\'on (AEI), Ministerio de Ciencia, Innovaci\'on y Universidades, Spain, under the project PID2024-155455NB-I00, within the 2024 Call for Knowledge Generation Projects.
    \end{acknowledgments}

    \bibliographystyle{apsrev4-2}
    \bibliography{trimmed_final}

\begin{thebibliography}{101}%
\makeatletter
\providecommand \@ifxundefined [1]{%
 \@ifx{#1\undefined}
}%
\providecommand \@ifnum [1]{%
 \ifnum #1\expandafter \@firstoftwo
 \else \expandafter \@secondoftwo
 \fi
}%
\providecommand \@ifx [1]{%
 \ifx #1\expandafter \@firstoftwo
 \else \expandafter \@secondoftwo
 \fi
}%
\providecommand \natexlab [1]{#1}%
\providecommand \enquote  [1]{``#1''}%
\providecommand \bibnamefont  [1]{#1}%
\providecommand \bibfnamefont [1]{#1}%
\providecommand \citenamefont [1]{#1}%
\providecommand \href@noop [0]{\@secondoftwo}%
\providecommand \href [0]{\begingroup \@sanitize@url \@href}%
\providecommand \@href[1]{\@@startlink{#1}\@@href}%
\providecommand \@@href[1]{\endgroup#1\@@endlink}%
\providecommand \@sanitize@url [0]{\catcode `\\12\catcode `\$12\catcode `\&12\catcode `\#12\catcode `\^12\catcode `\_12\catcode `\%12\relax}%
\providecommand \@@startlink[1]{}%
\providecommand \@@endlink[0]{}%
\providecommand \url  [0]{\begingroup\@sanitize@url \@url }%
\providecommand \@url [1]{\endgroup\@href {#1}{\urlprefix }}%
\providecommand \urlprefix  [0]{URL }%
\providecommand \Eprint [0]{\href }%
\providecommand \doibase [0]{https://doi.org/}%
\providecommand \selectlanguage [0]{\@gobble}%
\providecommand \bibinfo  [0]{\@secondoftwo}%
\providecommand \bibfield  [0]{\@secondoftwo}%
\providecommand \translation [1]{[#1]}%
\providecommand \BibitemOpen [0]{}%
\providecommand \bibitemStop [0]{}%
\providecommand \bibitemNoStop [0]{.\EOS\space}%
\providecommand \EOS [0]{\spacefactor3000\relax}%
\providecommand \BibitemShut  [1]{\csname bibitem#1\endcsname}%
\let\auto@bib@innerbib\@empty
\bibitem [{\citenamefont {Di~Valentino}\ \emph {et~al.}(2025{\natexlab{a}})\citenamefont {Di~Valentino}, \citenamefont {Said}, \citenamefont {Riess}, \citenamefont {Pollo}, \citenamefont {Poulin}, \citenamefont {Gómez-Valent}, \citenamefont {Weltman}, \citenamefont {Palmese}, \citenamefont {Huang}, \citenamefont {van~de Bruck} \emph {et~al.}}]{divalentino2025a}%
  \BibitemOpen
  \bibfield  {author} {\bibinfo {author} {\bibfnamefont {E.}~\bibnamefont {Di~Valentino}}, \bibinfo {author} {\bibfnamefont {J.~L.}\ \bibnamefont {Said}}, \bibinfo {author} {\bibfnamefont {A.}~\bibnamefont {Riess}}, \bibinfo {author} {\bibfnamefont {A.}~\bibnamefont {Pollo}}, \bibinfo {author} {\bibfnamefont {V.}~\bibnamefont {Poulin}}, \bibinfo {author} {\bibfnamefont {A.}~\bibnamefont {Gómez-Valent}}, \bibinfo {author} {\bibfnamefont {A.}~\bibnamefont {Weltman}}, \bibinfo {author} {\bibfnamefont {A.}~\bibnamefont {Palmese}}, \bibinfo {author} {\bibfnamefont {C.~D.}\ \bibnamefont {Huang}}, \bibinfo {author} {\bibfnamefont {C.}~\bibnamefont {van~de Bruck}}, \emph {et~al.},\ }\href {https://doi.org/10.1016/j.dark.2025.101965} {\bibfield  {journal} {\bibinfo  {journal} {Physics of the Dark Universe}\ }\textbf {\bibinfo {volume} {49}},\ \bibinfo {pages} {101965} (\bibinfo {year} {2025}{\natexlab{a}})},\ \bibinfo {note} {aDS Bibcode: 2025PDU....4901965D}\BibitemShut {NoStop}%
\bibitem [{\citenamefont {Adame}\ \emph {et~al.}(2025)\citenamefont {Adame}, \citenamefont {Aguilar}, \citenamefont {Ahlen}, \citenamefont {Alam}, \citenamefont {Alexander}, \citenamefont {Allende~Prieto}, \citenamefont {Alvarez}, \citenamefont {Alves}, \citenamefont {Anand}, \citenamefont {Andrade} \emph {et~al.}}]{adame2025}%
  \BibitemOpen
  \bibfield  {author} {\bibinfo {author} {\bibfnamefont {A.~G.}\ \bibnamefont {Adame}}, \bibinfo {author} {\bibfnamefont {J.}~\bibnamefont {Aguilar}}, \bibinfo {author} {\bibfnamefont {S.}~\bibnamefont {Ahlen}}, \bibinfo {author} {\bibfnamefont {S.}~\bibnamefont {Alam}}, \bibinfo {author} {\bibfnamefont {D.~M.}\ \bibnamefont {Alexander}}, \bibinfo {author} {\bibfnamefont {C.}~\bibnamefont {Allende~Prieto}}, \bibinfo {author} {\bibfnamefont {M.}~\bibnamefont {Alvarez}}, \bibinfo {author} {\bibfnamefont {O.}~\bibnamefont {Alves}}, \bibinfo {author} {\bibfnamefont {A.}~\bibnamefont {Anand}}, \bibinfo {author} {\bibfnamefont {U.}~\bibnamefont {Andrade}}, \emph {et~al.},\ }\href {https://doi.org/10.1088/1475-7516/2025/07/028} {\bibfield  {journal} {\bibinfo  {journal} {Journal of Cosmology and Astroparticle Physics}\ }\textbf {\bibinfo {volume} {2025}},\ \bibinfo {pages} {28}},\ \bibinfo {note} {aDS Bibcode: 2025JCAP...07..028A}\BibitemShut {NoStop}%
\bibitem [{\citenamefont {{Des Collaboration}}\ \emph {et~al.}(2026)\citenamefont {{Des Collaboration}}, \citenamefont {Abbott}, \citenamefont {Adamow}, \citenamefont {Aguena}, \citenamefont {Alarcon}, \citenamefont {Allam}, \citenamefont {Alves}, \citenamefont {Amon}, \citenamefont {Anbajagane}, \citenamefont {Andrade-Oliveira} \emph {et~al.}}]{descollaboration2026}%
  \BibitemOpen
  \bibfield  {author} {\bibinfo {author} {\bibnamefont {{Des Collaboration}}}, \bibinfo {author} {\bibfnamefont {T.~M.~C.}\ \bibnamefont {Abbott}}, \bibinfo {author} {\bibfnamefont {M.}~\bibnamefont {Adamow}}, \bibinfo {author} {\bibfnamefont {M.}~\bibnamefont {Aguena}}, \bibinfo {author} {\bibfnamefont {A.}~\bibnamefont {Alarcon}}, \bibinfo {author} {\bibfnamefont {S.~S.}\ \bibnamefont {Allam}}, \bibinfo {author} {\bibfnamefont {O.}~\bibnamefont {Alves}}, \bibinfo {author} {\bibfnamefont {A.}~\bibnamefont {Amon}}, \bibinfo {author} {\bibfnamefont {D.}~\bibnamefont {Anbajagane}}, \bibinfo {author} {\bibfnamefont {F.}~\bibnamefont {Andrade-Oliveira}}, \emph {et~al.},\ }\href {https://doi.org/10.48550/arXiv.2601.14559} {{\selectlanguage {en}\bibinfo {title} {Dark energy survey year 6 results: cosmological constraints from galaxy clustering and weak lensing}}} (\bibinfo {year} {2026}),\ \bibinfo {note} {aDS Bibcode: 2026arXiv260114559D}\BibitemShut {NoStop}%
\bibitem [{\citenamefont {Dalal}\ \emph {et~al.}(2023)\citenamefont {Dalal}, \citenamefont {Li}, \citenamefont {Nicola}, \citenamefont {Zuntz}, \citenamefont {Strauss}, \citenamefont {Sugiyama}, \citenamefont {Zhang}, \citenamefont {Rau}, \citenamefont {Mandelbaum}, \citenamefont {Takada} \emph {et~al.}}]{dalal2023}%
  \BibitemOpen
  \bibfield  {author} {\bibinfo {author} {\bibfnamefont {R.}~\bibnamefont {Dalal}}, \bibinfo {author} {\bibfnamefont {X.}~\bibnamefont {Li}}, \bibinfo {author} {\bibfnamefont {A.}~\bibnamefont {Nicola}}, \bibinfo {author} {\bibfnamefont {J.}~\bibnamefont {Zuntz}}, \bibinfo {author} {\bibfnamefont {M.~A.}\ \bibnamefont {Strauss}}, \bibinfo {author} {\bibfnamefont {S.}~\bibnamefont {Sugiyama}}, \bibinfo {author} {\bibfnamefont {T.}~\bibnamefont {Zhang}}, \bibinfo {author} {\bibfnamefont {M.~M.}\ \bibnamefont {Rau}}, \bibinfo {author} {\bibfnamefont {R.}~\bibnamefont {Mandelbaum}}, \bibinfo {author} {\bibfnamefont {M.}~\bibnamefont {Takada}}, \emph {et~al.},\ }\href {https://doi.org/10.1103/PhysRevD.108.123519} {\bibfield  {journal} {\bibinfo  {journal} {Physical Review D}\ }\textbf {\bibinfo {volume} {108}},\ \bibinfo {pages} {123519} (\bibinfo {year} {2023})},\ \bibinfo {note} {aDS Bibcode: 2023PhRvD.108l3519D}\BibitemShut {NoStop}%
\bibitem [{\citenamefont {Vagnozzi}(2023)}]{vagnozzi2023}%
  \BibitemOpen
  \bibfield  {author} {\bibinfo {author} {\bibfnamefont {S.}~\bibnamefont {Vagnozzi}},\ }\href {https://doi.org/10.3390/universe9090393} {\bibfield  {journal} {\bibinfo  {journal} {Universe}\ }\textbf {\bibinfo {volume} {9}},\ \bibinfo {pages} {393} (\bibinfo {year} {2023})},\ \bibinfo {note} {aDS Bibcode: 2023Univ....9..393V}\BibitemShut {NoStop}%
\bibitem [{\citenamefont {Turner}(2025)}]{turner2025}%
  \BibitemOpen
  \bibfield  {author} {\bibinfo {author} {\bibfnamefont {M.~S.}\ \bibnamefont {Turner}},\ }\href {https://doi.org/10.48550/arXiv.2510.05483} {\bibinfo {title} {Everyone wants something better than \${\textbackslash}{Lambda}\${CDM}}} (\bibinfo {year} {2025}),\ \bibinfo {note} {arXiv:2510.05483 [astro-ph]}\BibitemShut {NoStop}%
\bibitem [{\citenamefont {Knox}\ and\ \citenamefont {Millea}(2020)}]{knox2020}%
  \BibitemOpen
  \bibfield  {author} {\bibinfo {author} {\bibfnamefont {L.}~\bibnamefont {Knox}}\ and\ \bibinfo {author} {\bibfnamefont {M.}~\bibnamefont {Millea}},\ }\bibfield  {journal} {\bibinfo  {journal} {Physical Review D}\ }\textbf {\bibinfo {volume} {101}},\ \href {https://doi.org/10.1103/physrevd.101.043533} {10.1103/physrevd.101.043533} (\bibinfo {year} {2020})\BibitemShut {NoStop}%
\bibitem [{\citenamefont {Efstathiou}(2021)}]{efstathiou2021}%
  \BibitemOpen
  \bibfield  {author} {\bibinfo {author} {\bibfnamefont {G.}~\bibnamefont {Efstathiou}},\ }\href {https://doi.org/10.1093/mnras/stab1588} {\bibfield  {journal} {\bibinfo  {journal} {Monthly Notices of the Royal Astronomical Society}\ }\textbf {\bibinfo {volume} {505}},\ \bibinfo {pages} {3866} (\bibinfo {year} {2021})},\ \bibinfo {note} {aDS Bibcode: 2021MNRAS.505.3866E}\BibitemShut {NoStop}%
\bibitem [{\citenamefont {Di~Valentino}\ \emph {et~al.}(2025{\natexlab{b}})\citenamefont {Di~Valentino}, \citenamefont {Levi~Said}, \citenamefont {Riess}, \citenamefont {Pollo}, \citenamefont {Poulin}, \citenamefont {Gómez-Valent}, \citenamefont {Weltman}, \citenamefont {Palmese}, \citenamefont {Huang}, \citenamefont {van~de Bruck} \emph {et~al.}}]{divalentino2025}%
  \BibitemOpen
  \bibfield  {author} {\bibinfo {author} {\bibfnamefont {E.}~\bibnamefont {Di~Valentino}}, \bibinfo {author} {\bibfnamefont {J.}~\bibnamefont {Levi~Said}}, \bibinfo {author} {\bibfnamefont {A.}~\bibnamefont {Riess}}, \bibinfo {author} {\bibfnamefont {A.}~\bibnamefont {Pollo}}, \bibinfo {author} {\bibfnamefont {V.}~\bibnamefont {Poulin}}, \bibinfo {author} {\bibfnamefont {A.}~\bibnamefont {Gómez-Valent}}, \bibinfo {author} {\bibfnamefont {A.}~\bibnamefont {Weltman}}, \bibinfo {author} {\bibfnamefont {A.}~\bibnamefont {Palmese}}, \bibinfo {author} {\bibfnamefont {C.~D.}\ \bibnamefont {Huang}}, \bibinfo {author} {\bibfnamefont {C.}~\bibnamefont {van~de Bruck}}, \emph {et~al.},\ }\href {https://doi.org/10.48550/arXiv.2504.01669} {\bibinfo {title} {The {CosmoVerse} {White} {Paper}: {Addressing} observational tensions in cosmology with systematics and fundamental physics}} (\bibinfo {year} {2025}{\natexlab{b}}),\ \bibinfo {note} {aDS Bibcode: 2025arXiv250401669D}\BibitemShut {NoStop}%
\bibitem [{\citenamefont {Riess}\ \emph {et~al.}(2019)\citenamefont {Riess}, \citenamefont {Casertano}, \citenamefont {Yuan}, \citenamefont {Macri},\ and\ \citenamefont {Scolnic}}]{riess2019}%
  \BibitemOpen
  \bibfield  {author} {\bibinfo {author} {\bibfnamefont {A.~G.}\ \bibnamefont {Riess}}, \bibinfo {author} {\bibfnamefont {S.}~\bibnamefont {Casertano}}, \bibinfo {author} {\bibfnamefont {W.}~\bibnamefont {Yuan}}, \bibinfo {author} {\bibfnamefont {L.~M.}\ \bibnamefont {Macri}},\ and\ \bibinfo {author} {\bibfnamefont {D.}~\bibnamefont {Scolnic}},\ }\href {https://doi.org/10.3847/1538-4357/ab1422} {\bibfield  {journal} {\bibinfo  {journal} {Astrophysical Journal}\ }\textbf {\bibinfo {volume} {876}},\ \bibinfo {pages} {85} (\bibinfo {year} {2019})}\BibitemShut {NoStop}%
\bibitem [{\citenamefont {Riess}\ \emph {et~al.}(2021)\citenamefont {Riess}, \citenamefont {Casertano}, \citenamefont {Yuan}, \citenamefont {Bowers}, \citenamefont {Macri}, \citenamefont {Zinn},\ and\ \citenamefont {Scolnic}}]{riess2021}%
  \BibitemOpen
  \bibfield  {author} {\bibinfo {author} {\bibfnamefont {A.~G.}\ \bibnamefont {Riess}}, \bibinfo {author} {\bibfnamefont {S.}~\bibnamefont {Casertano}}, \bibinfo {author} {\bibfnamefont {W.}~\bibnamefont {Yuan}}, \bibinfo {author} {\bibfnamefont {J.~B.}\ \bibnamefont {Bowers}}, \bibinfo {author} {\bibfnamefont {L.}~\bibnamefont {Macri}}, \bibinfo {author} {\bibfnamefont {J.~C.}\ \bibnamefont {Zinn}},\ and\ \bibinfo {author} {\bibfnamefont {D.}~\bibnamefont {Scolnic}},\ }\href {https://doi.org/10.3847/2041-8213/abdbaf} {\bibfield  {journal} {\bibinfo  {journal} {Astrophysical Journal Letters}\ }\textbf {\bibinfo {volume} {908}},\ \bibinfo {pages} {L6} (\bibinfo {year} {2021})},\ \bibinfo {note} {aDS Bibcode: 2021ApJ...908L...6R}\BibitemShut {NoStop}%
\bibitem [{\citenamefont {Riess}\ \emph {et~al.}(2022)\citenamefont {Riess}, \citenamefont {Yuan}, \citenamefont {Macri}, \citenamefont {Scolnic}, \citenamefont {Brout}, \citenamefont {Casertano}, \citenamefont {Jones}, \citenamefont {Murakami}, \citenamefont {Anand}, \citenamefont {Breuval} \emph {et~al.}}]{riess2022}%
  \BibitemOpen
  \bibfield  {author} {\bibinfo {author} {\bibfnamefont {A.~G.}\ \bibnamefont {Riess}}, \bibinfo {author} {\bibfnamefont {W.}~\bibnamefont {Yuan}}, \bibinfo {author} {\bibfnamefont {L.~M.}\ \bibnamefont {Macri}}, \bibinfo {author} {\bibfnamefont {D.}~\bibnamefont {Scolnic}}, \bibinfo {author} {\bibfnamefont {D.}~\bibnamefont {Brout}}, \bibinfo {author} {\bibfnamefont {S.}~\bibnamefont {Casertano}}, \bibinfo {author} {\bibfnamefont {D.~O.}\ \bibnamefont {Jones}}, \bibinfo {author} {\bibfnamefont {Y.}~\bibnamefont {Murakami}}, \bibinfo {author} {\bibfnamefont {G.~S.}\ \bibnamefont {Anand}}, \bibinfo {author} {\bibfnamefont {L.}~\bibnamefont {Breuval}}, \emph {et~al.},\ }\href {https://doi.org/10.3847/2041-8213/ac5c5b} {\bibfield  {journal} {\bibinfo  {journal} {Astrophysical Journal Letters}\ }\textbf {\bibinfo {volume} {934}},\ \bibinfo {pages} {L7} (\bibinfo {year} {2022})}\BibitemShut {NoStop}%
\bibitem [{\citenamefont {Dainotti}\ \emph {et~al.}(2021)\citenamefont {Dainotti}, \citenamefont {De~Simone}, \citenamefont {Schiavone}, \citenamefont {Montani}, \citenamefont {Rinaldi},\ and\ \citenamefont {Lambiase}}]{dainotti2021}%
  \BibitemOpen
  \bibfield  {author} {\bibinfo {author} {\bibfnamefont {M.~G.}\ \bibnamefont {Dainotti}}, \bibinfo {author} {\bibfnamefont {B.}~\bibnamefont {De~Simone}}, \bibinfo {author} {\bibfnamefont {T.}~\bibnamefont {Schiavone}}, \bibinfo {author} {\bibfnamefont {G.}~\bibnamefont {Montani}}, \bibinfo {author} {\bibfnamefont {E.}~\bibnamefont {Rinaldi}},\ and\ \bibinfo {author} {\bibfnamefont {G.}~\bibnamefont {Lambiase}},\ }\href {https://doi.org/10.3847/1538-4357/abeb73} {\bibfield  {journal} {\bibinfo  {journal} {Astrophysical Journal}\ }\textbf {\bibinfo {volume} {912}},\ \bibinfo {pages} {150} (\bibinfo {year} {2021})}\BibitemShut {NoStop}%
\bibitem [{\citenamefont {Mörtsell}\ \emph {et~al.}(2022)\citenamefont {Mörtsell}, \citenamefont {Goobar}, \citenamefont {Johansson},\ and\ \citenamefont {Dhawan}}]{mortsell2022}%
  \BibitemOpen
  \bibfield  {author} {\bibinfo {author} {\bibfnamefont {E.}~\bibnamefont {Mörtsell}}, \bibinfo {author} {\bibfnamefont {A.}~\bibnamefont {Goobar}}, \bibinfo {author} {\bibfnamefont {J.}~\bibnamefont {Johansson}},\ and\ \bibinfo {author} {\bibfnamefont {S.}~\bibnamefont {Dhawan}},\ }\href {https://doi.org/10.3847/1538-4357/ac756e} {\bibfield  {journal} {\bibinfo  {journal} {Astrophysical Journal}\ }\textbf {\bibinfo {volume} {933}},\ \bibinfo {pages} {212} (\bibinfo {year} {2022})}\BibitemShut {NoStop}%
\bibitem [{\citenamefont {{H0Dn Collaboration}}\ \emph {et~al.}(2026)\citenamefont {{H0Dn Collaboration}}, \citenamefont {Casertano}, \citenamefont {Anand}, \citenamefont {Anderson}, \citenamefont {Beaton}, \citenamefont {Bhardwaj}, \citenamefont {Blakeslee}, \citenamefont {Boubel}, \citenamefont {Breuval}, \citenamefont {Brout} \emph {et~al.}}]{h0dncollaboration2026}%
  \BibitemOpen
  \bibfield  {author} {\bibinfo {author} {\bibnamefont {{H0Dn Collaboration}}}, \bibinfo {author} {\bibfnamefont {S.}~\bibnamefont {Casertano}}, \bibinfo {author} {\bibfnamefont {G.}~\bibnamefont {Anand}}, \bibinfo {author} {\bibfnamefont {R.~I.}\ \bibnamefont {Anderson}}, \bibinfo {author} {\bibfnamefont {R.}~\bibnamefont {Beaton}}, \bibinfo {author} {\bibfnamefont {A.}~\bibnamefont {Bhardwaj}}, \bibinfo {author} {\bibfnamefont {J.~P.}\ \bibnamefont {Blakeslee}}, \bibinfo {author} {\bibfnamefont {P.}~\bibnamefont {Boubel}}, \bibinfo {author} {\bibfnamefont {L.}~\bibnamefont {Breuval}}, \bibinfo {author} {\bibfnamefont {D.}~\bibnamefont {Brout}}, \emph {et~al.},\ }\href {https://doi.org/10.1051/0004-6361/202557993} {\bibfield  {journal} {\bibinfo  {journal} {Astronomy and Astrophysics}\ }\textbf {\bibinfo {volume} {708}},\ \bibinfo {pages} {A166} (\bibinfo {year} {2026})},\ \bibinfo {note} {aDS Bibcode: 2026A\&A...708A.166H}\BibitemShut {NoStop}%
\bibitem [{\citenamefont {Refsdal}(1964)}]{refsdal1964}%
  \BibitemOpen
  \bibfield  {author} {\bibinfo {author} {\bibfnamefont {S.}~\bibnamefont {Refsdal}},\ }\href {https://doi.org/10.1093/mnras/128.4.307} {\bibfield  {journal} {\bibinfo  {journal} {Monthly Notices of the Royal Astronomical Society}\ }\textbf {\bibinfo {volume} {128}},\ \bibinfo {pages} {307} (\bibinfo {year} {1964})},\ \bibinfo {note} {aDS Bibcode: 1964MNRAS.128..307R}\BibitemShut {NoStop}%
\bibitem [{\citenamefont {Millon}\ \emph {et~al.}(2020{\natexlab{a}})\citenamefont {Millon}, \citenamefont {Courbin}, \citenamefont {Bonvin}, \citenamefont {Paic}, \citenamefont {Meylan}, \citenamefont {Tewes}, \citenamefont {Sluse}, \citenamefont {Magain}, \citenamefont {Chan}, \citenamefont {Galan} \emph {et~al.}}]{millon2020}%
  \BibitemOpen
  \bibfield  {author} {\bibinfo {author} {\bibfnamefont {M.}~\bibnamefont {Millon}}, \bibinfo {author} {\bibfnamefont {F.}~\bibnamefont {Courbin}}, \bibinfo {author} {\bibfnamefont {V.}~\bibnamefont {Bonvin}}, \bibinfo {author} {\bibfnamefont {E.}~\bibnamefont {Paic}}, \bibinfo {author} {\bibfnamefont {G.}~\bibnamefont {Meylan}}, \bibinfo {author} {\bibfnamefont {M.}~\bibnamefont {Tewes}}, \bibinfo {author} {\bibfnamefont {D.}~\bibnamefont {Sluse}}, \bibinfo {author} {\bibfnamefont {P.}~\bibnamefont {Magain}}, \bibinfo {author} {\bibfnamefont {J.~H.~H.}\ \bibnamefont {Chan}}, \bibinfo {author} {\bibfnamefont {A.}~\bibnamefont {Galan}}, \emph {et~al.},\ }\href {https://doi.org/10.1051/0004-6361/202037740} {\bibfield  {journal} {\bibinfo  {journal} {Astronomy and Astrophysics}\ }\textbf {\bibinfo {volume} {640}},\ \bibinfo {pages} {A105} (\bibinfo {year} {2020}{\natexlab{a}})}\BibitemShut {NoStop}%
\bibitem [{\citenamefont {Bonvin}\ \emph {et~al.}(2019)\citenamefont {Bonvin}, \citenamefont {Millon}, \citenamefont {Chan}, \citenamefont {Courbin}, \citenamefont {Rusu}, \citenamefont {Sluse}, \citenamefont {Suyu}, \citenamefont {Wong}, \citenamefont {Fassnacht}, \citenamefont {Marshall} \emph {et~al.}}]{bonvin2019}%
  \BibitemOpen
  \bibfield  {author} {\bibinfo {author} {\bibfnamefont {V.}~\bibnamefont {Bonvin}}, \bibinfo {author} {\bibfnamefont {M.}~\bibnamefont {Millon}}, \bibinfo {author} {\bibfnamefont {J.~H.-H.}\ \bibnamefont {Chan}}, \bibinfo {author} {\bibfnamefont {F.}~\bibnamefont {Courbin}}, \bibinfo {author} {\bibfnamefont {C.~E.}\ \bibnamefont {Rusu}}, \bibinfo {author} {\bibfnamefont {D.}~\bibnamefont {Sluse}}, \bibinfo {author} {\bibfnamefont {S.~H.}\ \bibnamefont {Suyu}}, \bibinfo {author} {\bibfnamefont {K.~C.}\ \bibnamefont {Wong}}, \bibinfo {author} {\bibfnamefont {C.~D.}\ \bibnamefont {Fassnacht}}, \bibinfo {author} {\bibfnamefont {P.~J.}\ \bibnamefont {Marshall}}, \emph {et~al.},\ }\href {https://doi.org/10.1051/0004-6361/201935921} {\bibfield  {journal} {\bibinfo  {journal} {Astronomy and Astrophysics}\ }\textbf {\bibinfo {volume} {629}},\ \bibinfo {pages} {A97} (\bibinfo {year} {2019})}\BibitemShut {NoStop}%
\bibitem [{\citenamefont {Millon}\ \emph {et~al.}(2020{\natexlab{b}})\citenamefont {Millon}, \citenamefont {Courbin}, \citenamefont {Bonvin}, \citenamefont {Buckley-Geer}, \citenamefont {Fassnacht}, \citenamefont {Frieman}, \citenamefont {Marshall}, \citenamefont {Suyu}, \citenamefont {Treu}, \citenamefont {Anguita} \emph {et~al.}}]{millon2020a}%
  \BibitemOpen
  \bibfield  {author} {\bibinfo {author} {\bibfnamefont {M.}~\bibnamefont {Millon}}, \bibinfo {author} {\bibfnamefont {F.}~\bibnamefont {Courbin}}, \bibinfo {author} {\bibfnamefont {V.}~\bibnamefont {Bonvin}}, \bibinfo {author} {\bibfnamefont {E.}~\bibnamefont {Buckley-Geer}}, \bibinfo {author} {\bibfnamefont {C.~D.}\ \bibnamefont {Fassnacht}}, \bibinfo {author} {\bibfnamefont {J.}~\bibnamefont {Frieman}}, \bibinfo {author} {\bibfnamefont {P.~J.}\ \bibnamefont {Marshall}}, \bibinfo {author} {\bibfnamefont {S.~H.}\ \bibnamefont {Suyu}}, \bibinfo {author} {\bibfnamefont {T.}~\bibnamefont {Treu}}, \bibinfo {author} {\bibfnamefont {T.}~\bibnamefont {Anguita}}, \emph {et~al.},\ }\href {https://doi.org/10.1051/0004-6361/202038698} {\bibfield  {journal} {\bibinfo  {journal} {Astronomy and Astrophysics}\ }\textbf {\bibinfo {volume} {642}},\ \bibinfo {pages} {A193} (\bibinfo {year} {2020}{\natexlab{b}})},\ \bibinfo {note} {aDS Bibcode: 2020A\&A...642A.193M}\BibitemShut {NoStop}%
\bibitem [{\citenamefont {Dux}\ \emph {et~al.}(2025)\citenamefont {Dux}, \citenamefont {Millon}, \citenamefont {Galan}, \citenamefont {Paic}, \citenamefont {Lemon}, \citenamefont {Courbin}, \citenamefont {Bonvin}, \citenamefont {Anguita}, \citenamefont {Auger}, \citenamefont {Birrer} \emph {et~al.}}]{dux2025}%
  \BibitemOpen
  \bibfield  {author} {\bibinfo {author} {\bibfnamefont {F.}~\bibnamefont {Dux}}, \bibinfo {author} {\bibfnamefont {M.}~\bibnamefont {Millon}}, \bibinfo {author} {\bibfnamefont {A.}~\bibnamefont {Galan}}, \bibinfo {author} {\bibfnamefont {E.}~\bibnamefont {Paic}}, \bibinfo {author} {\bibfnamefont {C.}~\bibnamefont {Lemon}}, \bibinfo {author} {\bibfnamefont {F.}~\bibnamefont {Courbin}}, \bibinfo {author} {\bibfnamefont {V.}~\bibnamefont {Bonvin}}, \bibinfo {author} {\bibfnamefont {T.}~\bibnamefont {Anguita}}, \bibinfo {author} {\bibfnamefont {M.}~\bibnamefont {Auger}}, \bibinfo {author} {\bibfnamefont {S.}~\bibnamefont {Birrer}}, \emph {et~al.},\ }\href {https://doi.org/10.1051/0004-6361/202553807} {\bibfield  {journal} {\bibinfo  {journal} {Astronomy and Astrophysics}\ }\textbf {\bibinfo {volume} {697}},\ \bibinfo {pages} {A139} (\bibinfo {year} {2025})},\ \bibinfo {note} {aDS Bibcode: 2025A\&A...697A.139D}\BibitemShut {NoStop}%
\bibitem [{\citenamefont {Greene}\ \emph {et~al.}(2013)\citenamefont {Greene}, \citenamefont {Suyu}, \citenamefont {Treu}, \citenamefont {Hilbert}, \citenamefont {Auger}, \citenamefont {Collett}, \citenamefont {Marshall}, \citenamefont {Fassnacht}, \citenamefont {Blandford}, \citenamefont {Bradač} \emph {et~al.}}]{greene2013a}%
  \BibitemOpen
  \bibfield  {author} {\bibinfo {author} {\bibfnamefont {Z.~S.}\ \bibnamefont {Greene}}, \bibinfo {author} {\bibfnamefont {S.~H.}\ \bibnamefont {Suyu}}, \bibinfo {author} {\bibfnamefont {T.}~\bibnamefont {Treu}}, \bibinfo {author} {\bibfnamefont {S.}~\bibnamefont {Hilbert}}, \bibinfo {author} {\bibfnamefont {M.~W.}\ \bibnamefont {Auger}}, \bibinfo {author} {\bibfnamefont {T.~E.}\ \bibnamefont {Collett}}, \bibinfo {author} {\bibfnamefont {P.~J.}\ \bibnamefont {Marshall}}, \bibinfo {author} {\bibfnamefont {C.~D.}\ \bibnamefont {Fassnacht}}, \bibinfo {author} {\bibfnamefont {R.~D.}\ \bibnamefont {Blandford}}, \bibinfo {author} {\bibfnamefont {M.}~\bibnamefont {Bradač}}, \emph {et~al.},\ }\href {https://doi.org/10.1088/0004-637X/768/1/39} {\bibfield  {journal} {\bibinfo  {journal} {The Astrophysical Journal}\ }\textbf {\bibinfo {volume} {768}},\ \bibinfo {pages} {39} (\bibinfo {year} {2013})}\BibitemShut {NoStop}%
\bibitem [{\citenamefont {Rusu}\ \emph {et~al.}(2017)\citenamefont {Rusu}, \citenamefont {Fassnacht}, \citenamefont {Sluse}, \citenamefont {Hilbert}, \citenamefont {Wong}, \citenamefont {Huang}, \citenamefont {Suyu}, \citenamefont {Collett}, \citenamefont {Marshall}, \citenamefont {Treu} \emph {et~al.}}]{rusu2017a}%
  \BibitemOpen
  \bibfield  {author} {\bibinfo {author} {\bibfnamefont {C.~E.}\ \bibnamefont {Rusu}}, \bibinfo {author} {\bibfnamefont {C.~D.}\ \bibnamefont {Fassnacht}}, \bibinfo {author} {\bibfnamefont {D.}~\bibnamefont {Sluse}}, \bibinfo {author} {\bibfnamefont {S.}~\bibnamefont {Hilbert}}, \bibinfo {author} {\bibfnamefont {K.~C.}\ \bibnamefont {Wong}}, \bibinfo {author} {\bibfnamefont {K.-H.}\ \bibnamefont {Huang}}, \bibinfo {author} {\bibfnamefont {S.~H.}\ \bibnamefont {Suyu}}, \bibinfo {author} {\bibfnamefont {T.~E.}\ \bibnamefont {Collett}}, \bibinfo {author} {\bibfnamefont {P.~J.}\ \bibnamefont {Marshall}}, \bibinfo {author} {\bibfnamefont {T.}~\bibnamefont {Treu}}, \emph {et~al.},\ }\href {https://doi.org/10.1093/mnras/stx285} {\bibfield  {journal} {\bibinfo  {journal} {Monthly Notices of the Royal Astronomical Society}\ }\textbf {\bibinfo {volume} {467}},\ \bibinfo {pages} {4220} (\bibinfo {year} {2017})}\BibitemShut {NoStop}%
\bibitem [{\citenamefont {Wells}\ \emph {et~al.}(2023)\citenamefont {Wells}, \citenamefont {Fassnacht},\ and\ \citenamefont {Rusu}}]{wells2023}%
  \BibitemOpen
  \bibfield  {author} {\bibinfo {author} {\bibfnamefont {P.}~\bibnamefont {Wells}}, \bibinfo {author} {\bibfnamefont {C.~D.}\ \bibnamefont {Fassnacht}},\ and\ \bibinfo {author} {\bibfnamefont {C.~E.}\ \bibnamefont {Rusu}},\ }\href {https://doi.org/10.1051/0004-6361/202346093} {\bibfield  {journal} {\bibinfo  {journal} {Astronomy and Astrophysics}\ }\textbf {\bibinfo {volume} {676}},\ \bibinfo {pages} {A95} (\bibinfo {year} {2023})}\BibitemShut {NoStop}%
\bibitem [{\citenamefont {Wells}\ \emph {et~al.}(2024)\citenamefont {Wells}, \citenamefont {Fassnacht}, \citenamefont {Birrer},\ and\ \citenamefont {Williams}}]{wells2024}%
  \BibitemOpen
  \bibfield  {author} {\bibinfo {author} {\bibfnamefont {P.~R.}\ \bibnamefont {Wells}}, \bibinfo {author} {\bibfnamefont {C.~D.}\ \bibnamefont {Fassnacht}}, \bibinfo {author} {\bibfnamefont {S.}~\bibnamefont {Birrer}},\ and\ \bibinfo {author} {\bibfnamefont {D.}~\bibnamefont {Williams}},\ }\href {https://doi.org/10.1051/0004-6361/202450002} {\bibfield  {journal} {\bibinfo  {journal} {Astronomy \& Astrophysics}\ }\textbf {\bibinfo {volume} {689}},\ \bibinfo {pages} {A87} (\bibinfo {year} {2024})}\BibitemShut {NoStop}%
\bibitem [{\citenamefont {Knabel}\ \emph {et~al.}(2025{\natexlab{a}})\citenamefont {Knabel}, \citenamefont {Mozumdar}, \citenamefont {Shajib}, \citenamefont {Treu}, \citenamefont {Cappellari}, \citenamefont {Spiniello},\ and\ \citenamefont {Birrer}}]{knabel2025b}%
  \BibitemOpen
  \bibfield  {author} {\bibinfo {author} {\bibfnamefont {S.}~\bibnamefont {Knabel}}, \bibinfo {author} {\bibfnamefont {P.}~\bibnamefont {Mozumdar}}, \bibinfo {author} {\bibfnamefont {A.~J.}\ \bibnamefont {Shajib}}, \bibinfo {author} {\bibfnamefont {T.}~\bibnamefont {Treu}}, \bibinfo {author} {\bibfnamefont {M.}~\bibnamefont {Cappellari}}, \bibinfo {author} {\bibfnamefont {C.}~\bibnamefont {Spiniello}},\ and\ \bibinfo {author} {\bibfnamefont {S.}~\bibnamefont {Birrer}},\ }\href {https://doi.org/10.1051/0004-6361/202554229} {\bibfield  {journal} {\bibinfo  {journal} {Astronomy \& Astrophysics}\ }\textbf {\bibinfo {volume} {703}},\ \bibinfo {pages} {A117} (\bibinfo {year} {2025}{\natexlab{a}})}\BibitemShut {NoStop}%
\bibitem [{\citenamefont {Shajib}\ \emph {et~al.}(2026)\citenamefont {Shajib}, \citenamefont {Treu}, \citenamefont {Suyu}, \citenamefont {Law}, \citenamefont {Yıldırım}, \citenamefont {Cappellari}, \citenamefont {Galan}, \citenamefont {Knabel}, \citenamefont {Wang}, \citenamefont {Birrer} \emph {et~al.}}]{shajib2026}%
  \BibitemOpen
  \bibfield  {author} {\bibinfo {author} {\bibfnamefont {A.~J.}\ \bibnamefont {Shajib}}, \bibinfo {author} {\bibfnamefont {T.}~\bibnamefont {Treu}}, \bibinfo {author} {\bibfnamefont {S.~H.}\ \bibnamefont {Suyu}}, \bibinfo {author} {\bibfnamefont {D.}~\bibnamefont {Law}}, \bibinfo {author} {\bibfnamefont {A.}~\bibnamefont {Yıldırım}}, \bibinfo {author} {\bibfnamefont {M.}~\bibnamefont {Cappellari}}, \bibinfo {author} {\bibfnamefont {A.}~\bibnamefont {Galan}}, \bibinfo {author} {\bibfnamefont {S.}~\bibnamefont {Knabel}}, \bibinfo {author} {\bibfnamefont {H.}~\bibnamefont {Wang}}, \bibinfo {author} {\bibfnamefont {S.}~\bibnamefont {Birrer}}, \emph {et~al.},\ }\href {https://doi.org/10.1051/0004-6361/202556126} {\bibfield  {journal} {\bibinfo  {journal} {Astronomy \& Astrophysics}\ }\textbf {\bibinfo {volume} {707}},\ \bibinfo {pages} {A314} (\bibinfo {year} {2026})}\BibitemShut {NoStop}%
\bibitem [{\citenamefont {Shajib}\ \emph {et~al.}(2018)\citenamefont {Shajib}, \citenamefont {Treu},\ and\ \citenamefont {Agnello}}]{shajib2018}%
  \BibitemOpen
  \bibfield  {author} {\bibinfo {author} {\bibfnamefont {A.~J.}\ \bibnamefont {Shajib}}, \bibinfo {author} {\bibfnamefont {T.}~\bibnamefont {Treu}},\ and\ \bibinfo {author} {\bibfnamefont {A.}~\bibnamefont {Agnello}},\ }\href {https://doi.org/10.1093/mnras/stx2302} {\bibfield  {journal} {\bibinfo  {journal} {Monthly Notices of the Royal Astronomical Society}\ }\textbf {\bibinfo {volume} {473}},\ \bibinfo {pages} {210} (\bibinfo {year} {2018})}\BibitemShut {NoStop}%
\bibitem [{\citenamefont {Yıldırım}\ \emph {et~al.}(2020)\citenamefont {Yıldırım}, \citenamefont {Suyu},\ and\ \citenamefont {Halkola}}]{yildirim2020}%
  \BibitemOpen
  \bibfield  {author} {\bibinfo {author} {\bibfnamefont {A.}~\bibnamefont {Yıldırım}}, \bibinfo {author} {\bibfnamefont {S.~H.}\ \bibnamefont {Suyu}},\ and\ \bibinfo {author} {\bibfnamefont {A.}~\bibnamefont {Halkola}},\ }\href {https://doi.org/10.1093/mnras/staa498} {\bibfield  {journal} {\bibinfo  {journal} {Monthly Notices of the Royal Astronomical Society}\ }\textbf {\bibinfo {volume} {493}},\ \bibinfo {pages} {4783} (\bibinfo {year} {2020})}\BibitemShut {NoStop}%
\bibitem [{\citenamefont {Shajib}\ \emph {et~al.}(2023)\citenamefont {Shajib}, \citenamefont {Mozumdar}, \citenamefont {Chen}, \citenamefont {Treu}, \citenamefont {Cappellari}, \citenamefont {Knabel}, \citenamefont {Suyu}, \citenamefont {Bennert}, \citenamefont {Frieman}, \citenamefont {Sluse} \emph {et~al.}}]{shajib2023}%
  \BibitemOpen
  \bibfield  {author} {\bibinfo {author} {\bibfnamefont {A.~J.}\ \bibnamefont {Shajib}}, \bibinfo {author} {\bibfnamefont {P.}~\bibnamefont {Mozumdar}}, \bibinfo {author} {\bibfnamefont {G.~C.-F.}\ \bibnamefont {Chen}}, \bibinfo {author} {\bibfnamefont {T.}~\bibnamefont {Treu}}, \bibinfo {author} {\bibfnamefont {M.}~\bibnamefont {Cappellari}}, \bibinfo {author} {\bibfnamefont {S.}~\bibnamefont {Knabel}}, \bibinfo {author} {\bibfnamefont {S.~H.}\ \bibnamefont {Suyu}}, \bibinfo {author} {\bibfnamefont {V.~N.}\ \bibnamefont {Bennert}}, \bibinfo {author} {\bibfnamefont {J.~A.}\ \bibnamefont {Frieman}}, \bibinfo {author} {\bibfnamefont {D.}~\bibnamefont {Sluse}}, \emph {et~al.},\ }\href {https://doi.org/10.1051/0004-6361/202345878} {\bibfield  {journal} {\bibinfo  {journal} {Astronomy and Astrophysics}\ }\textbf {\bibinfo {volume} {673}},\ \bibinfo {pages} {A9} (\bibinfo {year} {2023})}\BibitemShut {NoStop}%
\bibitem [{\citenamefont {Paic}\ \emph {et~al.}(2026)\citenamefont {Paic}, \citenamefont {Courbin}, \citenamefont {Fassnacht}, \citenamefont {Galan}, \citenamefont {Millon}, \citenamefont {Sluse}, \citenamefont {Williams}, \citenamefont {Birrer}, \citenamefont {Buckley-Geer}, \citenamefont {Cappellari} \emph {et~al.}}]{paic2026}%
  \BibitemOpen
  \bibfield  {author} {\bibinfo {author} {\bibfnamefont {E.}~\bibnamefont {Paic}}, \bibinfo {author} {\bibfnamefont {F.}~\bibnamefont {Courbin}}, \bibinfo {author} {\bibfnamefont {C.~D.}\ \bibnamefont {Fassnacht}}, \bibinfo {author} {\bibfnamefont {A.}~\bibnamefont {Galan}}, \bibinfo {author} {\bibfnamefont {M.}~\bibnamefont {Millon}}, \bibinfo {author} {\bibfnamefont {D.}~\bibnamefont {Sluse}}, \bibinfo {author} {\bibfnamefont {D.~M.}\ \bibnamefont {Williams}}, \bibinfo {author} {\bibfnamefont {S.}~\bibnamefont {Birrer}}, \bibinfo {author} {\bibfnamefont {E.~J.}\ \bibnamefont {Buckley-Geer}}, \bibinfo {author} {\bibfnamefont {M.}~\bibnamefont {Cappellari}}, \emph {et~al.},\ }\href {https://doi.org/10.1051/0004-6361/202556411} {\bibfield  {journal} {\bibinfo  {journal} {Astronomy and Astrophysics}\ }\textbf {\bibinfo {volume} {706}},\ \bibinfo {pages} {A270} (\bibinfo {year} {2026})},\ \bibinfo {note} {aDS Bibcode: 2026A\&A...706A.270P}\BibitemShut {NoStop}%
\bibitem [{\citenamefont {Sheu}\ \emph {et~al.}(2026)\citenamefont {Sheu}, \citenamefont {Treu}, \citenamefont {Millon}, \citenamefont {Dux}, \citenamefont {Williams}, \citenamefont {Knabel}, \citenamefont {Birrer}, \citenamefont {Mozumdar}, \citenamefont {Queirolo}, \citenamefont {Shajib} \emph {et~al.}}]{sheu2026}%
  \BibitemOpen
  \bibfield  {author} {\bibinfo {author} {\bibfnamefont {W.}~\bibnamefont {Sheu}}, \bibinfo {author} {\bibfnamefont {T.}~\bibnamefont {Treu}}, \bibinfo {author} {\bibfnamefont {M.}~\bibnamefont {Millon}}, \bibinfo {author} {\bibfnamefont {F.}~\bibnamefont {Dux}}, \bibinfo {author} {\bibfnamefont {D.}~\bibnamefont {Williams}}, \bibinfo {author} {\bibfnamefont {S.}~\bibnamefont {Knabel}}, \bibinfo {author} {\bibfnamefont {S.}~\bibnamefont {Birrer}}, \bibinfo {author} {\bibfnamefont {P.}~\bibnamefont {Mozumdar}}, \bibinfo {author} {\bibfnamefont {G.}~\bibnamefont {Queirolo}}, \bibinfo {author} {\bibfnamefont {A.~J.}\ \bibnamefont {Shajib}}, \emph {et~al.},\ }\href {https://doi.org/10.48550/arXiv.2604.14145} {{\selectlanguage {en}\bibinfo {title} {{TDCOSMO} {XXV}: a "soup-to-nuts" 6.5\% \$h\_0\$ measurement \$-\$ strong lensing and dynamics with a maximally flexible mass sheet}}} (\bibinfo {year} {2026}),\ \bibinfo {note} {aDS Bibcode: 2026arXiv260414145S}\BibitemShut {NoStop}%
\bibitem [{\citenamefont {Suyu}\ \emph {et~al.}(2010)\citenamefont {Suyu}, \citenamefont {Marshall}, \citenamefont {Auger}, \citenamefont {Hilbert}, \citenamefont {Blandford}, \citenamefont {Koopmans}, \citenamefont {Fassnacht},\ and\ \citenamefont {Treu}}]{suyu2010}%
  \BibitemOpen
  \bibfield  {author} {\bibinfo {author} {\bibfnamefont {S.~H.}\ \bibnamefont {Suyu}}, \bibinfo {author} {\bibfnamefont {P.~J.}\ \bibnamefont {Marshall}}, \bibinfo {author} {\bibfnamefont {M.~W.}\ \bibnamefont {Auger}}, \bibinfo {author} {\bibfnamefont {S.}~\bibnamefont {Hilbert}}, \bibinfo {author} {\bibfnamefont {R.~D.}\ \bibnamefont {Blandford}}, \bibinfo {author} {\bibfnamefont {L.~V.~E.}\ \bibnamefont {Koopmans}}, \bibinfo {author} {\bibfnamefont {C.~D.}\ \bibnamefont {Fassnacht}},\ and\ \bibinfo {author} {\bibfnamefont {T.}~\bibnamefont {Treu}},\ }\href {https://doi.org/10.1088/0004-637x/711/1/201} {\bibfield  {journal} {\bibinfo  {journal} {Astrophysical Journal}\ }\textbf {\bibinfo {volume} {711}},\ \bibinfo {pages} {201} (\bibinfo {year} {2010})}\BibitemShut {NoStop}%
\bibitem [{\citenamefont {Chen}\ \emph {et~al.}(2019)\citenamefont {Chen}, \citenamefont {Fassnacht}, \citenamefont {Suyu}, \citenamefont {Rusu}, \citenamefont {Chan}, \citenamefont {Wong}, \citenamefont {Auger}, \citenamefont {Hilbert}, \citenamefont {Bonvin}, \citenamefont {Birrer} \emph {et~al.}}]{chen2019}%
  \BibitemOpen
  \bibfield  {author} {\bibinfo {author} {\bibfnamefont {G.~C.-F.}\ \bibnamefont {Chen}}, \bibinfo {author} {\bibfnamefont {C.~D.}\ \bibnamefont {Fassnacht}}, \bibinfo {author} {\bibfnamefont {S.~H.}\ \bibnamefont {Suyu}}, \bibinfo {author} {\bibfnamefont {C.~E.}\ \bibnamefont {Rusu}}, \bibinfo {author} {\bibfnamefont {J.~H.~H.}\ \bibnamefont {Chan}}, \bibinfo {author} {\bibfnamefont {K.~C.}\ \bibnamefont {Wong}}, \bibinfo {author} {\bibfnamefont {M.~W.}\ \bibnamefont {Auger}}, \bibinfo {author} {\bibfnamefont {S.}~\bibnamefont {Hilbert}}, \bibinfo {author} {\bibfnamefont {V.}~\bibnamefont {Bonvin}}, \bibinfo {author} {\bibfnamefont {S.}~\bibnamefont {Birrer}}, \emph {et~al.},\ }\href {https://doi.org/10.1093/mnras/stz2547} {\bibfield  {journal} {\bibinfo  {journal} {Monthly Notices of the Royal Astronomical Society}\ }\textbf {\bibinfo {volume} {490}},\ \bibinfo {pages} {1743} (\bibinfo {year} {2019})}\BibitemShut {NoStop}%
\bibitem [{\citenamefont {Williams}\ \emph {et~al.}(2025)\citenamefont {Williams}, \citenamefont {Treu}, \citenamefont {Birrer}, \citenamefont {Shajib}, \citenamefont {Wong}, \citenamefont {Morishita}, \citenamefont {Schmidt},\ and\ \citenamefont {Stiavelli}}]{williams2025}%
  \BibitemOpen
  \bibfield  {author} {\bibinfo {author} {\bibfnamefont {D.~M.}\ \bibnamefont {Williams}}, \bibinfo {author} {\bibfnamefont {T.}~\bibnamefont {Treu}}, \bibinfo {author} {\bibfnamefont {S.}~\bibnamefont {Birrer}}, \bibinfo {author} {\bibfnamefont {A.~J.}\ \bibnamefont {Shajib}}, \bibinfo {author} {\bibfnamefont {K.~C.}\ \bibnamefont {Wong}}, \bibinfo {author} {\bibfnamefont {T.}~\bibnamefont {Morishita}}, \bibinfo {author} {\bibfnamefont {T.}~\bibnamefont {Schmidt}},\ and\ \bibinfo {author} {\bibfnamefont {M.}~\bibnamefont {Stiavelli}},\ }\href {https://doi.org/10.48550/arXiv.2503.00099} {\bibinfo {title} {{TDCOSMO}: {XX}. {WFI2033}--4723, the {First} {Quadruply}-{Imaged} {Quasar} {Modeled} with {JWST} {Imaging}}} (\bibinfo {year} {2025}),\ \bibinfo {note} {arXiv:2503.00099 [astro-ph]}\BibitemShut {NoStop}%
\bibitem [{\citenamefont {{TDCOSMO Collaboration}}\ \emph {et~al.}(2025)\citenamefont {{TDCOSMO Collaboration}}, \citenamefont {Birrer}, \citenamefont {Buckley-Geer}, \citenamefont {Cappellari}, \citenamefont {Courbin}, \citenamefont {Dux}, \citenamefont {Fassnacht}, \citenamefont {Frieman}, \citenamefont {Galan}, \citenamefont {Gilman} \emph {et~al.}}]{collaboration2025}%
  \BibitemOpen
  \bibfield  {author} {\bibinfo {author} {\bibnamefont {{TDCOSMO Collaboration}}}, \bibinfo {author} {\bibfnamefont {S.}~\bibnamefont {Birrer}}, \bibinfo {author} {\bibfnamefont {E.~J.}\ \bibnamefont {Buckley-Geer}}, \bibinfo {author} {\bibfnamefont {M.}~\bibnamefont {Cappellari}}, \bibinfo {author} {\bibfnamefont {F.}~\bibnamefont {Courbin}}, \bibinfo {author} {\bibfnamefont {F.}~\bibnamefont {Dux}}, \bibinfo {author} {\bibfnamefont {C.~D.}\ \bibnamefont {Fassnacht}}, \bibinfo {author} {\bibfnamefont {J.~A.}\ \bibnamefont {Frieman}}, \bibinfo {author} {\bibfnamefont {A.}~\bibnamefont {Galan}}, \bibinfo {author} {\bibfnamefont {D.}~\bibnamefont {Gilman}}, \emph {et~al.},\ }\href {https://doi.org/10.48550/arXiv.2506.03023} {\bibinfo {title} {{TDCOSMO} 2025: {Cosmological} constraints from strong lensing time delays}} (\bibinfo {year} {2025}),\ \bibinfo {note} {arXiv:2506.03023 [astro-ph]}\BibitemShut {NoStop}%
\bibitem [{\citenamefont {Treu}\ and\ \citenamefont {Marshall}(2016)}]{treu2016}%
  \BibitemOpen
  \bibfield  {author} {\bibinfo {author} {\bibfnamefont {T.}~\bibnamefont {Treu}}\ and\ \bibinfo {author} {\bibfnamefont {P.~J.}\ \bibnamefont {Marshall}},\ }\href {https://doi.org/10.1007/s00159-016-0096-8} {\bibfield  {journal} {\bibinfo  {journal} {Astronomy and Astrophysics Review}\ }\textbf {\bibinfo {volume} {24}},\ \bibinfo {pages} {11} (\bibinfo {year} {2016})}\BibitemShut {NoStop}%
\bibitem [{\citenamefont {Suyu}\ \emph {et~al.}(2018)\citenamefont {Suyu}, \citenamefont {Chang}, \citenamefont {Courbin},\ and\ \citenamefont {Okumura}}]{suyu2018}%
  \BibitemOpen
  \bibfield  {author} {\bibinfo {author} {\bibfnamefont {S.~H.}\ \bibnamefont {Suyu}}, \bibinfo {author} {\bibfnamefont {T.-C.}\ \bibnamefont {Chang}}, \bibinfo {author} {\bibfnamefont {F.}~\bibnamefont {Courbin}},\ and\ \bibinfo {author} {\bibfnamefont {T.}~\bibnamefont {Okumura}},\ }\href {https://doi.org/10.1007/s11214-018-0524-3} {\bibfield  {journal} {\bibinfo  {journal} {Space Science Reviews}\ }\textbf {\bibinfo {volume} {214}},\ \bibinfo {pages} {91} (\bibinfo {year} {2018})}\BibitemShut {NoStop}%
\bibitem [{\citenamefont {Birrer}\ \emph {et~al.}(2024)\citenamefont {Birrer}, \citenamefont {Millon}, \citenamefont {Sluse}, \citenamefont {Shajib}, \citenamefont {Courbin}, \citenamefont {Erickson}, \citenamefont {Koopmans}, \citenamefont {Suyu},\ and\ \citenamefont {Treu}}]{birrer2024}%
  \BibitemOpen
  \bibfield  {author} {\bibinfo {author} {\bibfnamefont {S.}~\bibnamefont {Birrer}}, \bibinfo {author} {\bibfnamefont {M.}~\bibnamefont {Millon}}, \bibinfo {author} {\bibfnamefont {D.}~\bibnamefont {Sluse}}, \bibinfo {author} {\bibfnamefont {A.~J.}\ \bibnamefont {Shajib}}, \bibinfo {author} {\bibfnamefont {F.}~\bibnamefont {Courbin}}, \bibinfo {author} {\bibfnamefont {S.}~\bibnamefont {Erickson}}, \bibinfo {author} {\bibfnamefont {L.~V.~E.}\ \bibnamefont {Koopmans}}, \bibinfo {author} {\bibfnamefont {S.~H.}\ \bibnamefont {Suyu}},\ and\ \bibinfo {author} {\bibfnamefont {T.}~\bibnamefont {Treu}},\ }\href {https://doi.org/10.1007/s11214-024-01079-w} {\bibfield  {journal} {\bibinfo  {journal} {Space Science Reviews}\ }\textbf {\bibinfo {volume} {220}},\ \bibinfo {pages} {48} (\bibinfo {year} {2024})},\ \bibinfo {note} {aDS Bibcode: 2024SSRv..220...48B}\BibitemShut {NoStop}%
\bibitem [{\citenamefont {Blandford}\ and\ \citenamefont {Narayan}(1986)}]{blandford1986}%
  \BibitemOpen
  \bibfield  {author} {\bibinfo {author} {\bibfnamefont {R.}~\bibnamefont {Blandford}}\ and\ \bibinfo {author} {\bibfnamefont {R.}~\bibnamefont {Narayan}},\ }\href {https://doi.org/10.1086/164709} {\bibfield  {journal} {\bibinfo  {journal} {Astrophysical Journal}\ }\textbf {\bibinfo {volume} {310}},\ \bibinfo {pages} {568} (\bibinfo {year} {1986})},\ \bibinfo {note} {aDS Bibcode: 1986ApJ...310..568B}\BibitemShut {NoStop}%
\bibitem [{\citenamefont {Kovner}(1987)}]{kovner1987}%
  \BibitemOpen
  \bibfield  {author} {\bibinfo {author} {\bibfnamefont {I.}~\bibnamefont {Kovner}},\ }\href {https://doi.org/10.1086/165179} {\bibfield  {journal} {\bibinfo  {journal} {Astrophysical Journal}\ }\textbf {\bibinfo {volume} {316}},\ \bibinfo {pages} {52} (\bibinfo {year} {1987})},\ \bibinfo {note} {aDS Bibcode: 1987ApJ...316...52K}\BibitemShut {NoStop}%
\bibitem [{\citenamefont {Schneider}\ \emph {et~al.}(1992)\citenamefont {Schneider}, \citenamefont {Ehlers},\ and\ \citenamefont {Falco}}]{schneider1992}%
  \BibitemOpen
  \bibfield  {author} {\bibinfo {author} {\bibfnamefont {P.}~\bibnamefont {Schneider}}, \bibinfo {author} {\bibfnamefont {J.}~\bibnamefont {Ehlers}},\ and\ \bibinfo {author} {\bibfnamefont {E.~E.}\ \bibnamefont {Falco}},\ }\href {https://doi.org/10.1007/978-3-662-03758-4} {\emph {\bibinfo {title} {Gravitational {Lenses}}}}\ (\bibinfo {year} {1992})\ \bibinfo {note} {publication Title: Gravitational Lenses ADS Bibcode: 1992grle.book.....S}\BibitemShut {NoStop}%
\bibitem [{\citenamefont {Collett}\ and\ \citenamefont {Auger}(2014)}]{collett2014}%
  \BibitemOpen
  \bibfield  {author} {\bibinfo {author} {\bibfnamefont {T.~E.}\ \bibnamefont {Collett}}\ and\ \bibinfo {author} {\bibfnamefont {M.~W.}\ \bibnamefont {Auger}},\ }\href {https://doi.org/10.1093/mnras/stu1190} {\bibfield  {journal} {\bibinfo  {journal} {Monthly Notices of the Royal Astronomical Society}\ }\textbf {\bibinfo {volume} {443}},\ \bibinfo {pages} {969} (\bibinfo {year} {2014})},\ \bibinfo {note} {aDS Bibcode: 2014MNRAS.443..969C}\BibitemShut {NoStop}%
\bibitem [{\citenamefont {McCully}\ \emph {et~al.}(2014)\citenamefont {McCully}, \citenamefont {Keeton}, \citenamefont {Wong},\ and\ \citenamefont {Zabludoff}}]{mccully2014}%
  \BibitemOpen
  \bibfield  {author} {\bibinfo {author} {\bibfnamefont {C.}~\bibnamefont {McCully}}, \bibinfo {author} {\bibfnamefont {C.~R.}\ \bibnamefont {Keeton}}, \bibinfo {author} {\bibfnamefont {K.~C.}\ \bibnamefont {Wong}},\ and\ \bibinfo {author} {\bibfnamefont {A.~I.}\ \bibnamefont {Zabludoff}},\ }\href {https://doi.org/10.1093/mnras/stu1316} {\bibfield  {journal} {\bibinfo  {journal} {Monthly Notices of the Royal Astronomical Society}\ }\textbf {\bibinfo {volume} {443}},\ \bibinfo {pages} {3631} (\bibinfo {year} {2014})}\BibitemShut {NoStop}%
\bibitem [{\citenamefont {Wong}\ \emph {et~al.}(2017)\citenamefont {Wong}, \citenamefont {Suyu}, \citenamefont {Auger}, \citenamefont {Bonvin}, \citenamefont {Courbin}, \citenamefont {Fassnacht}, \citenamefont {Halkola}, \citenamefont {Rusu}, \citenamefont {Sluse}, \citenamefont {Sonnenfeld} \emph {et~al.}}]{wong2017}%
  \BibitemOpen
  \bibfield  {author} {\bibinfo {author} {\bibfnamefont {K.~C.}\ \bibnamefont {Wong}}, \bibinfo {author} {\bibfnamefont {S.~H.}\ \bibnamefont {Suyu}}, \bibinfo {author} {\bibfnamefont {M.~W.}\ \bibnamefont {Auger}}, \bibinfo {author} {\bibfnamefont {V.}~\bibnamefont {Bonvin}}, \bibinfo {author} {\bibfnamefont {F.}~\bibnamefont {Courbin}}, \bibinfo {author} {\bibfnamefont {C.~D.}\ \bibnamefont {Fassnacht}}, \bibinfo {author} {\bibfnamefont {A.}~\bibnamefont {Halkola}}, \bibinfo {author} {\bibfnamefont {C.~E.}\ \bibnamefont {Rusu}}, \bibinfo {author} {\bibfnamefont {D.}~\bibnamefont {Sluse}}, \bibinfo {author} {\bibfnamefont {A.}~\bibnamefont {Sonnenfeld}}, \emph {et~al.},\ }\href {https://doi.org/10.1093/mnras/stw3077} {\bibfield  {journal} {\bibinfo  {journal} {Monthly Notices of the Royal Astronomical Society}\ }\textbf {\bibinfo {volume} {465}},\ \bibinfo {pages} {4895} (\bibinfo {year} {2017})}\BibitemShut {NoStop}%
\bibitem [{\citenamefont {Rusu}\ \emph {et~al.}(2020)\citenamefont {Rusu}, \citenamefont {Wong}, \citenamefont {Bonvin}, \citenamefont {Sluse}, \citenamefont {Suyu}, \citenamefont {Fassnacht}, \citenamefont {Chan}, \citenamefont {Hilbert}, \citenamefont {Auger}, \citenamefont {Sonnenfeld} \emph {et~al.}}]{rusu2020}%
  \BibitemOpen
  \bibfield  {author} {\bibinfo {author} {\bibfnamefont {C.~E.}\ \bibnamefont {Rusu}}, \bibinfo {author} {\bibfnamefont {K.~C.}\ \bibnamefont {Wong}}, \bibinfo {author} {\bibfnamefont {V.}~\bibnamefont {Bonvin}}, \bibinfo {author} {\bibfnamefont {D.}~\bibnamefont {Sluse}}, \bibinfo {author} {\bibfnamefont {S.~H.}\ \bibnamefont {Suyu}}, \bibinfo {author} {\bibfnamefont {C.~D.}\ \bibnamefont {Fassnacht}}, \bibinfo {author} {\bibfnamefont {J.~H.~H.}\ \bibnamefont {Chan}}, \bibinfo {author} {\bibfnamefont {S.}~\bibnamefont {Hilbert}}, \bibinfo {author} {\bibfnamefont {M.~W.}\ \bibnamefont {Auger}}, \bibinfo {author} {\bibfnamefont {A.}~\bibnamefont {Sonnenfeld}}, \emph {et~al.},\ }\href {https://doi.org/10.1093/mnras/stz3451} {\bibfield  {journal} {\bibinfo  {journal} {Monthly Notices of the Royal Astronomical Society}\ }\textbf {\bibinfo {volume} {498}},\ \bibinfo {pages} {1440} (\bibinfo {year} {2020})}\BibitemShut {NoStop}%
\bibitem [{\citenamefont {Shajib}\ \emph {et~al.}(2020)\citenamefont {Shajib}, \citenamefont {Birrer}, \citenamefont {Treu}, \citenamefont {Agnello}, \citenamefont {Buckley-Geer}, \citenamefont {Chan}, \citenamefont {Christensen}, \citenamefont {Lemon}, \citenamefont {Lin}, \citenamefont {Millon} \emph {et~al.}}]{shajib2020}%
  \BibitemOpen
  \bibfield  {author} {\bibinfo {author} {\bibfnamefont {A.~J.}\ \bibnamefont {Shajib}}, \bibinfo {author} {\bibfnamefont {S.}~\bibnamefont {Birrer}}, \bibinfo {author} {\bibfnamefont {T.}~\bibnamefont {Treu}}, \bibinfo {author} {\bibfnamefont {A.}~\bibnamefont {Agnello}}, \bibinfo {author} {\bibfnamefont {E.~J.}\ \bibnamefont {Buckley-Geer}}, \bibinfo {author} {\bibfnamefont {J.~H.~H.}\ \bibnamefont {Chan}}, \bibinfo {author} {\bibfnamefont {L.}~\bibnamefont {Christensen}}, \bibinfo {author} {\bibfnamefont {C.}~\bibnamefont {Lemon}}, \bibinfo {author} {\bibfnamefont {H.}~\bibnamefont {Lin}}, \bibinfo {author} {\bibfnamefont {M.}~\bibnamefont {Millon}}, \emph {et~al.},\ }\href {https://doi.org/10.1093/mnras/staa828} {\bibfield  {journal} {\bibinfo  {journal} {Monthly Notices of the Royal Astronomical Society}\ }\textbf {\bibinfo {volume} {494}},\ \bibinfo {pages} {6072} (\bibinfo {year} {2020})}\BibitemShut {NoStop}%
\bibitem [{\citenamefont {Falco}\ \emph {et~al.}(1985)\citenamefont {Falco}, \citenamefont {Gorenstein},\ and\ \citenamefont {Shapiro}}]{falco1985}%
  \BibitemOpen
  \bibfield  {author} {\bibinfo {author} {\bibfnamefont {E.~E.}\ \bibnamefont {Falco}}, \bibinfo {author} {\bibfnamefont {M.~V.}\ \bibnamefont {Gorenstein}},\ and\ \bibinfo {author} {\bibfnamefont {I.~I.}\ \bibnamefont {Shapiro}},\ }\href {https://doi.org/10.1086/184422} {\bibfield  {journal} {\bibinfo  {journal} {Astrophysical Journal}\ }\textbf {\bibinfo {volume} {289}},\ \bibinfo {pages} {L1} (\bibinfo {year} {1985})},\ \bibinfo {note} {aDS Bibcode: 1985ApJ...289L...1F}\BibitemShut {NoStop}%
\bibitem [{\citenamefont {Suyu}\ and\ \citenamefont {Halkola}(2010)}]{suyu2010b}%
  \BibitemOpen
  \bibfield  {author} {\bibinfo {author} {\bibfnamefont {S.~H.}\ \bibnamefont {Suyu}}\ and\ \bibinfo {author} {\bibfnamefont {A.}~\bibnamefont {Halkola}},\ }\href {https://doi.org/10.1051/0004-6361/201015481} {\bibfield  {journal} {\bibinfo  {journal} {Astronomy and Astrophysics}\ }\textbf {\bibinfo {volume} {524}},\ \bibinfo {pages} {A94} (\bibinfo {year} {2010})}\BibitemShut {NoStop}%
\bibitem [{\citenamefont {Fassnacht}\ \emph {et~al.}(2011)\citenamefont {Fassnacht}, \citenamefont {Koopmans},\ and\ \citenamefont {Wong}}]{fassnacht2011}%
  \BibitemOpen
  \bibfield  {author} {\bibinfo {author} {\bibfnamefont {C.~D.}\ \bibnamefont {Fassnacht}}, \bibinfo {author} {\bibfnamefont {L.~V.~E.}\ \bibnamefont {Koopmans}},\ and\ \bibinfo {author} {\bibfnamefont {K.~C.}\ \bibnamefont {Wong}},\ }\href {https://doi.org/10.1111/j.1365-2966.2010.17591.x} {\bibfield  {journal} {\bibinfo  {journal} {Monthly Notices of the Royal Astronomical Society}\ }\textbf {\bibinfo {volume} {410}},\ \bibinfo {pages} {2167} (\bibinfo {year} {2011})}\BibitemShut {NoStop}%
\bibitem [{\citenamefont {{Euclid Collaboration}}\ \emph {et~al.}(2025)\citenamefont {{Euclid Collaboration}}, \citenamefont {{Castander}}, \citenamefont {{Fosalba}}, \citenamefont {{Stadel}}, \citenamefont {{Potter}}, \citenamefont {{Carretero}}, \citenamefont {{Tallada-Cresp{\'\i}}}, \citenamefont {{Pozzetti}}, \citenamefont {{Bolzonella}}, \citenamefont {{Mamon}} \emph {et~al.}}]{Castander_2025}%
  \BibitemOpen
  \bibfield  {author} {\bibinfo {author} {\bibnamefont {{Euclid Collaboration}}}, \bibinfo {author} {\bibfnamefont {F.~J.}\ \bibnamefont {{Castander}}}, \bibinfo {author} {\bibfnamefont {P.}~\bibnamefont {{Fosalba}}}, \bibinfo {author} {\bibfnamefont {J.}~\bibnamefont {{Stadel}}}, \bibinfo {author} {\bibfnamefont {D.}~\bibnamefont {{Potter}}}, \bibinfo {author} {\bibfnamefont {J.}~\bibnamefont {{Carretero}}}, \bibinfo {author} {\bibfnamefont {P.}~\bibnamefont {{Tallada-Cresp{\'\i}}}}, \bibinfo {author} {\bibfnamefont {L.}~\bibnamefont {{Pozzetti}}}, \bibinfo {author} {\bibfnamefont {M.}~\bibnamefont {{Bolzonella}}}, \bibinfo {author} {\bibfnamefont {G.~A.}\ \bibnamefont {{Mamon}}}, \emph {et~al.},\ }\href {https://doi.org/10.1051/0004-6361/202450853} {\bibfield  {journal} {\bibinfo  {journal} {\aap}\ }\textbf {\bibinfo {volume} {697}},\ \bibinfo {eid} {A5} (\bibinfo {year} {2025})},\ \Eprint {https://arxiv.org/abs/2405.13495} {arXiv:2405.13495 [astro-ph.CO]} \BibitemShut {NoStop}%
\bibitem [{\citenamefont {Schneider}\ and\ \citenamefont {Sluse}(2013)}]{schneider2013}%
  \BibitemOpen
  \bibfield  {author} {\bibinfo {author} {\bibfnamefont {P.}~\bibnamefont {Schneider}}\ and\ \bibinfo {author} {\bibfnamefont {D.}~\bibnamefont {Sluse}},\ }\href {https://doi.org/10.1051/0004-6361/201321882} {\bibfield  {journal} {\bibinfo  {journal} {Astronomy and Astrophysics}\ }\textbf {\bibinfo {volume} {559}},\ \bibinfo {pages} {A37} (\bibinfo {year} {2013})},\ \bibinfo {note} {aDS Bibcode: 2013A\&A...559A..37S}\BibitemShut {NoStop}%
\bibitem [{\citenamefont {Birrer}\ \emph {et~al.}(2020)\citenamefont {Birrer}, \citenamefont {Shajib}, \citenamefont {Galan}, \citenamefont {Millon}, \citenamefont {Treu}, \citenamefont {Agnello}, \citenamefont {Auger}, \citenamefont {Chen}, \citenamefont {Christensen}, \citenamefont {Collett} \emph {et~al.}}]{birrer2020}%
  \BibitemOpen
  \bibfield  {author} {\bibinfo {author} {\bibfnamefont {S.}~\bibnamefont {Birrer}}, \bibinfo {author} {\bibfnamefont {A.~J.}\ \bibnamefont {Shajib}}, \bibinfo {author} {\bibfnamefont {A.}~\bibnamefont {Galan}}, \bibinfo {author} {\bibfnamefont {M.}~\bibnamefont {Millon}}, \bibinfo {author} {\bibfnamefont {T.}~\bibnamefont {Treu}}, \bibinfo {author} {\bibfnamefont {A.}~\bibnamefont {Agnello}}, \bibinfo {author} {\bibfnamefont {M.}~\bibnamefont {Auger}}, \bibinfo {author} {\bibfnamefont {G.~C.-F.}\ \bibnamefont {Chen}}, \bibinfo {author} {\bibfnamefont {L.}~\bibnamefont {Christensen}}, \bibinfo {author} {\bibfnamefont {T.}~\bibnamefont {Collett}}, \emph {et~al.},\ }\href {https://doi.org/10.1051/0004-6361/202038861} {\bibfield  {journal} {\bibinfo  {journal} {Astronomy and Astrophysics}\ }\textbf {\bibinfo {volume} {643}},\ \bibinfo {pages} {A165} (\bibinfo {year} {2020})}\BibitemShut {NoStop}%
\bibitem [{\citenamefont {Bonvin}\ \emph {et~al.}(2017)\citenamefont {Bonvin}, \citenamefont {Courbin}, \citenamefont {Suyu}, \citenamefont {Marshall}, \citenamefont {Rusu}, \citenamefont {Sluse}, \citenamefont {Tewes}, \citenamefont {Wong}, \citenamefont {Collett}, \citenamefont {Fassnacht} \emph {et~al.}}]{bonvin2017}%
  \BibitemOpen
  \bibfield  {author} {\bibinfo {author} {\bibfnamefont {V.}~\bibnamefont {Bonvin}}, \bibinfo {author} {\bibfnamefont {F.}~\bibnamefont {Courbin}}, \bibinfo {author} {\bibfnamefont {S.~H.}\ \bibnamefont {Suyu}}, \bibinfo {author} {\bibfnamefont {P.~J.}\ \bibnamefont {Marshall}}, \bibinfo {author} {\bibfnamefont {C.~E.}\ \bibnamefont {Rusu}}, \bibinfo {author} {\bibfnamefont {D.}~\bibnamefont {Sluse}}, \bibinfo {author} {\bibfnamefont {M.}~\bibnamefont {Tewes}}, \bibinfo {author} {\bibfnamefont {K.~C.}\ \bibnamefont {Wong}}, \bibinfo {author} {\bibfnamefont {T.}~\bibnamefont {Collett}}, \bibinfo {author} {\bibfnamefont {C.~D.}\ \bibnamefont {Fassnacht}}, \emph {et~al.},\ }\href {https://doi.org/10.1093/mnras/stw3006} {\bibfield  {journal} {\bibinfo  {journal} {Monthly Notices of the Royal Astronomical Society}\ }\textbf {\bibinfo {volume} {465}},\ \bibinfo {pages} {4914} (\bibinfo {year} {2017})},\ \bibinfo {note} {aDS Bibcode: 2017MNRAS.465.4914B}\BibitemShut {NoStop}%
\bibitem [{\citenamefont {Treu}\ and\ \citenamefont {Koopmans}(2002)}]{treu2002a}%
  \BibitemOpen
  \bibfield  {author} {\bibinfo {author} {\bibfnamefont {T.}~\bibnamefont {Treu}}\ and\ \bibinfo {author} {\bibfnamefont {L.~V.~E.}\ \bibnamefont {Koopmans}},\ }\href {https://doi.org/10.1046/j.1365-8711.2002.06107.x} {\bibfield  {journal} {\bibinfo  {journal} {Monthly Notices of the Royal Astronomical Society}\ }\textbf {\bibinfo {volume} {337}},\ \bibinfo {pages} {L6} (\bibinfo {year} {2002})},\ \bibinfo {note} {aDS Bibcode: 2002MNRAS.337L...6T}\BibitemShut {NoStop}%
\bibitem [{\citenamefont {Knabel}\ \emph {et~al.}(2025{\natexlab{b}})\citenamefont {Knabel}, \citenamefont {Treu}, \citenamefont {Cappellari}, \citenamefont {Shajib}, \citenamefont {Chen}, \citenamefont {Birrer},\ and\ \citenamefont {Bennert}}]{knabel2025a}%
  \BibitemOpen
  \bibfield  {author} {\bibinfo {author} {\bibfnamefont {S.}~\bibnamefont {Knabel}}, \bibinfo {author} {\bibfnamefont {T.}~\bibnamefont {Treu}}, \bibinfo {author} {\bibfnamefont {M.}~\bibnamefont {Cappellari}}, \bibinfo {author} {\bibfnamefont {A.~J.}\ \bibnamefont {Shajib}}, \bibinfo {author} {\bibfnamefont {C.-F.}\ \bibnamefont {Chen}}, \bibinfo {author} {\bibfnamefont {S.}~\bibnamefont {Birrer}},\ and\ \bibinfo {author} {\bibfnamefont {V.~N.}\ \bibnamefont {Bennert}},\ }\href {https://doi.org/10.48550/arXiv.2409.10631} {{\selectlanguage {en}\bibinfo {title} {Spatially resolved kinematics of {SLACS} lens galaxies. {I}: data and kinematic classification}}} (\bibinfo {year} {2025}{\natexlab{b}})\BibitemShut {NoStop}%
\bibitem [{\citenamefont {Binney}\ and\ \citenamefont {Mamon}(1982)}]{binney1982}%
  \BibitemOpen
  \bibfield  {author} {\bibinfo {author} {\bibfnamefont {J.}~\bibnamefont {Binney}}\ and\ \bibinfo {author} {\bibfnamefont {G.~A.}\ \bibnamefont {Mamon}},\ }\href {https://doi.org/10.1093/mnras/200.2.361} {\bibfield  {journal} {\bibinfo  {journal} {Monthly Notices of the Royal Astronomical Society}\ }\textbf {\bibinfo {volume} {200}},\ \bibinfo {pages} {361} (\bibinfo {year} {1982})},\ \bibinfo {note} {aDS Bibcode: 1982MNRAS.200..361B}\BibitemShut {NoStop}%
\bibitem [{\citenamefont {Merritt}(1985{\natexlab{a}})}]{merritt1985}%
  \BibitemOpen
  \bibfield  {author} {\bibinfo {author} {\bibfnamefont {D.}~\bibnamefont {Merritt}},\ }\href {https://doi.org/10.1086/113810} {\bibfield  {journal} {\bibinfo  {journal} {Astronomical Journal}\ }\textbf {\bibinfo {volume} {90}},\ \bibinfo {pages} {1027} (\bibinfo {year} {1985}{\natexlab{a}})},\ \bibinfo {note} {aDS Bibcode: 1985AJ.....90.1027M}\BibitemShut {NoStop}%
\bibitem [{\citenamefont {Cappellari}\ \emph {et~al.}(2007)\citenamefont {Cappellari}, \citenamefont {Emsellem}, \citenamefont {Bacon}, \citenamefont {Bureau}, \citenamefont {Davies}, \citenamefont {de~Zeeuw}, \citenamefont {Falcón-Barroso}, \citenamefont {Krajnović}, \citenamefont {Kuntschner}, \citenamefont {McDermid} \emph {et~al.}}]{cappellari2007}%
  \BibitemOpen
  \bibfield  {author} {\bibinfo {author} {\bibfnamefont {M.}~\bibnamefont {Cappellari}}, \bibinfo {author} {\bibfnamefont {E.}~\bibnamefont {Emsellem}}, \bibinfo {author} {\bibfnamefont {R.}~\bibnamefont {Bacon}}, \bibinfo {author} {\bibfnamefont {M.}~\bibnamefont {Bureau}}, \bibinfo {author} {\bibfnamefont {R.~L.}\ \bibnamefont {Davies}}, \bibinfo {author} {\bibfnamefont {P.~T.}\ \bibnamefont {de~Zeeuw}}, \bibinfo {author} {\bibfnamefont {J.}~\bibnamefont {Falcón-Barroso}}, \bibinfo {author} {\bibfnamefont {D.}~\bibnamefont {Krajnović}}, \bibinfo {author} {\bibfnamefont {H.}~\bibnamefont {Kuntschner}}, \bibinfo {author} {\bibfnamefont {R.~M.}\ \bibnamefont {McDermid}}, \emph {et~al.},\ }\href {https://doi.org/10.1111/j.1365-2966.2007.11963.x} {\bibfield  {journal} {\bibinfo  {journal} {Monthly Notices of the Royal Astronomical Society}\ }\textbf {\bibinfo {volume} {379}},\ \bibinfo {pages} {418} (\bibinfo {year} {2007})},\ \bibinfo {note} {aDS Bibcode: 2007MNRAS.379..418C}\BibitemShut {NoStop}%
\bibitem [{\citenamefont {Cappellari}(2020)}]{cappellari2020}%
  \BibitemOpen
  \bibfield  {author} {\bibinfo {author} {\bibfnamefont {M.}~\bibnamefont {Cappellari}},\ }\href {https://doi.org/10.1093/mnras/staa959} {\bibfield  {journal} {\bibinfo  {journal} {Monthly Notices of the Royal Astronomical Society}\ }\textbf {\bibinfo {volume} {494}},\ \bibinfo {pages} {4819} (\bibinfo {year} {2020})}\BibitemShut {NoStop}%
\bibitem [{\citenamefont {Zhu}\ \emph {et~al.}(2023)\citenamefont {Zhu}, \citenamefont {Lu}, \citenamefont {Cappellari}, \citenamefont {Li}, \citenamefont {Mao},\ and\ \citenamefont {Gao}}]{zhu2023}%
  \BibitemOpen
  \bibfield  {author} {\bibinfo {author} {\bibfnamefont {K.}~\bibnamefont {Zhu}}, \bibinfo {author} {\bibfnamefont {S.}~\bibnamefont {Lu}}, \bibinfo {author} {\bibfnamefont {M.}~\bibnamefont {Cappellari}}, \bibinfo {author} {\bibfnamefont {R.}~\bibnamefont {Li}}, \bibinfo {author} {\bibfnamefont {S.}~\bibnamefont {Mao}},\ and\ \bibinfo {author} {\bibfnamefont {L.}~\bibnamefont {Gao}},\ }\href {https://doi.org/10.1093/mnras/stad1299} {\bibfield  {journal} {\bibinfo  {journal} {Monthly Notices of the Royal Astronomical Society}\ }\textbf {\bibinfo {volume} {522}},\ \bibinfo {pages} {6326} (\bibinfo {year} {2023})}\BibitemShut {NoStop}%
\bibitem [{\citenamefont {Cappellari}(2023)}]{cappellari2023}%
  \BibitemOpen
  \bibfield  {author} {\bibinfo {author} {\bibfnamefont {M.}~\bibnamefont {Cappellari}},\ }\href {https://doi.org/10.1093/mnras/stad2597} {\bibfield  {journal} {\bibinfo  {journal} {Monthly Notices of the Royal Astronomical Society}\ }\textbf {\bibinfo {volume} {526}},\ \bibinfo {pages} {3273} (\bibinfo {year} {2023})},\ \bibinfo {note} {aDS Bibcode: 2023MNRAS.526.3273C}\BibitemShut {NoStop}%
\bibitem [{\citenamefont {Verma}\ and\ \citenamefont {Minor}(2026)}]{verma2026}%
  \BibitemOpen
  \bibfield  {author} {\bibinfo {author} {\bibfnamefont {V.}~\bibnamefont {Verma}}\ and\ \bibinfo {author} {\bibfnamefont {Q.}~\bibnamefont {Minor}},\ }\href {https://doi.org/10.3847/1538-4357/ae4722} {\bibfield  {journal} {\bibinfo  {journal} {The Astrophysical Journal}\ }\textbf {\bibinfo {volume} {1000}},\ \bibinfo {pages} {264} (\bibinfo {year} {2026})}\BibitemShut {NoStop}%
\bibitem [{\citenamefont {Binney}\ and\ \citenamefont {Tremaine}(1987)}]{binney1987}%
  \BibitemOpen
  \bibfield  {author} {\bibinfo {author} {\bibfnamefont {J.}~\bibnamefont {Binney}}\ and\ \bibinfo {author} {\bibfnamefont {S.}~\bibnamefont {Tremaine}},\ }\href {https://ui.adsabs.harvard.edu/abs/1987gady.book.....B} {{\selectlanguage {en}\emph {\bibinfo {title} {Galactic dynamics}}}}\ (\bibinfo {year} {1987})\ \bibinfo {note} {aDS Bibcode: 1987gady.book.....B Publication Title: Princeton}\BibitemShut {NoStop}%
\bibitem [{\citenamefont {Morgan}\ \emph {et~al.}(2004)\citenamefont {Morgan}, \citenamefont {Caldwell}, \citenamefont {Schechter}, \citenamefont {Dressler}, \citenamefont {Egami},\ and\ \citenamefont {Rix}}]{morgan2004}%
  \BibitemOpen
  \bibfield  {author} {\bibinfo {author} {\bibfnamefont {N.~D.}\ \bibnamefont {Morgan}}, \bibinfo {author} {\bibfnamefont {J.~A.~R.}\ \bibnamefont {Caldwell}}, \bibinfo {author} {\bibfnamefont {P.~L.}\ \bibnamefont {Schechter}}, \bibinfo {author} {\bibfnamefont {A.}~\bibnamefont {Dressler}}, \bibinfo {author} {\bibfnamefont {E.}~\bibnamefont {Egami}},\ and\ \bibinfo {author} {\bibfnamefont {H.-W.}\ \bibnamefont {Rix}},\ }\href {https://doi.org/10.1086/383295} {\bibfield  {journal} {\bibinfo  {journal} {Astronomical Journal}\ }\textbf {\bibinfo {volume} {127}},\ \bibinfo {pages} {2617} (\bibinfo {year} {2004})}\BibitemShut {NoStop}%
\bibitem [{\citenamefont {Sluse}\ \emph {et~al.}(2012)\citenamefont {Sluse}, \citenamefont {Hutsemékers}, \citenamefont {Courbin}, \citenamefont {Meylan},\ and\ \citenamefont {Wambsganss}}]{sluse2012}%
  \BibitemOpen
  \bibfield  {author} {\bibinfo {author} {\bibfnamefont {D.}~\bibnamefont {Sluse}}, \bibinfo {author} {\bibfnamefont {D.}~\bibnamefont {Hutsemékers}}, \bibinfo {author} {\bibfnamefont {F.}~\bibnamefont {Courbin}}, \bibinfo {author} {\bibfnamefont {G.}~\bibnamefont {Meylan}},\ and\ \bibinfo {author} {\bibfnamefont {J.}~\bibnamefont {Wambsganss}},\ }\href {https://doi.org/10.1051/0004-6361/201219125} {\bibfield  {journal} {\bibinfo  {journal} {Astronomy and Astrophysics}\ }\textbf {\bibinfo {volume} {544}},\ \bibinfo {pages} {A62} (\bibinfo {year} {2012})}\BibitemShut {NoStop}%
\bibitem [{\citenamefont {Sluse}\ \emph {et~al.}(2019)\citenamefont {Sluse}, \citenamefont {Rusu}, \citenamefont {Fassnacht}, \citenamefont {Sonnenfeld}, \citenamefont {Richard}, \citenamefont {Auger}, \citenamefont {Coccato}, \citenamefont {Wong}, \citenamefont {Suyu}, \citenamefont {Treu} \emph {et~al.}}]{sluse2019}%
  \BibitemOpen
  \bibfield  {author} {\bibinfo {author} {\bibfnamefont {D.}~\bibnamefont {Sluse}}, \bibinfo {author} {\bibfnamefont {C.~E.}\ \bibnamefont {Rusu}}, \bibinfo {author} {\bibfnamefont {C.~D.}\ \bibnamefont {Fassnacht}}, \bibinfo {author} {\bibfnamefont {A.}~\bibnamefont {Sonnenfeld}}, \bibinfo {author} {\bibfnamefont {J.}~\bibnamefont {Richard}}, \bibinfo {author} {\bibfnamefont {M.~W.}\ \bibnamefont {Auger}}, \bibinfo {author} {\bibfnamefont {L.}~\bibnamefont {Coccato}}, \bibinfo {author} {\bibfnamefont {K.~C.}\ \bibnamefont {Wong}}, \bibinfo {author} {\bibfnamefont {S.~H.}\ \bibnamefont {Suyu}}, \bibinfo {author} {\bibfnamefont {T.}~\bibnamefont {Treu}}, \emph {et~al.},\ }\href {https://doi.org/10.1093/mnras/stz2483} {\bibfield  {journal} {\bibinfo  {journal} {Monthly Notices of the Royal Astronomical Society}\ }\textbf {\bibinfo {volume} {490}},\ \bibinfo {pages} {613} (\bibinfo {year} {2019})}\BibitemShut {NoStop}%
\bibitem [{\citenamefont {Wisotzki}\ \emph {et~al.}(2002)\citenamefont {Wisotzki}, \citenamefont {Schechter}, \citenamefont {Bradt}, \citenamefont {Heinmüller},\ and\ \citenamefont {Reimers}}]{wisotzki2002}%
  \BibitemOpen
  \bibfield  {author} {\bibinfo {author} {\bibfnamefont {L.}~\bibnamefont {Wisotzki}}, \bibinfo {author} {\bibfnamefont {P.~L.}\ \bibnamefont {Schechter}}, \bibinfo {author} {\bibfnamefont {H.~V.}\ \bibnamefont {Bradt}}, \bibinfo {author} {\bibfnamefont {J.}~\bibnamefont {Heinmüller}},\ and\ \bibinfo {author} {\bibfnamefont {D.}~\bibnamefont {Reimers}},\ }\href {https://doi.org/10.1051/0004-6361:20021213} {\bibfield  {journal} {\bibinfo  {journal} {Astronomy and Astrophysics}\ }\textbf {\bibinfo {volume} {395}},\ \bibinfo {pages} {17} (\bibinfo {year} {2002})},\ \bibinfo {note} {aDS Bibcode: 2002A\&A...395...17W}\BibitemShut {NoStop}%
\bibitem [{\citenamefont {Morgan}\ \emph {et~al.}(2005)\citenamefont {Morgan}, \citenamefont {Kochanek}, \citenamefont {Pevunova},\ and\ \citenamefont {Schechter}}]{morgan2005}%
  \BibitemOpen
  \bibfield  {author} {\bibinfo {author} {\bibfnamefont {N.~D.}\ \bibnamefont {Morgan}}, \bibinfo {author} {\bibfnamefont {C.~S.}\ \bibnamefont {Kochanek}}, \bibinfo {author} {\bibfnamefont {O.}~\bibnamefont {Pevunova}},\ and\ \bibinfo {author} {\bibfnamefont {P.~L.}\ \bibnamefont {Schechter}},\ }\href {https://doi.org/10.1086/430145} {\bibfield  {journal} {\bibinfo  {journal} {The Astronomical Journal}\ }\textbf {\bibinfo {volume} {129}},\ \bibinfo {pages} {2531} (\bibinfo {year} {2005})},\ \bibinfo {note} {aDS Bibcode: 2005AJ....129.2531M}\BibitemShut {NoStop}%
\bibitem [{\citenamefont {Momcheva}\ \emph {et~al.}(2006)\citenamefont {Momcheva}, \citenamefont {Williams}, \citenamefont {Keeton},\ and\ \citenamefont {Zabludoff}}]{momcheva2006}%
  \BibitemOpen
  \bibfield  {author} {\bibinfo {author} {\bibfnamefont {I.}~\bibnamefont {Momcheva}}, \bibinfo {author} {\bibfnamefont {K.}~\bibnamefont {Williams}}, \bibinfo {author} {\bibfnamefont {C.}~\bibnamefont {Keeton}},\ and\ \bibinfo {author} {\bibfnamefont {A.}~\bibnamefont {Zabludoff}},\ }\href {https://doi.org/10.1086/500382} {\bibfield  {journal} {\bibinfo  {journal} {Astrophysical Journal}\ }\textbf {\bibinfo {volume} {641}},\ \bibinfo {pages} {169} (\bibinfo {year} {2006})},\ \bibinfo {note} {aDS Bibcode: 2006ApJ...641..169M}\BibitemShut {NoStop}%
\bibitem [{\citenamefont {Wong}\ \emph {et~al.}(2011)\citenamefont {Wong}, \citenamefont {Keeton}, \citenamefont {Williams}, \citenamefont {Momcheva},\ and\ \citenamefont {Zabludoff}}]{wong2010}%
  \BibitemOpen
  \bibfield  {author} {\bibinfo {author} {\bibfnamefont {K.~C.}\ \bibnamefont {Wong}}, \bibinfo {author} {\bibfnamefont {C.~R.}\ \bibnamefont {Keeton}}, \bibinfo {author} {\bibfnamefont {K.~A.}\ \bibnamefont {Williams}}, \bibinfo {author} {\bibfnamefont {I.~G.}\ \bibnamefont {Momcheva}},\ and\ \bibinfo {author} {\bibfnamefont {A.~I.}\ \bibnamefont {Zabludoff}},\ }\href {https://doi.org/10.1088/0004-637x/726/2/84} {\bibfield  {journal} {\bibinfo  {journal} {Astrophysical Journal}\ }\textbf {\bibinfo {volume} {726}},\ \bibinfo {pages} {84} (\bibinfo {year} {2011})}\BibitemShut {NoStop}%
\bibitem [{\citenamefont {Wilson}\ \emph {et~al.}(2016)\citenamefont {Wilson}, \citenamefont {Zabludoff}, \citenamefont {Ammons}, \citenamefont {Momcheva}, \citenamefont {Williams},\ and\ \citenamefont {Keeton}}]{wilson2016}%
  \BibitemOpen
  \bibfield  {author} {\bibinfo {author} {\bibfnamefont {M.~L.}\ \bibnamefont {Wilson}}, \bibinfo {author} {\bibfnamefont {A.~I.}\ \bibnamefont {Zabludoff}}, \bibinfo {author} {\bibfnamefont {S.~M.}\ \bibnamefont {Ammons}}, \bibinfo {author} {\bibfnamefont {I.~G.}\ \bibnamefont {Momcheva}}, \bibinfo {author} {\bibfnamefont {K.~A.}\ \bibnamefont {Williams}},\ and\ \bibinfo {author} {\bibfnamefont {C.~R.}\ \bibnamefont {Keeton}},\ }\href {https://doi.org/10.3847/1538-4357/833/2/194} {\bibfield  {journal} {\bibinfo  {journal} {Astrophysical Journal}\ }\textbf {\bibinfo {volume} {833}},\ \bibinfo {pages} {194} (\bibinfo {year} {2016})},\ \bibinfo {note} {aDS Bibcode: 2016ApJ...833..194W}\BibitemShut {NoStop}%
\bibitem [{\citenamefont {Sluse}\ \emph {et~al.}(2017)\citenamefont {Sluse}, \citenamefont {Sonnenfeld}, \citenamefont {Rumbaugh}, \citenamefont {Rusu}, \citenamefont {Fassnacht}, \citenamefont {Treu}, \citenamefont {Suyu}, \citenamefont {Wong}, \citenamefont {Auger}, \citenamefont {Bonvin} \emph {et~al.}}]{sluse2017}%
  \BibitemOpen
  \bibfield  {author} {\bibinfo {author} {\bibfnamefont {D.}~\bibnamefont {Sluse}}, \bibinfo {author} {\bibfnamefont {A.}~\bibnamefont {Sonnenfeld}}, \bibinfo {author} {\bibfnamefont {N.}~\bibnamefont {Rumbaugh}}, \bibinfo {author} {\bibfnamefont {C.~E.}\ \bibnamefont {Rusu}}, \bibinfo {author} {\bibfnamefont {C.~D.}\ \bibnamefont {Fassnacht}}, \bibinfo {author} {\bibfnamefont {T.}~\bibnamefont {Treu}}, \bibinfo {author} {\bibfnamefont {S.~H.}\ \bibnamefont {Suyu}}, \bibinfo {author} {\bibfnamefont {K.~C.}\ \bibnamefont {Wong}}, \bibinfo {author} {\bibfnamefont {M.~W.}\ \bibnamefont {Auger}}, \bibinfo {author} {\bibfnamefont {V.}~\bibnamefont {Bonvin}}, \emph {et~al.},\ }\href {https://doi.org/10.1093/mnras/stx1484} {\bibfield  {journal} {\bibinfo  {journal} {Monthly Notices of the Royal Astronomical Society}\ }\textbf {\bibinfo {volume} {470}},\ \bibinfo {pages} {4838} (\bibinfo {year} {2017})}\BibitemShut {NoStop}%
\bibitem [{\citenamefont {Mozumdar}\ \emph {et~al.}(2025)\citenamefont {Mozumdar}, \citenamefont {Cappellari}, \citenamefont {Fassnacht},\ and\ \citenamefont {Treu}}]{mozumdar2025}%
  \BibitemOpen
  \bibfield  {author} {\bibinfo {author} {\bibfnamefont {P.}~\bibnamefont {Mozumdar}}, \bibinfo {author} {\bibfnamefont {M.}~\bibnamefont {Cappellari}}, \bibinfo {author} {\bibfnamefont {C.~D.}\ \bibnamefont {Fassnacht}},\ and\ \bibinfo {author} {\bibfnamefont {T.}~\bibnamefont {Treu}},\ }\href {https://doi.org/10.48550/arXiv.2510.23863} {{\selectlanguage {en}\bibinfo {title} {{MAGNUS} {I}: a {MUSE}-{DEEP} sample of early-type galaxies at intermediate redshift}}} (\bibinfo {year} {2025}),\ \bibinfo {note} {aDS Bibcode: 2025arXiv251023863M}\BibitemShut {NoStop}%
\bibitem [{\citenamefont {Valdes}\ \emph {et~al.}(2004)\citenamefont {Valdes}, \citenamefont {Gupta}, \citenamefont {Rose}, \citenamefont {Singh},\ and\ \citenamefont {Bell}}]{valdes2004}%
  \BibitemOpen
  \bibfield  {author} {\bibinfo {author} {\bibfnamefont {F.}~\bibnamefont {Valdes}}, \bibinfo {author} {\bibfnamefont {R.}~\bibnamefont {Gupta}}, \bibinfo {author} {\bibfnamefont {J.~A.}\ \bibnamefont {Rose}}, \bibinfo {author} {\bibfnamefont {H.~P.}\ \bibnamefont {Singh}},\ and\ \bibinfo {author} {\bibfnamefont {D.~J.}\ \bibnamefont {Bell}},\ }\href {https://doi.org/10.1086/386343} {\bibfield  {journal} {\bibinfo  {journal} {The Astrophysical Journal Supplement Series}\ }\textbf {\bibinfo {volume} {152}},\ \bibinfo {pages} {251} (\bibinfo {year} {2004})},\ \bibinfo {note} {aDS Bibcode: 2004ApJS..152..251V}\BibitemShut {NoStop}%
\bibitem [{\citenamefont {Falcón-Barroso}\ \emph {et~al.}(2011)\citenamefont {Falcón-Barroso}, \citenamefont {Sánchez-Blázquez}, \citenamefont {Vazdekis}, \citenamefont {Ricciardelli}, \citenamefont {Cardiel}, \citenamefont {Cenarro}, \citenamefont {Gorgas},\ and\ \citenamefont {Peletier}}]{falcon-barroso2011}%
  \BibitemOpen
  \bibfield  {author} {\bibinfo {author} {\bibfnamefont {J.}~\bibnamefont {Falcón-Barroso}}, \bibinfo {author} {\bibfnamefont {P.}~\bibnamefont {Sánchez-Blázquez}}, \bibinfo {author} {\bibfnamefont {A.}~\bibnamefont {Vazdekis}}, \bibinfo {author} {\bibfnamefont {E.}~\bibnamefont {Ricciardelli}}, \bibinfo {author} {\bibfnamefont {N.}~\bibnamefont {Cardiel}}, \bibinfo {author} {\bibfnamefont {A.~J.}\ \bibnamefont {Cenarro}}, \bibinfo {author} {\bibfnamefont {J.}~\bibnamefont {Gorgas}},\ and\ \bibinfo {author} {\bibfnamefont {R.~F.}\ \bibnamefont {Peletier}},\ }\href {https://doi.org/10.1051/0004-6361/201116842} {\bibfield  {journal} {\bibinfo  {journal} {Astronomy and Astrophysics}\ }\textbf {\bibinfo {volume} {532}},\ \bibinfo {pages} {A95} (\bibinfo {year} {2011})},\ \bibinfo {note} {aDS Bibcode: 2011A\&A...532A..95F}\BibitemShut {NoStop}%
\bibitem [{\citenamefont {Verro}\ \emph {et~al.}(2022)\citenamefont {Verro}, \citenamefont {Trager}, \citenamefont {Peletier}, \citenamefont {Lançon}, \citenamefont {Gonneau}, \citenamefont {Vazdekis}, \citenamefont {Prugniel}, \citenamefont {Chen}, \citenamefont {Coelho}, \citenamefont {Sánchez-Blázquez} \emph {et~al.}}]{verro2022}%
  \BibitemOpen
  \bibfield  {author} {\bibinfo {author} {\bibfnamefont {K.}~\bibnamefont {Verro}}, \bibinfo {author} {\bibfnamefont {S.~C.}\ \bibnamefont {Trager}}, \bibinfo {author} {\bibfnamefont {R.~F.}\ \bibnamefont {Peletier}}, \bibinfo {author} {\bibfnamefont {A.}~\bibnamefont {Lançon}}, \bibinfo {author} {\bibfnamefont {A.}~\bibnamefont {Gonneau}}, \bibinfo {author} {\bibfnamefont {A.}~\bibnamefont {Vazdekis}}, \bibinfo {author} {\bibfnamefont {P.}~\bibnamefont {Prugniel}}, \bibinfo {author} {\bibfnamefont {Y.-P.}\ \bibnamefont {Chen}}, \bibinfo {author} {\bibfnamefont {P.~R.~T.}\ \bibnamefont {Coelho}}, \bibinfo {author} {\bibfnamefont {P.}~\bibnamefont {Sánchez-Blázquez}}, \emph {et~al.},\ }\href {https://doi.org/10.1051/0004-6361/202142388} {\bibfield  {journal} {\bibinfo  {journal} {Astronomy and Astrophysics}\ }\textbf {\bibinfo {volume} {660}},\ \bibinfo {pages} {A34} (\bibinfo {year} {2022})},\ \bibinfo {note} {aDS Bibcode: 2022A\&A...660A..34V}\BibitemShut {NoStop}%
\bibitem [{\citenamefont {Weymann}\ \emph {et~al.}(1980)\citenamefont {Weymann}, \citenamefont {Latham}, \citenamefont {Angel}, \citenamefont {Green}, \citenamefont {Liebert}, \citenamefont {Turnshek}, \citenamefont {Turnshek},\ and\ \citenamefont {Tyson}}]{weymann1980}%
  \BibitemOpen
  \bibfield  {author} {\bibinfo {author} {\bibfnamefont {R.~J.}\ \bibnamefont {Weymann}}, \bibinfo {author} {\bibfnamefont {D.}~\bibnamefont {Latham}}, \bibinfo {author} {\bibfnamefont {J.~R.~P.}\ \bibnamefont {Angel}}, \bibinfo {author} {\bibfnamefont {R.~F.}\ \bibnamefont {Green}}, \bibinfo {author} {\bibfnamefont {J.~W.}\ \bibnamefont {Liebert}}, \bibinfo {author} {\bibfnamefont {D.~A.}\ \bibnamefont {Turnshek}}, \bibinfo {author} {\bibfnamefont {D.~E.}\ \bibnamefont {Turnshek}},\ and\ \bibinfo {author} {\bibfnamefont {J.~A.}\ \bibnamefont {Tyson}},\ }\href {https://doi.org/10.1038/285641a0} {\bibfield  {journal} {\bibinfo  {journal} {Nature}\ }\textbf {\bibinfo {volume} {285}},\ \bibinfo {pages} {641} (\bibinfo {year} {1980})},\ \bibinfo {note} {aDS Bibcode: 1980Natur.285..641W}\BibitemShut {NoStop}%
\bibitem [{\citenamefont {Tonry}(1997)}]{tonry1997}%
  \BibitemOpen
  \bibfield  {author} {\bibinfo {author} {\bibfnamefont {J.~L.}\ \bibnamefont {Tonry}},\ }\href {https://doi.org/10.1086/300170} {\bibinfo {title} {Redshifts of the {Gravitational} {Lenses} {B1422}+231 and {PG1115}+080}} (\bibinfo {year} {1997}),\ \bibinfo {note} {arXiv:astro-ph/9706199}\BibitemShut {NoStop}%
\bibitem [{\citenamefont {Henry}\ and\ \citenamefont {Heasley}(1986)}]{henry1986}%
  \BibitemOpen
  \bibfield  {author} {\bibinfo {author} {\bibfnamefont {J.~P.}\ \bibnamefont {Henry}}\ and\ \bibinfo {author} {\bibfnamefont {J.~N.}\ \bibnamefont {Heasley}},\ }\href {https://doi.org/10.1038/321139a0} {\bibfield  {journal} {\bibinfo  {journal} {Nature}\ }\textbf {\bibinfo {volume} {321}},\ \bibinfo {pages} {139} (\bibinfo {year} {1986})}\BibitemShut {NoStop}%
\bibitem [{\citenamefont {Christian}\ \emph {et~al.}(1987)\citenamefont {Christian}, \citenamefont {Crabtree},\ and\ \citenamefont {Waddell}}]{christian1987}%
  \BibitemOpen
  \bibfield  {author} {\bibinfo {author} {\bibfnamefont {C.~A.}\ \bibnamefont {Christian}}, \bibinfo {author} {\bibfnamefont {D.}~\bibnamefont {Crabtree}},\ and\ \bibinfo {author} {\bibfnamefont {P.}~\bibnamefont {Waddell}},\ }\href {https://doi.org/10.1086/164847} {\bibfield  {journal} {\bibinfo  {journal} {The Astrophysical Journal}\ }\textbf {\bibinfo {volume} {312}},\ \bibinfo {pages} {45} (\bibinfo {year} {1987})},\ \bibinfo {note} {aDS Bibcode: 1987ApJ...312...45C}\BibitemShut {NoStop}%
\bibitem [{\citenamefont {Bonvin}\ \emph {et~al.}(2018)\citenamefont {Bonvin}, \citenamefont {Chan}, \citenamefont {Millon}, \citenamefont {Rojas}, \citenamefont {Courbin}, \citenamefont {Chen}, \citenamefont {Fassnacht}, \citenamefont {Paic}, \citenamefont {Tewes}, \citenamefont {Chao} \emph {et~al.}}]{bonvin2018}%
  \BibitemOpen
  \bibfield  {author} {\bibinfo {author} {\bibfnamefont {V.}~\bibnamefont {Bonvin}}, \bibinfo {author} {\bibfnamefont {J.~H.~H.}\ \bibnamefont {Chan}}, \bibinfo {author} {\bibfnamefont {M.}~\bibnamefont {Millon}}, \bibinfo {author} {\bibfnamefont {K.}~\bibnamefont {Rojas}}, \bibinfo {author} {\bibfnamefont {F.}~\bibnamefont {Courbin}}, \bibinfo {author} {\bibfnamefont {G.~C.-F.}\ \bibnamefont {Chen}}, \bibinfo {author} {\bibfnamefont {C.~D.}\ \bibnamefont {Fassnacht}}, \bibinfo {author} {\bibfnamefont {E.}~\bibnamefont {Paic}}, \bibinfo {author} {\bibfnamefont {M.}~\bibnamefont {Tewes}}, \bibinfo {author} {\bibfnamefont {D.~C.-Y.}\ \bibnamefont {Chao}}, \emph {et~al.},\ }\href {https://doi.org/10.1051/0004-6361/201833287} {\bibfield  {journal} {\bibinfo  {journal} {Astronomy \& Astrophysics}\ }\textbf {\bibinfo {volume} {616}},\ \bibinfo {pages} {A183} (\bibinfo {year} {2018})}\BibitemShut {NoStop}%
\bibitem [{\citenamefont {Birrer}\ \emph {et~al.}(2015)\citenamefont {Birrer}, \citenamefont {Amara},\ and\ \citenamefont {Refregier}}]{birrer2015}%
  \BibitemOpen
  \bibfield  {author} {\bibinfo {author} {\bibfnamefont {S.}~\bibnamefont {Birrer}}, \bibinfo {author} {\bibfnamefont {A.}~\bibnamefont {Amara}},\ and\ \bibinfo {author} {\bibfnamefont {A.}~\bibnamefont {Refregier}},\ }\href {https://doi.org/10.1088/0004-637x/813/2/102} {\bibfield  {journal} {\bibinfo  {journal} {Astrophysical Journal}\ }\textbf {\bibinfo {volume} {813}},\ \bibinfo {pages} {102} (\bibinfo {year} {2015})},\ \bibinfo {note} {aDS Bibcode: 2015ApJ...813..102B}\BibitemShut {NoStop}%
\bibitem [{\citenamefont {Birrer}\ and\ \citenamefont {Amara}(2018)}]{birrer2018}%
  \BibitemOpen
  \bibfield  {author} {\bibinfo {author} {\bibfnamefont {S.}~\bibnamefont {Birrer}}\ and\ \bibinfo {author} {\bibfnamefont {A.}~\bibnamefont {Amara}},\ }\href {https://doi.org/10.1016/j.dark.2018.11.002} {\bibfield  {journal} {\bibinfo  {journal} {Physics of the Dark Universe}\ }\textbf {\bibinfo {volume} {22}},\ \bibinfo {pages} {189} (\bibinfo {year} {2018})}\BibitemShut {NoStop}%
\bibitem [{\citenamefont {Birrer}\ \emph {et~al.}(2021)\citenamefont {Birrer}, \citenamefont {Shajib}, \citenamefont {Gilman}, \citenamefont {Galan}, \citenamefont {Aalbers}, \citenamefont {Millon}, \citenamefont {Morgan}, \citenamefont {Pagano}, \citenamefont {Park}, \citenamefont {Teodori} \emph {et~al.}}]{birrer2021}%
  \BibitemOpen
  \bibfield  {author} {\bibinfo {author} {\bibfnamefont {S.}~\bibnamefont {Birrer}}, \bibinfo {author} {\bibfnamefont {A.}~\bibnamefont {Shajib}}, \bibinfo {author} {\bibfnamefont {D.}~\bibnamefont {Gilman}}, \bibinfo {author} {\bibfnamefont {A.}~\bibnamefont {Galan}}, \bibinfo {author} {\bibfnamefont {J.}~\bibnamefont {Aalbers}}, \bibinfo {author} {\bibfnamefont {M.}~\bibnamefont {Millon}}, \bibinfo {author} {\bibfnamefont {R.}~\bibnamefont {Morgan}}, \bibinfo {author} {\bibfnamefont {G.}~\bibnamefont {Pagano}}, \bibinfo {author} {\bibfnamefont {J.}~\bibnamefont {Park}}, \bibinfo {author} {\bibfnamefont {L.}~\bibnamefont {Teodori}}, \emph {et~al.},\ }\href {https://doi.org/10.21105/joss.03283} {\bibfield  {journal} {\bibinfo  {journal} {Journal of Open Source Software}\ }\textbf {\bibinfo {volume} {6}},\ \bibinfo {pages} {3283} (\bibinfo {year} {2021})}\BibitemShut {NoStop}%
\bibitem [{\citenamefont {Barkana}(1998)}]{barkana1998}%
  \BibitemOpen
  \bibfield  {author} {\bibinfo {author} {\bibfnamefont {R.}~\bibnamefont {Barkana}},\ }\href {https://doi.org/10.1086/305950} {\bibfield  {journal} {\bibinfo  {journal} {Astrophysical Journal}\ }\textbf {\bibinfo {volume} {502}},\ \bibinfo {pages} {531} (\bibinfo {year} {1998})}\BibitemShut {NoStop}%
\bibitem [{\citenamefont {Schmidt}\ \emph {et~al.}(2022)\citenamefont {Schmidt}, \citenamefont {Treu}, \citenamefont {Birrer}, \citenamefont {Shajib}, \citenamefont {Lemon}, \citenamefont {Millon}, \citenamefont {Sluse}, \citenamefont {Agnello}, \citenamefont {Anguita}, \citenamefont {Auger-Williams} \emph {et~al.}}]{schmidt2022}%
  \BibitemOpen
  \bibfield  {author} {\bibinfo {author} {\bibfnamefont {T.}~\bibnamefont {Schmidt}}, \bibinfo {author} {\bibfnamefont {T.}~\bibnamefont {Treu}}, \bibinfo {author} {\bibfnamefont {S.}~\bibnamefont {Birrer}}, \bibinfo {author} {\bibfnamefont {A.~J.}\ \bibnamefont {Shajib}}, \bibinfo {author} {\bibfnamefont {C.}~\bibnamefont {Lemon}}, \bibinfo {author} {\bibfnamefont {M.}~\bibnamefont {Millon}}, \bibinfo {author} {\bibfnamefont {D.}~\bibnamefont {Sluse}}, \bibinfo {author} {\bibfnamefont {A.}~\bibnamefont {Agnello}}, \bibinfo {author} {\bibfnamefont {T.}~\bibnamefont {Anguita}}, \bibinfo {author} {\bibfnamefont {M.~W.}\ \bibnamefont {Auger-Williams}}, \emph {et~al.},\ }\href {https://doi.org/10.1093/mnras/stac2235} {\bibfield  {journal} {\bibinfo  {journal} {Monthly Notices of the Royal Astronomical Society}\ }\textbf {\bibinfo {volume} {518}},\ \bibinfo {pages} {1260} (\bibinfo {year} {2022})}\BibitemShut {NoStop}%
\bibitem [{\citenamefont {Sersic}(1968)}]{sersic1968}%
  \BibitemOpen
  \bibfield  {author} {\bibinfo {author} {\bibfnamefont {J.~L.}\ \bibnamefont {Sersic}},\ }\href {https://ui.adsabs.harvard.edu/abs/1968adga.book.....S} {\emph {\bibinfo {title} {Atlas de {Galaxias} {Australes}}}}\ (\bibinfo {year} {1968})\ \bibinfo {note} {publication Title: Cordoba ADS Bibcode: 1968adga.book.....S}\BibitemShut {NoStop}%
\bibitem [{\citenamefont {Refregier}(2003)}]{refregier2003}%
  \BibitemOpen
  \bibfield  {author} {\bibinfo {author} {\bibfnamefont {A.}~\bibnamefont {Refregier}},\ }\href {https://doi.org/10.1046/j.1365-8711.2003.05901.x} {\bibfield  {journal} {\bibinfo  {journal} {Monthly Notices of the Royal Astronomical Society}\ }\textbf {\bibinfo {volume} {338}},\ \bibinfo {pages} {35} (\bibinfo {year} {2003})}\BibitemShut {NoStop}%
\bibitem [{\citenamefont {McCully}\ \emph {et~al.}(2017)\citenamefont {McCully}, \citenamefont {Keeton}, \citenamefont {Wong},\ and\ \citenamefont {Zabludoff}}]{mccully2017}%
  \BibitemOpen
  \bibfield  {author} {\bibinfo {author} {\bibfnamefont {C.}~\bibnamefont {McCully}}, \bibinfo {author} {\bibfnamefont {C.~R.}\ \bibnamefont {Keeton}}, \bibinfo {author} {\bibfnamefont {K.~C.}\ \bibnamefont {Wong}},\ and\ \bibinfo {author} {\bibfnamefont {A.~I.}\ \bibnamefont {Zabludoff}},\ }\href {https://doi.org/10.3847/1538-4357/836/1/141} {\bibfield  {journal} {\bibinfo  {journal} {Astrophysical Journal}\ }\textbf {\bibinfo {volume} {836}},\ \bibinfo {pages} {141} (\bibinfo {year} {2017})}\BibitemShut {NoStop}%
\bibitem [{\citenamefont {Treu}\ \emph {et~al.}(2006)\citenamefont {Treu}, \citenamefont {Koopmans}, \citenamefont {Bolton}, \citenamefont {Burles},\ and\ \citenamefont {Moustakas}}]{treu2006}%
  \BibitemOpen
  \bibfield  {author} {\bibinfo {author} {\bibfnamefont {T.}~\bibnamefont {Treu}}, \bibinfo {author} {\bibfnamefont {L.~V.}\ \bibnamefont {Koopmans}}, \bibinfo {author} {\bibfnamefont {A.~S.}\ \bibnamefont {Bolton}}, \bibinfo {author} {\bibfnamefont {S.}~\bibnamefont {Burles}},\ and\ \bibinfo {author} {\bibfnamefont {L.~A.}\ \bibnamefont {Moustakas}},\ }\href {https://doi.org/10.1086/500124} {\bibfield  {journal} {\bibinfo  {journal} {Astrophysical Journal}\ }\textbf {\bibinfo {volume} {640}},\ \bibinfo {pages} {662} (\bibinfo {year} {2006})}\BibitemShut {NoStop}%
\bibitem [{\citenamefont {Navarro}\ \emph {et~al.}(1996)\citenamefont {Navarro}, \citenamefont {Frenk},\ and\ \citenamefont {White}}]{navarro1996}%
  \BibitemOpen
  \bibfield  {author} {\bibinfo {author} {\bibfnamefont {J.~F.}\ \bibnamefont {Navarro}}, \bibinfo {author} {\bibfnamefont {C.~S.}\ \bibnamefont {Frenk}},\ and\ \bibinfo {author} {\bibfnamefont {S.~D.~M.}\ \bibnamefont {White}},\ }\href {https://doi.org/10.1086/177173} {\bibfield  {journal} {\bibinfo  {journal} {The Astrophysical Journal}\ }\textbf {\bibinfo {volume} {462}},\ \bibinfo {pages} {563} (\bibinfo {year} {1996})},\ \bibinfo {note} {aDS Bibcode: 1996ApJ...462..563N}\BibitemShut {NoStop}%
\bibitem [{\citenamefont {Golse}\ and\ \citenamefont {Kneib}(2002)}]{golse2002}%
  \BibitemOpen
  \bibfield  {author} {\bibinfo {author} {\bibfnamefont {G.}~\bibnamefont {Golse}}\ and\ \bibinfo {author} {\bibfnamefont {J.-P.}\ \bibnamefont {Kneib}},\ }\href {https://doi.org/10.1051/0004-6361:20020639} {\bibfield  {journal} {\bibinfo  {journal} {Astronomy \& Astrophysics}\ }\textbf {\bibinfo {volume} {390}},\ \bibinfo {pages} {821} (\bibinfo {year} {2002})}\BibitemShut {NoStop}%
\bibitem [{\citenamefont {Kennedy}\ and\ \citenamefont {Eberhart}(1995)}]{kennedy1995}%
  \BibitemOpen
  \bibfield  {author} {\bibinfo {author} {\bibfnamefont {J.}~\bibnamefont {Kennedy}}\ and\ \bibinfo {author} {\bibfnamefont {R.}~\bibnamefont {Eberhart}},\ }in\ \href {https://doi.org/10.1109/ICNN.1995.488968} {{\selectlanguage {en}\emph {\bibinfo {booktitle} {Proceedings of {ICNN}'95 - {International} {Conference} on {Neural} {Networks}}}}},\ Vol.~\bibinfo {volume} {4}\ (\bibinfo {year} {1995})\ pp.\ \bibinfo {pages} {1942--1948 vol.4}\BibitemShut {NoStop}%
\bibitem [{\citenamefont {Foreman-Mackey}\ \emph {et~al.}(2013)\citenamefont {Foreman-Mackey}, \citenamefont {Hogg}, \citenamefont {Lang},\ and\ \citenamefont {Goodman}}]{foreman-mackey2013}%
  \BibitemOpen
  \bibfield  {author} {\bibinfo {author} {\bibfnamefont {D.}~\bibnamefont {Foreman-Mackey}}, \bibinfo {author} {\bibfnamefont {D.~W.}\ \bibnamefont {Hogg}}, \bibinfo {author} {\bibfnamefont {D.}~\bibnamefont {Lang}},\ and\ \bibinfo {author} {\bibfnamefont {J.}~\bibnamefont {Goodman}},\ }\href {https://doi.org/10.1086/670067} {\bibfield  {journal} {\bibinfo  {journal} {Publications of the Astronomical Society of the Pacific}\ }\textbf {\bibinfo {volume} {125}},\ \bibinfo {pages} {306} (\bibinfo {year} {2013})},\ \bibinfo {note} {aDS Bibcode: 2013PASP..125..306F}\BibitemShut {NoStop}%
\bibitem [{\citenamefont {Etherington}\ \emph {et~al.}(2024)\citenamefont {Etherington}, \citenamefont {Nightingale}, \citenamefont {Massey}, \citenamefont {Tam}, \citenamefont {Cao}, \citenamefont {Niemiec}, \citenamefont {He}, \citenamefont {Robertson}, \citenamefont {Li}, \citenamefont {Amvrosiadis} \emph {et~al.}}]{etherington2023}%
  \BibitemOpen
  \bibfield  {author} {\bibinfo {author} {\bibfnamefont {A.}~\bibnamefont {Etherington}}, \bibinfo {author} {\bibfnamefont {J.~W.}\ \bibnamefont {Nightingale}}, \bibinfo {author} {\bibfnamefont {R.}~\bibnamefont {Massey}}, \bibinfo {author} {\bibfnamefont {S.-I.}\ \bibnamefont {Tam}}, \bibinfo {author} {\bibfnamefont {X.}~\bibnamefont {Cao}}, \bibinfo {author} {\bibfnamefont {A.}~\bibnamefont {Niemiec}}, \bibinfo {author} {\bibfnamefont {Q.}~\bibnamefont {He}}, \bibinfo {author} {\bibfnamefont {A.}~\bibnamefont {Robertson}}, \bibinfo {author} {\bibfnamefont {R.}~\bibnamefont {Li}}, \bibinfo {author} {\bibfnamefont {A.}~\bibnamefont {Amvrosiadis}}, \emph {et~al.},\ }\href {https://doi.org/10.1093/mnras/stae1375} {\bibfield  {journal} {\bibinfo  {journal} {Monthly Notices of the Royal Astronomical Society}\ }\textbf {\bibinfo {volume} {531}},\ \bibinfo {pages} {3684} (\bibinfo {year} {2024})}\BibitemShut {NoStop}%
\bibitem [{\citenamefont {Birrer}\ \emph {et~al.}(2019)\citenamefont {Birrer}, \citenamefont {Treu}, \citenamefont {Rusu}, \citenamefont {Bonvin}, \citenamefont {Fassnacht}, \citenamefont {Chan}, \citenamefont {Agnello}, \citenamefont {Shajib}, \citenamefont {Chen}, \citenamefont {Auger} \emph {et~al.}}]{birrer2019a}%
  \BibitemOpen
  \bibfield  {author} {\bibinfo {author} {\bibfnamefont {S.}~\bibnamefont {Birrer}}, \bibinfo {author} {\bibfnamefont {T.}~\bibnamefont {Treu}}, \bibinfo {author} {\bibfnamefont {C.~E.}\ \bibnamefont {Rusu}}, \bibinfo {author} {\bibfnamefont {V.}~\bibnamefont {Bonvin}}, \bibinfo {author} {\bibfnamefont {C.~D.}\ \bibnamefont {Fassnacht}}, \bibinfo {author} {\bibfnamefont {J.~H.~H.}\ \bibnamefont {Chan}}, \bibinfo {author} {\bibfnamefont {A.}~\bibnamefont {Agnello}}, \bibinfo {author} {\bibfnamefont {A.~J.}\ \bibnamefont {Shajib}}, \bibinfo {author} {\bibfnamefont {G.~C.-F.}\ \bibnamefont {Chen}}, \bibinfo {author} {\bibfnamefont {M.}~\bibnamefont {Auger}}, \emph {et~al.},\ }\href {https://doi.org/10.1093/mnras/stz200} {\bibfield  {journal} {\bibinfo  {journal} {Monthly Notices of the Royal Astronomical Society}\ }\textbf {\bibinfo {volume} {484}},\ \bibinfo {pages} {4726} (\bibinfo {year} {2019})}\BibitemShut {NoStop}%
\bibitem [{\citenamefont {Millon}\ \emph {et~al.}(2020{\natexlab{c}})\citenamefont {Millon}, \citenamefont {Galan}, \citenamefont {Courbin}, \citenamefont {Treu}, \citenamefont {Suyu}, \citenamefont {Ding}, \citenamefont {Birrer}, \citenamefont {Chen}, \citenamefont {Shajib}, \citenamefont {Sluse} \emph {et~al.}}]{millon2020b}%
  \BibitemOpen
  \bibfield  {author} {\bibinfo {author} {\bibfnamefont {M.}~\bibnamefont {Millon}}, \bibinfo {author} {\bibfnamefont {A.}~\bibnamefont {Galan}}, \bibinfo {author} {\bibfnamefont {F.}~\bibnamefont {Courbin}}, \bibinfo {author} {\bibfnamefont {T.}~\bibnamefont {Treu}}, \bibinfo {author} {\bibfnamefont {S.~H.}\ \bibnamefont {Suyu}}, \bibinfo {author} {\bibfnamefont {X.}~\bibnamefont {Ding}}, \bibinfo {author} {\bibfnamefont {S.}~\bibnamefont {Birrer}}, \bibinfo {author} {\bibfnamefont {G.~C.-F.}\ \bibnamefont {Chen}}, \bibinfo {author} {\bibfnamefont {A.~J.}\ \bibnamefont {Shajib}}, \bibinfo {author} {\bibfnamefont {D.}~\bibnamefont {Sluse}}, \emph {et~al.},\ }\href {https://doi.org/10.1051/0004-6361/201937351} {\bibfield  {journal} {\bibinfo  {journal} {Astronomy and Astrophysics}\ }\textbf {\bibinfo {volume} {639}},\ \bibinfo {pages} {A101} (\bibinfo {year} {2020}{\natexlab{c}})},\ \bibinfo {note} {aDS Bibcode: 2020A\&A...639A.101M}\BibitemShut {NoStop}%
\bibitem [{\citenamefont {Birrer}\ and\ \citenamefont {Treu}(2019)}]{birrer2019}%
  \BibitemOpen
  \bibfield  {author} {\bibinfo {author} {\bibfnamefont {S.}~\bibnamefont {Birrer}}\ and\ \bibinfo {author} {\bibfnamefont {T.}~\bibnamefont {Treu}},\ }\href {https://doi.org/10.1093/mnras/stz2254} {\bibfield  {journal} {\bibinfo  {journal} {Monthly Notices of the Royal Astronomical Society}\ }\textbf {\bibinfo {volume} {489}},\ \bibinfo {pages} {2097} (\bibinfo {year} {2019})}\BibitemShut {NoStop}%
\bibitem [{\citenamefont {Osipkov}(1979)}]{osipkov1979}%
  \BibitemOpen
  \bibfield  {author} {\bibinfo {author} {\bibfnamefont {L.~P.}\ \bibnamefont {Osipkov}},\ }\href {https://ui.adsabs.harvard.edu/abs/1979SvAL....5...42O} {\bibfield  {journal} {\bibinfo  {journal} {Soviet Astronomy Letters}\ }\textbf {\bibinfo {volume} {5}},\ \bibinfo {pages} {42} (\bibinfo {year} {1979})},\ \bibinfo {note} {aDS Bibcode: 1979SvAL....5...42O}\BibitemShut {NoStop}%
\bibitem [{\citenamefont {Merritt}(1985{\natexlab{b}})}]{merritt1985a}%
  \BibitemOpen
  \bibfield  {author} {\bibinfo {author} {\bibfnamefont {D.}~\bibnamefont {Merritt}},\ }\href {https://doi.org/10.1093/mnras/214.1.25p} {\bibfield  {journal} {\bibinfo  {journal} {Monthly Notices of the Royal Astronomical Society}\ }\textbf {\bibinfo {volume} {214}},\ \bibinfo {pages} {25P} (\bibinfo {year} {1985}{\natexlab{b}})},\ \bibinfo {note} {aDS Bibcode: 1985MNRAS.214P..25M}\BibitemShut {NoStop}%
\bibitem [{\citenamefont {{Suyu}}\ and\ \citenamefont {{Halkola}}(2010)}]{GLEE}%
  \BibitemOpen
  \bibfield  {author} {\bibinfo {author} {\bibfnamefont {S.~H.}\ \bibnamefont {{Suyu}}}\ and\ \bibinfo {author} {\bibfnamefont {A.}~\bibnamefont {{Halkola}}},\ }\href {https://doi.org/10.1051/0004-6361/201015481} {\bibfield  {journal} {\bibinfo  {journal} {\aap}\ }\textbf {\bibinfo {volume} {524}},\ \bibinfo {eid} {A94} (\bibinfo {year} {2010})},\ \Eprint {https://arxiv.org/abs/1007.4815} {arXiv:1007.4815 [astro-ph.CO]} \BibitemShut {NoStop}%
\end{thebibliography}%

\appendix

\section{Parameter results}
\label{sec:param_results}

Best fit parameters are summarized in Table~\ref{tab:combined_final_parameters}.

\begin{table}[ht]
    \centering
    \begin{threeparttable}
    \caption{Weighted posterior parameter estimates for the three JWST-modeled quads.}
    \renewcommand{\arraystretch}{1.5}
    \begin{tabular}{l|ccc}
    \hline
    Parameter & WFI2033 & HE0435 & PG1115 \\
    \hline
    $\theta_{E,\mathrm{PEMD}}$ [$''$] & 0.987\lowup{0.003}{0.003} & 1.134\lowup{0.004}{0.003} & 0.948\lowup{0.014}{0.015} \\
    $\gamma_\mathrm{PEMD}$ & 1.85\lowup{0.02}{0.01} & 1.966\lowup{0.014}{0.012} & 2.19\lowup{0.02}{0.03} \\
    $q_\mathrm{PEMD}$ & 0.782\lowup{0.008}{0.010} & 0.830\lowup{0.003}{0.004} & 0.934\lowup{0.005}{0.002} \\
    $\phi_\mathrm{PEMD}$ [deg] & 31.1\lowup{0.8}{1.0} & 163.7\lowup{0.2}{0.2} & 109\lowup{4}{3} \\
    $\gamma_\mathrm{ext}$ & 0.086\lowup{0.003}{0.003} & 0.021\lowup{0.002}{0.001} & 0.11\lowup{0.01}{0.02} \\
    $\phi_\mathrm{ext}$ [deg] & 93\lowup{4}{5} & 86\lowup{10}{10} & 151\lowup{4}{3} \\
    \hline
    \multicolumn{4}{l}{Perturber Einstein radii $\theta_{E}$ [$''$]} \\
    & X: 0.101\lowup{0.007}{0.005} & G2: 0.41\lowup{0.01}{0.02} & G1: 1.5\lowup{0.3}{0.3} \\
    & G2: 0.73\lowup{0.02}{0.02} & G3: 0.164\lowup{0.005}{0.006} & G2: 0.38\lowup{0.08}{0.07} \\
    & G3: 0.104\lowup{0.003}{0.003} & G4: 0.227\lowup{0.007}{0.008} & -- \\
    & G7: 0.459\lowup{0.013}{0.012} & G5: 0.381\lowup{0.012}{0.014} & -- \\
    \hline
    \end{tabular}
    \begin{tablenotes}
    \item Reported values are medians with errors corresponding to the 16th and 84th percentiles. Perturber rows list each system's companion galaxies under their own modeling labels and carry no cross-system correspondence.
    \end{tablenotes}
    \label{tab:combined_final_parameters}
    \end{threeparttable}
\end{table}

\section{Systematics and Final Model Weights}

Systematic and final model weights are summarized in Tables~\ref{tab:model_weights_he0435} and~\ref{tab:model_weights_pg1115}.

\begin{table*}
\centering
\begin{threeparttable}
\caption{HE0435: final models sorted by total weight (kinematics weight $\times$ modeling/BIC weight).}
\label{tab:model_weights_he0435}
\begin{tabular}{cccccccc}
\toprule
Main Deflector & Source & Perturbers & PSF Type & Mask (asec) & $W_\mathrm{BIC}$ & $W_\mathrm{Kin}$ & Total Weight \\
\midrule
SPEMD + 2 SERSIC & 25 $n_\mathrm{max}$ + SERSIC & G1--G5 & STARRED & 2.1'' & 1.000 & 0.719 & 1.000 \\
SPEMD + 2 SERSIC & 23 $n_\mathrm{max}$ + SERSIC & G1--G5 & STARRED & 2.1'' & 0.271 & 0.994 & 0.349 \\
SPEMD + 2 SERSIC & 27 $n_\mathrm{max}$ + SERSIC & G1--G5 & STARRED & 2.1'' & 0.099 & 0.857 & 0.113 \\
SPEMD + 2 SERSIC & 25 $n_\mathrm{max}$ + SERSIC & G1--G5 & PSFr & 2.1'' & 0.000 & 0.764 & 0.000 \\
SPEMD + 2 SERSIC & 27 $n_\mathrm{max}$ + SERSIC & G1--G5 & PSFr & 2.1'' & 0.000 & 0.767 & 0.000 \\
SPEMD + 2 SERSIC & 23 $n_\mathrm{max}$ + SERSIC & G1--G5 & PSFr & 2.1'' & 0.000 & 0.896 & 0.000 \\
SPEMD + 2 SERSIC & 25 $n_\mathrm{max}$ + SERSIC & G1--G5 & STARRED & 2.2'' & 0.000 & 0.734 & 0.000 \\
SPEMD + 2 SERSIC & 23 $n_\mathrm{max}$ + SERSIC & G1--G5 & STARRED & 2.2'' & 0.000 & 0.960 & 0.000 \\
SPEMD + 2 SERSIC & 27 $n_\mathrm{max}$ + SERSIC & G1--G5 & STARRED & 2.2'' & 0.000 & 0.978 & 0.000 \\
SPEMD + 2 SERSIC & 25 $n_\mathrm{max}$ + SERSIC & G1--G5 & PSFr & 2.2'' & 0.000 & 0.684 & 0.000 \\
SPEMD + 2 SERSIC & 23 $n_\mathrm{max}$ + SERSIC & G1--G5 & PSFr & 2.2'' & 0.000 & 1.000 & 0.000 \\
SPEMD + 2 SERSIC & 27 $n_\mathrm{max}$ + SERSIC & G1--G5 & PSFr & 2.2'' & 0.000 & 0.948 & 0.000 \\
\bottomrule
\end{tabular}
\begin{tablenotes}
\item All weights normalized to the model with maximum weight. $W_\mathrm{BIC}$ is the BIC-derived modeling weight; $W_\mathrm{Kin}$ is the kinematics weight. See \Cref{subsec:comb_sys}.
\end{tablenotes}
\end{threeparttable}
\end{table*}

\begin{table*}
\centering
\begin{threeparttable}
\caption{PG1115: final models sorted by total weight (kinematics weight $\times$ modeling/BIC weight).}
\label{tab:model_weights_pg1115}
\begin{tabular}{cccccccc}
\toprule
Main Deflector & Source & Perturbers & PSF Type & Mask (asec) & $W_\mathrm{BIC}$ & $W_\mathrm{Kin}$ & Total Weight \\
\midrule
SPEMD + 2 SERSIC & SERSIC & G1+G2+NFW & PSFr & 2.4'' & 1.000 & 0.426 & 1.000 \\
SPEMD + 2 SERSIC + Gaussian & SERSIC & G1+G2+NFW & PSFr & 2.4'' & 0.403 & 0.104 & 0.101 \\
SPEMD + 2 SERSIC + Gaussian & SERSIC & G1+G2+NFW & PSFr & 2.3'' & 0.442 & 0.061 & 0.068 \\
SPEMD + 2 SERSIC & SERSIC & G1+G2+NFW & PSFr & 2.4'' & 0.007 & 0.074 & 0.002 \\
SPEMD + 2 SERSIC & SERSIC & G1+G2+NFW & STARRED & 2.3'' & 0.001 & 0.328 & 0.001 \\
SPEMD + 2 SERSIC + Gaussian & SERSIC & G1+G2+NFW & PSFr & 2.3'' & 0.006 & 0.000 & 0.000 \\
SPEMD + 2 SERSIC & SERSIC & G1+G2+NFW & PSFr & 2.3'' & 0.000 & 1.000 & 0.000 \\
SPEMD + 2 SERSIC & SERSIC & G1+G2+NFW & STARRED & 2.3'' & 0.000 & 0.759 & 0.000 \\
SPEMD + 2 SERSIC + Gaussian & SERSIC & G1+G2+NFW & STARRED & 2.4'' & 0.000 & 0.004 & 0.000 \\
SPEMD + 2 SERSIC + Gaussian & SERSIC & G1+G2+NFW & STARRED & 2.3'' & 0.000 & 0.000 & 0.000 \\
SPEMD + 2 SERSIC & SERSIC & G1+G2+NFW & STARRED & 2.4'' & 0.000 & 0.008 & 0.000 \\
SPEMD + 2 SERSIC + Gaussian & SERSIC & G1+G2+NFW & STARRED & 2.3'' & 0.000 & 0.000 & 0.000 \\
SPEMD + 2 SERSIC + Gaussian & SERSIC & G1+G2+NFW & STARRED & 2.4'' & 0.000 & 0.000 & 0.000 \\
SPEMD + 2 SERSIC + Gaussian & SERSIC & G1+G2+NFW & PSFr & 2.4'' & 0.000 & 0.000 & 0.000 \\
SPEMD + 2 SERSIC & SERSIC & G1+G2+NFW & STARRED & 2.4'' & 0.000 & 0.000 & 0.000 \\
\bottomrule
\end{tabular}
\begin{tablenotes}
\item Same as \cref{tab:model_weights_he0435}, but for PG1115.
\end{tablenotes}
\end{threeparttable}
\end{table*}

\section{Parameter Priors}

Model parameter priors are summarized in Tables~\ref{tab:priors_he0435} and~\ref{tab:priors_pg1115}.

\begin{table*}[htbp]
\centering
\begin{threeparttable}
\caption{HE0435: model parameter priors (some values rounded).}
\label{tab:priors_he0435}
\begin{tabular}{llllll}
\toprule
Object & Component & Parameter & Distribution & Initial Position & Step Size ($1\sigma$) \\
\midrule
Main Deflector (D) ($z=0.4546$) & Mass: PEMD & $\gamma$ & $U(1.5, 2.5)$ & 1.987 & 0.15 \\
 & Mass: PEMD & $\theta_\mathrm{E}$ & $N(0.9286, 0.04931)$ & 0.9286 & 0.04931 \\
 & Mass: PEMD & $e_1$ & $N(0, 0.2)$ & -0.07396 & 0.03 \\
 & Mass: PEMD & $e_2$ & $N(0, 0.2)$ & -0.04911 & 0.02 \\
 & Mass: PEMD & $(x, y)$ & $U(-3, 3)$ & (0.01714, 0.02465) & 0.004 \\
\midrule
External Shear ($z=0.4546$) & Mass: SHEAR & $\gamma_\mathrm{ext,1}$ & $U(-0.5, 0.5)$ & 0.02307 & 0.02 \\
 & Mass: SHEAR & $\gamma_\mathrm{ext,2}$ & $U(-0.5, 0.5)$ & 0.00582 & 0.01 \\
\midrule
Galaxy G1 ($z=0.7821$) & Mass: SIS & $\theta_\mathrm{E}$ & $N(0.2261, 0.04951)$\tnote{d} & scaled &  \\
\midrule
Galaxy G2 ($z=0.7806$) & Mass: SIS & $\theta_\mathrm{E}$ & $N(0.2644, 0.06075)$\tnote{d} & scaled &  \\
\midrule
Galaxy G3 ($z=0.4190$) & Mass: SIS & $\theta_\mathrm{E}$ & $N(0.1064, 0.03974)$\tnote{d} & scaled &  \\
\midrule
Galaxy G4 ($z=0.4568$) & Mass: SIS & $\theta_\mathrm{E}$ & $N(0.1469, 0.04897)$\tnote{d} & scaled &  \\
\midrule
Galaxy G5 ($z=0.7792$) & Mass: SIS & $\theta_\mathrm{E}$ & $N(0.2474, 0.08303)$\tnote{d} & scaled &  \\
\midrule
Main Deflector (D) ($z=0.4546$) & Light: Bulge S\'ersic & $R_\mathrm{Sersic}$ & $U(0.1, 2)$ & 1.069 & 0.2 \\
 & Light: Bulge S\'ersic & $n_\mathrm{Sersic}$ & $U(2.5, 6)$ & 4.923 & 0.15 \\
 & Light: Bulge S\'ersic & $e_1$ & $U(-0.2, 0.2)$ & -0.105 & 0.03 \\
 & Light: Bulge S\'ersic & $e_2$ & $U(-0.2, 0.2)$ & -0.02889 & 0.06 \\
 & Light: Bulge S\'ersic & $(x, y)$ & $U(-1, 1)$ & (0.02182, 0.01865) & 0.003 \\
 & Light: Disk S\'ersic & $R_\mathrm{Sersic}$ & $U(0.001, 5)$ & 1.924 & 0.15 \\
 & Light: Disk S\'ersic & $e_1$ & $U(-0.5, 0.5)$ & -0.1546 & 0.13 \\
 & Light: Disk S\'ersic & $e_2$ & $U(-0.5, 0.5)$ & -0.2687 & 0.2 \\
 & Light: Disk S\'ersic & $(x, y)$ & $U(-1, 1)$ & (0.02182, 0.01865) & 0.003 \\
\midrule
Quasar Host ($z=1.6930$) & Light: S\'ersic & $R_\mathrm{Sersic}$ & $U(0.05, 0.75)$ & 0.4513 & 0.1 \\
 & Light: S\'ersic & $n_\mathrm{Sersic}$ & $U(0.75, 6)$ & 3.743 & 1 \\
 & Light: S\'ersic & $e_1$ & $U(-0.3, 0.3)$ & 0.1977 & 0.08 \\
 & Light: S\'ersic & $e_2$ & $U(-0.3, 0.3)$ & 0.0789 & 0.05 \\
 & Light: S\'ersic & $(x, y)$ & $U(-2.604, 1.396)$ & (-0.6044, -0.5404) & 0.2 \\
 & Light: Shapelets & $\beta$ & $U(0.032, 0.3)$ & 0.07522 & 0.15 \\
\midrule
Quasar Images ($z=0.4546$) & Light: Point Sources & A $(x,y)$ & $U(\mathrm{Init} \pm 0.3)$ & (1.187, 0.5912) & 0.02 \\
 & Light: Point Sources & B $(x,y)$ & $U(\mathrm{Init} \pm 0.3)$ & (-0.2907, 1.144) & 0.02 \\
 & Light: Point Sources & C $(x,y)$ & $U(\mathrm{Init} \pm 0.3)$ & (-1.282, -0.01092) & 0.02 \\
 & Light: Point Sources & D $(x,y)$ & $U(\mathrm{Init} \pm 0.3)$ & (0.2511, -1.022) & 0.02 \\
 \midrule
Astrometric flexibility ($z=0.4546$) & Correction & $\delta x_\mathrm{image}$ & $N(0, 0.003)$, $\pm5\sigma$ & 0 & 0.003 \\
 & Correction & $\delta y_\mathrm{image}$ & $N(0, 0.003)$, $\pm5\sigma$ & 0 & 0.003 \\
\bottomrule
\end{tabular}
\begin{tablenotes}\footnotesize
\item[d] Einstein radius given a prior but jointly scaled with the other perturbers via a shared factor, passing the velocity-dispersion constraint while limiting degeneracies.
\end{tablenotes}
\end{threeparttable}
\end{table*}

\begin{table*}[htbp]
    \centering
    \begin{threeparttable}
    \caption{PG1115: model parameter priors (some values rounded).}
    \label{tab:priors_pg1115}
    \begin{tabular}{llllll}
    \toprule
    Object & Component & Parameter & Distribution & Initial Position & Step Size ($1\sigma$) \\
    \midrule
    Main Deflector (G) ($z=0.3098$) & Mass: PEMD & $\gamma$ & $U(1.5, 2.75)$ & 2 & 0.15 \\
     & Mass: PEMD & $\theta_\mathrm{E}$ & $N(1.191, 0.07065)$ & 1.191 & 0.07065 \\
     & Mass: PEMD & $e_1$ & $N(0, 0.2)$ & 0 & 0.03 \\
     & Mass: PEMD & $e_2$ & $N(0, 0.2)$ & 0 & 0.03 \\
     & Mass: PEMD & $(x, y)$ & $U(-0.5, 0.5)$ & (0, 0) & 0.01 \\
    \midrule
    External Shear ($z=0.3098$) & Mass: SHEAR & $\gamma_\mathrm{ext,1}$ & $U(-0.5, 0.5)$ & 0 & 0.02 \\
     & Mass: SHEAR & $\gamma_\mathrm{ext,2}$ & $U(-0.5, 0.5)$ & 0 & 0.02 \\
    \midrule
    Galaxy G1 ($z=0.3098$) & Mass: SIS & $\theta_\mathrm{E}$ & $N(1.401, 0.219)$\tnote{d} & scaled &  \\
    \midrule
    Galaxy G2 ($z=0.3123$) & Mass: SIS & $\theta_\mathrm{E}$ & $N(0.3605, 0.3327)$\tnote{d} & scaled &  \\
    \midrule
    Group Halo ($z=0.3098$) & Mass: NFW & $R_s$ & $N(38.19, 11.9)$ & 23.94 & 11.9 \\
     & Mass: NFW & $\alpha_{R_s}$ & $N(2.651, 0.926)$ & 2.651 & 0.5 \\
     & Mass: NFW & $(x, y)$ & $N(-7.986, 23.39)$ & (-44.33, -15.89) & 20 \\
    \midrule
    Main Deflector (G) ($z=0.3098$) & Light: Bulge S\'ersic & $R_\mathrm{Sersic}$ & $U(0.1, 2)$ & 0.5 & 0.2 \\
     & Light: Bulge S\'ersic & $n_\mathrm{Sersic}$ & $U(2.5, 6)$ & 4 & 0.2 \\
     & Light: Bulge S\'ersic & $e_1$ & $U(-0.5, 0.5)$ & 0 & 0.1 \\
     & Light: Bulge S\'ersic & $e_2$ & $U(-0.5, 0.5)$ & 0 & 0.1 \\
     & Light: Bulge S\'ersic & $(x, y)$ & $U(-0.5, 0.5)$ & (0, 0) & 0.003 \\
     & Light: Disk S\'ersic & $R_\mathrm{Sersic}$ & $U(0.5, 5)$ & 1 & 0.2 \\
     & Light: Disk S\'ersic & $n_\mathrm{Sersic}$ & $U(0.5, 2.5)$ & 1 & 0.2 \\
     & Light: Disk S\'ersic & $e_1$ & $U(-0.5, 0.5)$ & 0 & 0.1 \\
     & Light: Disk S\'ersic & $e_2$ & $U(-0.5, 0.5)$ & 0 & 0.1 \\
     & Light: Disk S\'ersic & $(x, y)$ & $U(-0.5, 0.5)$ & (0, 0) & 0.003 \\
    \midrule
    Quasar Host ($z=1.7270$) & Light: S\'ersic & $R_\mathrm{Sersic}$ & $U(0.05, 2)$ & 0.5 & 0.1 \\
     & Light: S\'ersic & $n_\mathrm{Sersic}$ & $U(0.75, 6)$ & 4 & 1 \\
     & Light: S\'ersic & $e_1$ & $U(-0.3, 0.3)$ & 0 & 0.1 \\
     & Light: S\'ersic & $e_2$ & $U(-0.3, 0.3)$ & 0 & 0.1 \\
     & Light: S\'ersic & $(x, y)$ & $U(-2, 2)$ & (0, 0) & 0.2 \\
     & Light: Shapelets & $\beta$ & $U(0.032, 0.3)$ & 0.1 & 0.05 \\
    \midrule
    Quasar Images ($z=0.3098$) & Light: Point Sources & A1 $(x,y)$ & $U(\mathrm{Init} \pm 0.3)$ & (0.9684, -0.6917) & 0.02 \\
     & Light: Point Sources & A2 $(x,y)$ & $U(\mathrm{Init} \pm 0.3)$ & (1.118, -0.2278) & 0.02 \\
     & Light: Point Sources & B $(x,y)$ & $U(\mathrm{Init} \pm 0.3)$ & (-0.6958, -0.6207) & 0.02 \\
     & Light: Point Sources & C $(x,y)$ & $U(\mathrm{Init} \pm 0.3)$ & (-0.3474, 1.345) & 0.02 \\
    \midrule
    Astrometric flexibility ($z=0.3098$) & Correction & $\delta x_\mathrm{image}$ & $N(0, 0.003)$, $\pm5\sigma$ & 0 & 0.003 \\
     & Correction & $\delta y_\mathrm{image}$ & $N(0, 0.003)$, $\pm5\sigma$ & 0 & 0.003 \\
    \bottomrule
    \end{tabular}
    \begin{tablenotes}\footnotesize
    \item[d] Einstein radius given a prior but jointly scaled with the other perturbers via a shared factor, passing the velocity-dispersion constraint while limiting degeneracies.
    \end{tablenotes}
    \end{threeparttable}
\end{table*}

\section{STARRED versus PSFr comparison}
\label{sec:starred_vs_psfr_corner}

\begin{figure*}[htbp!]
	\centering
	\includegraphics[width=\textwidth]{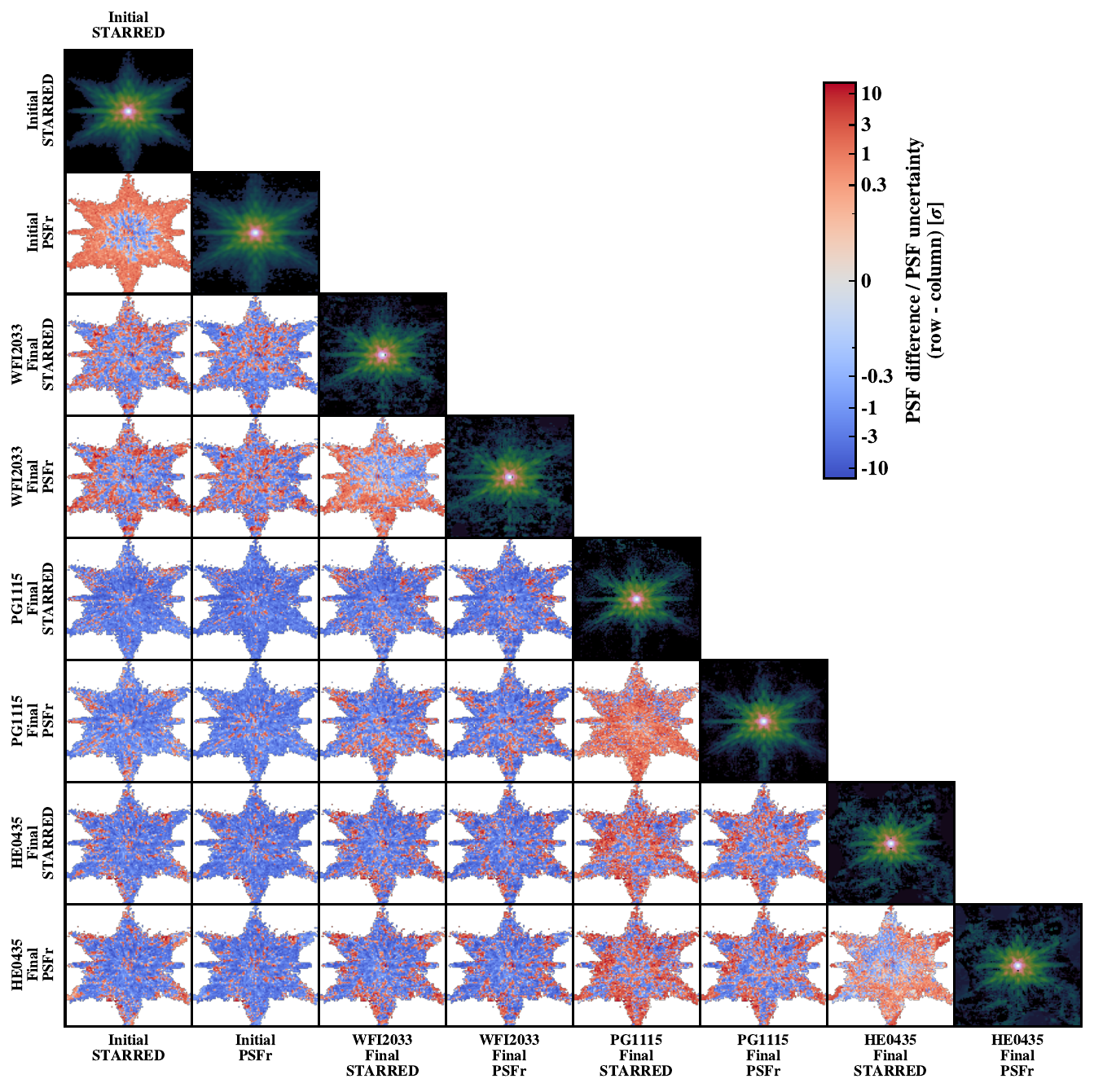}
	\caption{Comparison between the initial PSFs generated from stars in the field of WFI2033 and the final, iterated PSFs from the best model from each system for each initial PSF type (STARRED and PSFr). The figure is in units of uncertainty, empirically estimated by fitting stars in the field and comparing residuals.}
	\label{fig:psf_comparison}
\end{figure*}

\begin{figure*}[htbp!]
	\centering
	\includegraphics[width=\textwidth]{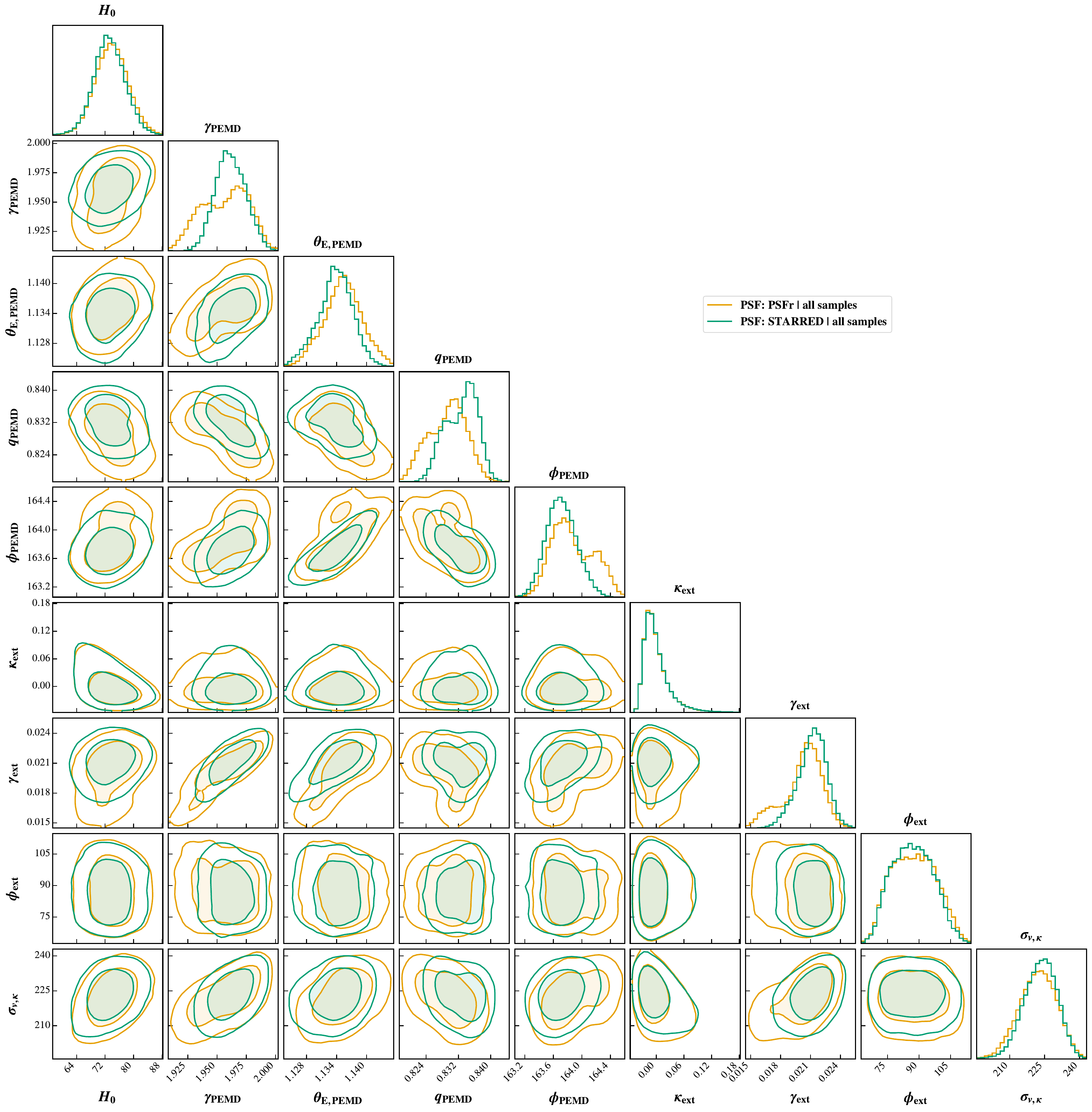}
	\caption{Same as \cref{fig:he0435_corner}, except comparing HE0435 models using an initial STARRED PSF (in green) vs PSFr PSF (yellow).}
	\label{fig:he0435_corner_psf}
\end{figure*}

Unlike WFI2033, which found a minor discrepancy between the STARRED and PSFr models, HE0435 finds the two PSF results entirely consistent (\Cref{fig:he0435_corner_psf}). However, we reiterate that the models using STARRED fit the imaging data so well that the corresponding models using PSFr were assigned zero imaging weight in comparison.

\cref{fig:he0435_corner_psf} shows corner plots of the \helens parameters, comparing models using PSFr vs STARRED to generate their initial PSF models. The two samples are assigned weights separately.

\begin{figure*}[htbp!]
	\centering
	\includegraphics[width=\textwidth]{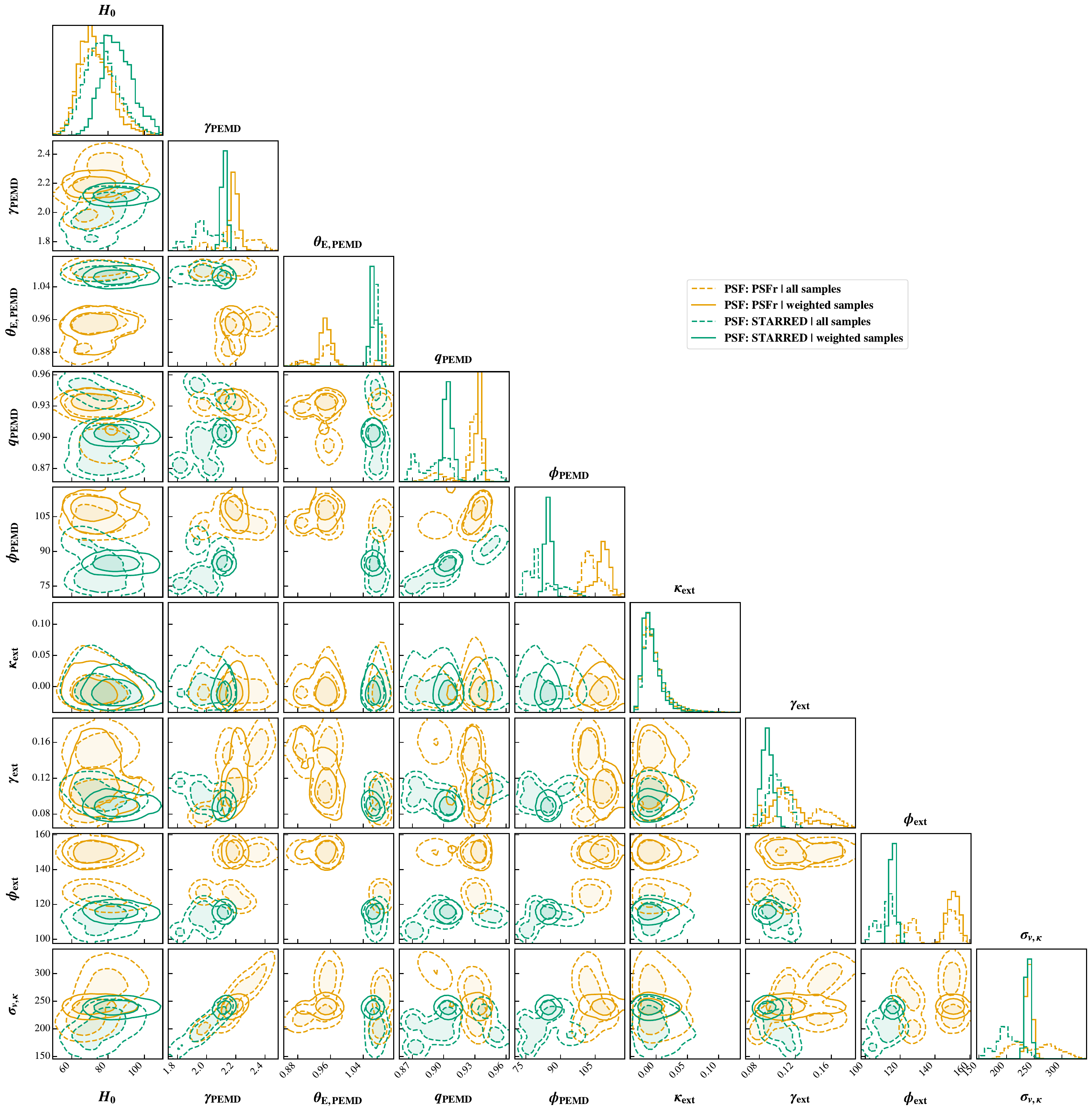}
	\caption{Same as \cref{fig:he0435_corner_psf} but for PG1115. We note model weighting here was dominated by predicted velocity dispersions, which leads to the slice in $\sigma_{v,\kappa}$ (model-predicted velocity dispersion, taking into account the external convergence). This constraint will be relaxed by fitting for $\lint$ in our TDCOSMO 2026 milestone paper, allowing models to effectively vary their velocity dispersion via an internal mass sheet.}
	\label{fig:pg1115_corner_psf}
\end{figure*}

\cref{fig:pg1115_corner_psf} shows the same comparison for PG1115. This is the first system for which the PSFr model is preferred, and we find disagreement in perturber constraints (namely in the group halo scale radius and concentration parameters), along with a slight disagreement in the Einstein radius.

\section{Source complexity trends}
\label{sec:source_complex}

\begin{figure*}[htbp!]
	\centering
	\includegraphics[width=\textwidth]{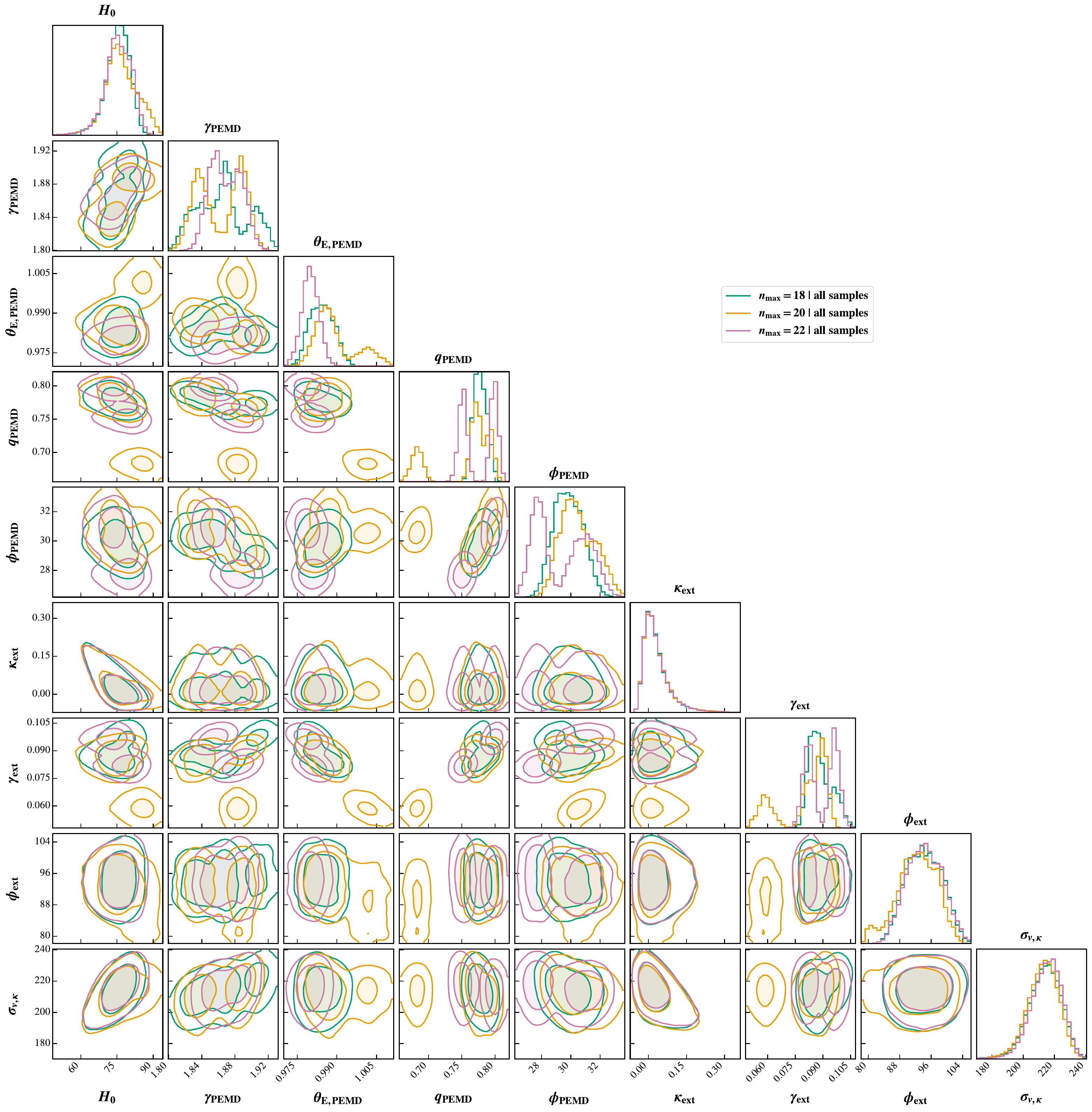}
	\caption{Same as \cref{fig:he0435_corner}, except comparing WFI2033 models by model complexity, with the least complex (maximum shapelet order $n_{\mathrm{max}}=18$) in green, middling complexity in yellow ($n_{\mathrm{max}}=20$), and most complex in red ($n_{\mathrm{max}}=22$). Trends are most noticeable in the Einstein radius, power-law slope, and axis ratio, but have no impact on H$_0$.}
	\label{fig:wfi2033_corner_shapelets}
    \end{figure*}

\begin{figure*}[htbp!]
	\centering
	\includegraphics[width=\textwidth]{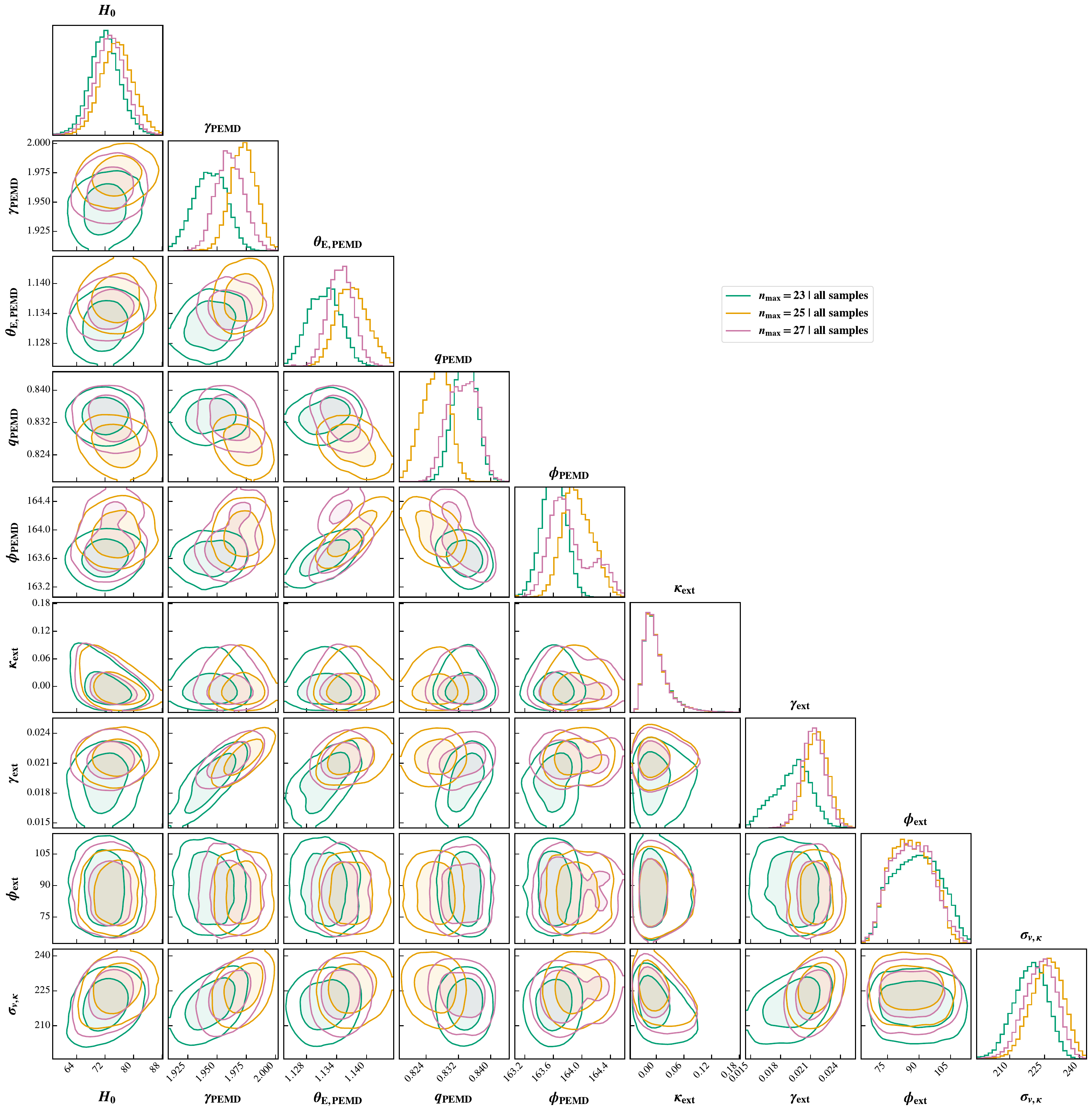}
	\caption{Same as \cref{fig:wfi2033_corner_shapelets}, except for HE0435 models. Discrepancies are noticeable in the same parameters, but once again remaining consistent within H$_0$.}
	\label{fig:he0435_corner_shapelets}
    \end{figure*}

Source complexity trends are summarized in Figures~\ref{fig:wfi2033_corner_shapelets} and \ref{fig:he0435_corner_shapelets}.

\section{How model-predicted external shear absorbs internal mass structure}
\label{sec:appendix_shear_interpretation}

Here we provide an example explanation for why the external shear parameter in standard lens models (e.g.\ PEMD+$\gamma_{\rm ext}$) does not, in general, correspond to the true tidal shear from the environment, but instead can absorb mismatches between the assumed and true mass distributions.

We begin by expanding the lensing potential in polar coordinates,
\begin{equation}
    \psi(r,\theta)
    =
    \psi_0(r)
    +
    \sum_{m=1}^{\infty}
    \psi_m(r)\cos\!\left[m(\theta-\theta_m)\right],
\end{equation}
where $\psi_0(r)$ is the azimuthally averaged component and $\psi_m(r)$ describes angular structure of order $m$. The $m=2$ term corresponds to the quadrupole, which captures both internal ellipticity and external tidal shear.

External shear contributes a pure quadrupole term of the form
\begin{equation}
    \psi_{\rm shear}(r,\theta)
    =
    \frac{\gamma_{\rm ext}}{2}
    r^2
    \cos\!\left[2(\theta-\theta_\gamma)\right].
\end{equation}

Strong-lensing constraints are localized near the Einstein radius $R_{\rm Ein}$. Writing
\begin{equation}
    r = R_{\rm Ein} + \delta r,
\end{equation}
we expand the true quadrupole term as
\begin{equation}
    \psi_{2,\rm true}(r)
    \approx
    \psi_{2,\rm true}(R_{\rm Ein})
    +
    \psi'_{2,\rm true}(R_{\rm Ein})\,\delta r
    +
    \frac{1}{2}\psi''_{2,\rm true}(R_{\rm Ein})\,\delta r^2.
\end{equation}

Similarly, the quadrupole allowed by the model (e.g.\ from an elliptical power-law profile) has a restricted radial dependence,
\begin{equation}
    \psi_{2,\rm model}(r)
    \approx
    \psi_{2,\rm model}(R_{\rm Ein})
    +
    \psi'_{2,\rm model}(R_{\rm Ein})\,\delta r + \cdots.
\end{equation}

The mismatch between the true and modeled quadrupole is then
\begin{equation}
    \Delta\psi_2(r)
    =
    \psi_{2,\rm true}(r) - \psi_{2,\rm model}(r),
\end{equation}
which, near $R_{\rm Ein}$, can be approximated as a low-order polynomial in $\delta r$. Importantly, strong-lensing observables probe only a narrow radial range around  $R_{\rm Ein}$, so only the first few terms in this expansion are constrained by the data.

Expanding the external shear potential around $R_{\rm Ein}$ gives
\begin{equation}
    \psi_{\rm shear}
    =
    \frac{\gamma_{\rm ext}}{2}
    \left(
        R_{\rm Ein}^2
        + 2 R_{\rm Ein}\,\delta r
        + \delta r^2
    \right)
    \cos 2(\theta-\theta_\gamma).
\end{equation}
This has the same functional form: a constant, linear, and quadratic dependence on $\delta r$ multiplying a quadrupole angular term.

Therefore, by adjusting $\gamma_{\rm ext}$ (and its orientation), the shear term can reproduce the leading-order behavior of any smooth quadrupole mismatch $\Delta \psi_2(r)$ within the narrow annulus probed by the images. This reproduction is indifferent to the cause, be it radial variations in ellipticity, deviations from power-law profiles, or projections of higher-order multipoles into effective quadrupoles near $R_{\rm Ein}$.

\section{Outdated Results}
\label{sec:appendix_outdated}

\begin{figure*}[ht]
		\centering
		\includegraphics[width=\textwidth]{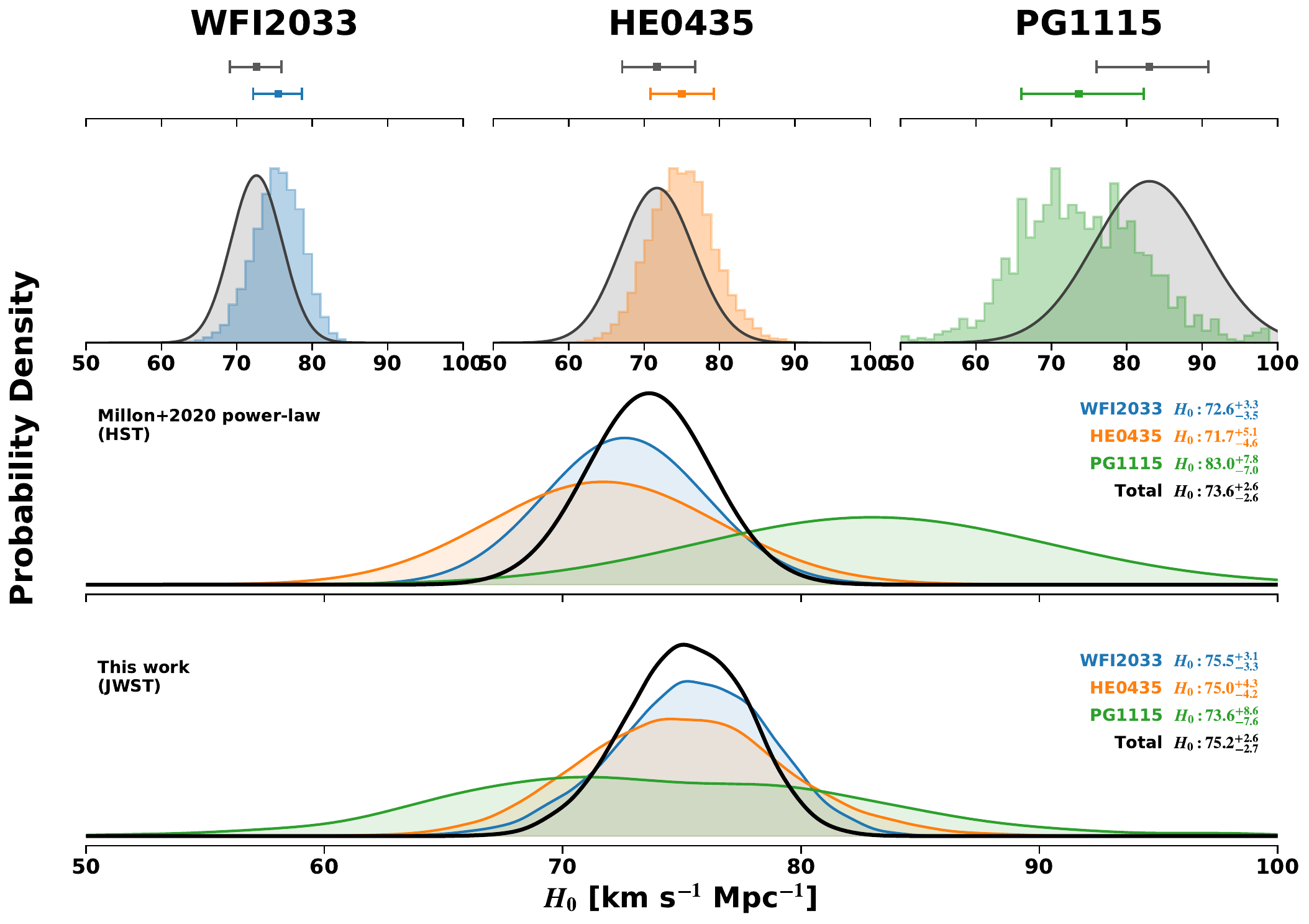}
		\caption{Same as \Cref{fig:H0_update_comparison}; however, these results use the original unblinded values, obtained before a bug was identified in auxiliary data not produced as part of this analysis—specifically, the aperture-integrated stellar velocity dispersion of the main lens.}
		\label{fig:H0_update_comparison_wrong}
	\end{figure*}

For completeness and transparency, \Cref{fig:H0_update_comparison_wrong} is the version of Figure~\ref{fig:H0_update_comparison} that was obtained after unblinding and prior to correction of the coding error in the auxiliary stellar velocity dispersion measurements.

\end{document}